\pdfoutput=1
\documentclass[aps,prb,superscriptaddress,preprintnumbers,
showpacs,a4,twoside,twocolumn]{revtex4}

\usepackage{bm}
\usepackage{graphicx}
\usepackage{amsfonts}
\usepackage{amsmath}
\usepackage{amssymb}

\begin{document}

\title{Bosons in one-dimensional incommensurate superlattices}

\author{Tommaso Roscilde}
\affiliation{Max-Planck-Institut f\"ur Quantenoptik, Hans-Kopfermann-strasse 1,
85748 Garching, Germany}

\begin{abstract}
We investigate numerically the zero-temperature physics of 
the one-dimensional Bose-Hubbard model 
in an incommensurate cosine potential, recently realized in experiments
with cold
bosons in optical superlattices [L. Fallani \emph{et al.},
Phys. Rev. Lett. {\bf 98}, 130404, (2007)]. 
An incommensurate cosine potential has intermediate properties between
a truly periodic and a fully random potential, displaying  
a characteristic length scale (the quasi-period) 
which is shown to set a finite lower bound to 
the excitation energy of the system at special incommensurate fillings. 
This leads to the emergence of gapped incommensurate band-insulator (IBI) 
phases along with gapless Bose-glass (BG) phases for strong quasi-periodic
potential, both for hardcore and softcore bosons. 
Enriching the spatial features of the potential
by the addition of a second incommensurate component appears
to remove the IBI regions, stabilizing a continuous BG 
phase over an extended parameter range.
Moreover we discuss the validity of the local-density approximation
in presence of a parabolic trap, clarifying the notion of 
a \emph{local} BG phase in a trapped system; we investigate
the behavior of first- and second-order coherence upon increasing
the strength of the quasi-periodic potential; and we discuss 
the \emph{ab-initio} derivation of the Bose-Hubbard Hamiltonian with
quasi-periodic potential starting from the microscopic Hamiltonian
of bosons in an incommensurate superlattice. 
\end{abstract}
\maketitle

 Localization effects in random potentials represent a striking 
manifestation of the wavelike nature of quantum particles,
and they are particularly pronounced in dimensions less than three,
where, in absence of interaction, the suppression of transport 
due to disorder occurs at all energy scales \cite{Thouless74,KramerMK93}.
A particularly intriguing aspect of 
localization phenomena is their subtle interplay with 
particle-particle interaction. In the following we will
specify our discussion to \emph{bosons}.  The 
screening effect of weak interactions can lead to a transition
from an localized (Anderson) insulator to a superfluid, 
namely to the persistence of superfluidity 
in a disordered potential \cite{LeeG90,GiamarchiS88}. Systems on a commensurately
filled lattice without disorder exhibit a gapped Mott-insulating (MI)
phase for strong interactions, competing with a superfluid (SF) phase
as the interaction is reduced; in presence of disorder, the gapless
\emph{Bose-glass} (BG) insulator competes with the MI
when the disorder strength becomes comparable with the Mott
gap \cite{GiamarchiS88,Fisheretal89}.
 
 Recent experiments on trapped cold atoms have demonstrated
the ability of loading an atomic Bose gas in a well controlled
disordered potential \cite{Lyeetal05, Clementetal05, Fortetal05,
Schulteetal05,  Fallanietal07, Lyeetal07}. 
In particular, the regime of strong repulsion
in a strong \emph{pseudo}-disordered potential has been  
achieved by loading cold bosons in a deep one-dimensional 
\emph{quasi-periodic} (QP) 
optical superlattice, formed by two standing waves with incommensurate 
wavelenghts \cite{Fallanietal07}. In the insulating phase of
the system, Bragg spectroscopy reveals an excitation spectrum which 
strongly differs from that of a MI, showing that
the QP potential introduces many new states in the spectral Mott
gaps, caused by the strong repulsion. This
feature indirectly suggests that the insulating phase measured in
the experiments could be a BG; yet, direct
experimental inspection in the low-energy spectrum, which
is decisive to distinguish a BG insulator from a
more conventional MI, is currently not available. 

On the theoretical side, the study of  
 tight-binding models in QP potentials has a long history
\cite{Hofstadter76,AubryA79,Kohmotoetal83, Ostlundetal83,
Sokoloff85}, 
connected to the fundamental question of the fate
of Anderson localization in quasi-random potentials. 
Although Bloch's theorem does not apply to QP potentials,
for one-dimensional systems it was soon realized that the 
single-particle spectrum of the system is composed of bands,
made of either
\emph{all} extended or \emph{all} localized states depending on the 
potential strength. Therefore, filling such a system with spinless
fermions leads to the appearence of incommensurate band-insulator
(IBI) phases alternating with metallic phases for weak QP potential,
and with Anderson-insulating ones for strong QP potential.
Remarkably, the very same picture can be exactly recovered for
\emph{bosons} in the hardcore repulsive limit 
\cite{Reyetal06, ScarolaD06}, with the following
correspondence between fermionic and bosonic phases:
``metallic" $\to$ ``superfluid"  and ``Anderson insulator" $\to$ 
``Bose glass". 

  In this paper we provide a systematic study of one-dimensional
bosons in a QP potential. We explicitly map out
the phase diagram of the system in the hardcore limit
through exact diagonalization,
reconstructing the tight alternation of the SF, IBI and BG phases
upon changing the chemical potential. 
Making use of quantum Monte Carlo we can then
move on to the \emph{softcore} case which is most appropriate
for the experiments \cite{Fallanietal07}. There we find that
a strong QP potential, equalling in strength the inter-particle
repulsion, completely removes the MI phase, 
leaving space for various insulating phases at fractional fillings: 
similarly to the hardcore case,
slight changes of the model parameters (chemical potential
and hopping amplitude) can drive the system
from BG to IBI in a tight alternation.
Therefore pseudo-disorder created by a single
incommensurate potential component leads to a significantly different 
phase diagram than in the case of a truly random potential
\cite{Scalettaretal91,Rapschetal99,Prokofevetal98}, where
a continuous BG appears, becoming the only insulating
phase for strong enough disorder. The fundamental differences between 
a truly random potential and a QP one are explained quantitatively
within a generalized atomic-limit approximation, which
captures how the statistics of spatial structures
in the potential reflects itself in the spectral properties. On the 
basis of this observation, we extend the degrees
of freedom of the QP potential beyond what has
been so far explored in experiments \cite{Fallanietal07},
in search for the minimal setup which would lead to
a similar behavior to the case of a truly random 
potential. We find that the addition of a further
incommensurate component, leading to a 2-color QP potential,  
causes the removal of the IBI phases, 
and hence to the appearence of a BG persisting 
over an extended filling range. 

 Moreover we discuss the behavior of the system in presence
of a parabolic trap, mimicking the experimental situation
encountered in current optical-lattice setups. The local-density
approximation (LDA), generally successful in the
absence of a QP potential,
is critically discussed in its presence. On the one hand
the density profile and local particle-number fluctuations are 
correctly predicted by LDA; on the other hand the identification
of the local behavior in the trapped system with that of the bulk
system at the same chemical potential does not generally hold, due 
to the severeness of the finite-size effects imposed by the trap
in presence of the QP potential. At the same time we show that 
a 2-color QP potential can realize a uniform BG phase in the
trap, on average over the fluctuations of the spatial phases
of the potential components. Finally we make a closer connection
to optical lattice experiments by investigating behavior of the currently 
accessible observables, which are shown to reveal the full evolution of phase and 
density correlations in the system upon changing the strength of 
the QP potential, and we discuss in details the ab-initio
derivation of the one-dimensional Bose-Hubbard Hamiltionian in the
case of an incommensurate optical superlattice. 

 This paper is structured as follows. Section \ref{sec.model}
 introduces the Bose-Hubbard model in an incommensurate
 cosine potential, and the numerical
 methods used in the paper; Section \ref{sec.hardcore} discusses
 the phase diagram in the hardcore limit; 
 the phase diagram of the softcore model is then presented
 in Section \ref{sec.softcore}; Section \ref{sec.truepseudo}
 discusses the emergence of the IBI in 
 presence of the QP potential; Section \ref{sec.LDA} focuses
 on the local-density approximation in the trapped system; 
 Section \ref{sec.observables} shows results for the experimental
 observables; and  Section \ref{sec.abinitio} presents the
 ab-initio derivation
 of the Bose-Hubbard Hamiltonian. Summary and conclusions
 are presented in Section \ref{sec.summary}.

\section{Model and methods}
\label{sec.model}

 We investigate the one-dimensional Bose-Hubbard model in an incommensurate
 cosine potential, with Hamiltonian
 
 \begin{eqnarray}
 {\cal H}_0 = \sum_{i=1}^{L}&& \Big\{-J \left(b_i b_{i+1}^{\dagger} + {\rm h.c.}\right) 
 +\frac{U}{2} n_i (n_i-1) \nonumber \\
 && + V_2~g(i;\alpha,\phi)~ n_i - \mu n_i\Big\}
 \label{e.hamiltonian}
 \end{eqnarray} 
 with 
 \begin{equation}
 g(i;\alpha,\phi) = \cos^2(\pi\alpha~i+\phi)-\frac{1}{2} = \frac{\cos(2\pi\alpha~i+2\phi)}{2} ,
 \label{e.pot}
 \end{equation}
where $b_i$, $b_i^{\dagger}$ are bosonic operators, $J$ is the 
hopping amplitude, $U$ is the on-site repulsion, 
$V_2$ is the strength of the QP potential, and $\mu$ the 
chemical potential. The incommensuration
parameter $0<\alpha<1$ is in principle an \emph{irrational} number;
making use of periodic boundary conditions, it is appropriate to
take $\alpha = N_{\rm cycles}/L$ (with non-integer $L/N_{\rm cycles}$) 
so that the incommensurate cosine potential has a period     
exactly equal to the lattice size $L$. 
 
 The realization of the Bose-Hubbard model by loading cold
 bosons in the lowest band of an optical lattice \cite{Jakschetal98} 
 has been demonstrated by now in a large variety of experiments
 \cite{Greineretal02,Stoeferleetal04,Spielmanetal07}. More recently
 Fallani \emph{et al.} \cite{Fallanietal07} have been able to 
 add a QP potential to a one-dimensional optical lattice
 by application of a secondary standing wave, with an incommensurate
 wavelength $\lambda_2$ with respect to that ($\lambda_1$) of the 
 primary lattice. If the intensity of the secondary wave is 
 significantly weaker than that of the primary lattice, the 
 main effect of the  $\lambda_2$-lattice is to 
 modulate the depth of the potential wells of the primary 
 $\lambda_1$-lattice, as modeled by the $V_2$-term in Eq. \eqref{e.hamiltonian}
 (see Section \ref{sec.abinitio} for a thorough discussion of this
 point). We choose the representation
of the incommensurate cosine potential as in Eq.~\eqref{e.pot} 
such that the potential strength $V_2$ reflects directly the 
intensity of a secondary standing wave 
at wavelength $\lambda_2 = \alpha \lambda_1$: given that 
the first, dominant standing wave ``discretizes" the 
space to points $x_i = (\lambda_1/2)~i$, the secondary standing 
wave creates a potential on such points which is proportional to
$\cos^2[(2\pi/\lambda_1+\phi) x_i] = \cos^2(\alpha\pi i+\phi)$ as in 
Eq.~\eqref{e.pot}. The wavelength relation defines the incommensuration
 parameter, $\alpha = \lambda_1/\lambda_2 = 830.7/1076.8 = 0.77137...$~.
 In the following we will use different rational approximants thereof
 ($\alpha \approx 97/126$, $830/1076$).
 
  A fundamental limit of the Bose-Hubbard model 
  is the hardcore case $U \to \infty$, in which double occupancy
  is strictly forbidden, and which has been experimentally
  demonstrated in Ref.~\onlinecite{Paredesetal04}.  
  In the hardcore limit the constraint of 
  forbidden double occupancy can be directly incorporated
  in the operator algebra by introducing on-site 
  anticommutation rules for the \emph{hardcore-boson} operators, 
  $\{a_i, a_i^{\dagger}\} = 1$, 
  $\{a_i^{(\dagger)}, a_i^{(\dagger)}\} = 0$, coexisting
  with ordinary bosonic commutation rules between different sites. 
  Hardcore bosons can be exactly mapped onto spinless fermions 
  $f_i$, $f_i^{\dagger}$ through the Jordan-Wigner transformation
  \cite{LSM61}:
  
\begin{equation}
\label{Jordan-Wigner transformation}
a_i^\dagger=f_i^\dagger\prod_{k=1}^{i-1}e^{-i\pi f_k^\dagger f_k}, 
\hspace{0.5 cm} a_i=\prod_{k=1}^{i-1}e^{i\pi f_k^\dagger f_k}f_i, 
\end{equation}
  
   so that the resulting Hamiltonian is that of free fermions
  in a potential:
   
   \begin{eqnarray}
 {\cal H} &=& \sum_{i=1}^{L-1} \Big[-J \left(f_i f_{i+1}^{\dagger} + 
 {\rm h.c.}\right) 
 + V_2~g(i;\alpha,\phi)~ m_i - \mu m_i\Big] \nonumber \\
 && ~~~~~~ - J \left(e^{i\theta} f_L f_{1}^{\dagger}+{\rm h.c.} \right).
 \label{e.fermiham}
 \end{eqnarray} 
  
  The boundary phase term resulting from the non-local nature
  of the Jordan-Wigner transformation reads 
  $\theta = [(N + 1) \pm 1] \pi$, where $N$ is the particle
  number and the sign $+$ ($-$) applies to periodic (antiperiodic)
  boundary conditions.   
    
    Once reduced to spinless fermions, the hardcore boson problem
   is exactly solvable through simple single-particle 
   diagonalization. In the following we will focus our attention
   on two fundamental quantities, namely the 
   \emph{superfluid density}, estimated through the 
   energy difference between periodic $(+)$ and antiperiodic $(-)$ 
   boundary conditions,   
   \begin{equation}
    \rho_s = \frac{L}{J\pi ^2}\left(E^{(-)}-E^{(+)}\right),
   \label{e.rhos}
   \end{equation}
   and the compressibility $\kappa = dn/d\mu$, where
   $n=N/L$ is the particle density.

   In the softcore case of the original Hamiltonian 
   Eq.~\eqref{e.hamiltonian} an exact solution is not
  available. In this case we make use of 
  quantum Monte Carlo (QMC),
  based on the Stochastic Series Expansion formulation
  and on the directed-loop
  algorithm \cite{SSE}. The simulations
  are systematically done at temperatures low enough
  (typically $T\sim J/L$) so as to 
  remove significant thermal effects. The truncation of the local 
  Hilbert space up to a maximum bosonic occupancy $n_{\rm max}$
  is accurately done to avoid significant truncation errors
  (we went up to $n_{\rm max}=6$ for the largest
  fillings considered, $n\approx 3$). When averaging over different
  realizations of the QP potential, we have considered 
  typical samples of $100-300$ realizations. 
  The QMC method we use is
  formulated in the Fock-state basis, so that all quantities
  which are diagonal in that basis 
 (density, compressibility,
  etc.) are straightforwardly evaluated. The superfluid density
  is estimated through the fluctuations of the winding number
   $W$ in the worldline configurations produced during the simulation
   \cite{PollockC87},
  $\rho_s = \langle W^2 \rangle L/(2\beta J)$.  Finally we 
  calculate the momentum distribution
  \begin{equation}
  n(k) = \frac{1}{L} \sum_{lm} e^{iq(l-m)}
  \langle b_l^{\dagger} b_m \rangle  
  \label{e.nk}
  \end{equation}
  obtained through the statistics of the directed-loop update
  \cite{Wesseletal04}.  
   Most of the QMC results of the paper are based on a standard
  \emph{grand-canonical} algorithm, already well documented
  in the literature \cite{SSE,Wesseletal04}. In Sections
  \ref{sec.LDA}, \ref{sec.observables} and \ref{sec.abinitio} we also 
  present results obtained with
  a novel \emph{canonical} algorithm with multiple directed 
  loops, designed for the calculation of higher-order off-diagonal
  correlators. The description of the algorithm is beyond the 
  scope of the present paper, and we postpone it to a forthcoming
  publication.

\begin{figure}[h]
\begin{center}
\includegraphics[
     width=80mm,angle=270]{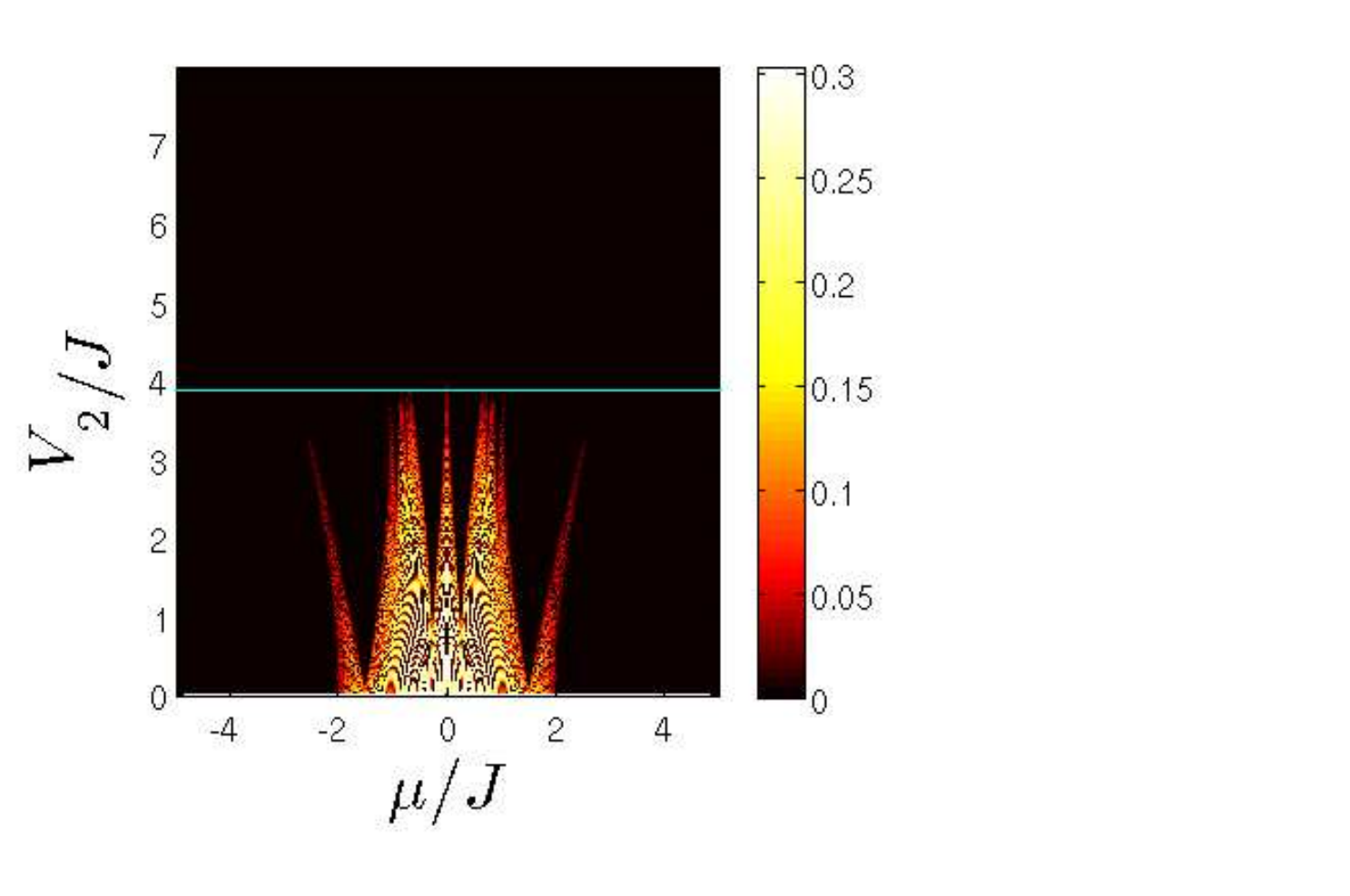} 
 \includegraphics[
     width=82mm,angle=270]{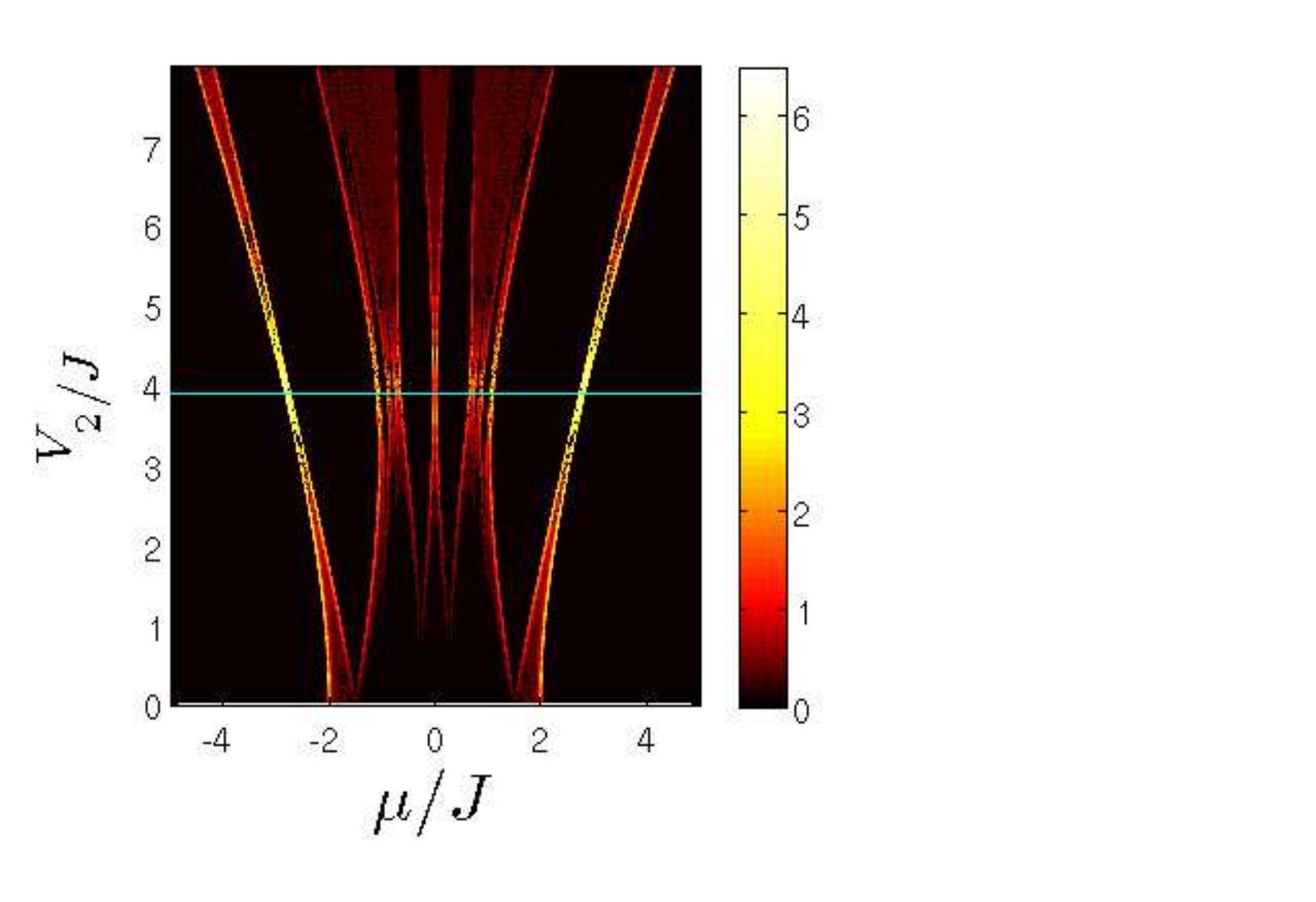}    
\caption{(Color online) Superfluid density (upper panel) and 
compressibility (lower panel)
for hardcore bosons in a QP superlattice;
here the system size is $L=1076$ and the incommensuration
parameter is $\alpha=830/1076$. The cyan horizontal
line marks the critical value $V_2/J=4$ for the localization
transition. The ``zebra''-like features in the
superfluid density are finite-size effects.}
\label{f.hardcore}
\end{center}
\end{figure}

\section{Hardcore limit}
\label{sec.hardcore}

  In this section we present exact results for the hardcore 
case. The study of hardcore bosons in QP potentials has been 
recently initiated by Refs.~\onlinecite{Reyetal06,ScarolaD06}. 
Remarkably the exact mapping onto spinless fermions, 
Eq.~\eqref{e.fermiham}, allows one to exploit the large body
of results produced several years ago in the context
of tight-binding models in a QP potential \cite{Sokoloff85}. 
 In particular, focusing on localization phenomena, 
a fundamental result is the so-called Aubry-Andr\'e conjecture 
\cite{AubryA79, Sokoloff85} stating that for $V_2 < 4J$ 
all single-particle states are extended, while for 
$V_2 > 4J$ all states are localized. The single-particle
spectrum is organized in \emph{bands}, which have 
a fractal support\cite{Hofstadter76} (Cantor set) at the 
critical value $V_2 = 4J$.

 \begin{figure}[h]
\begin{center}
\includegraphics[
     width=50mm,angle=270]{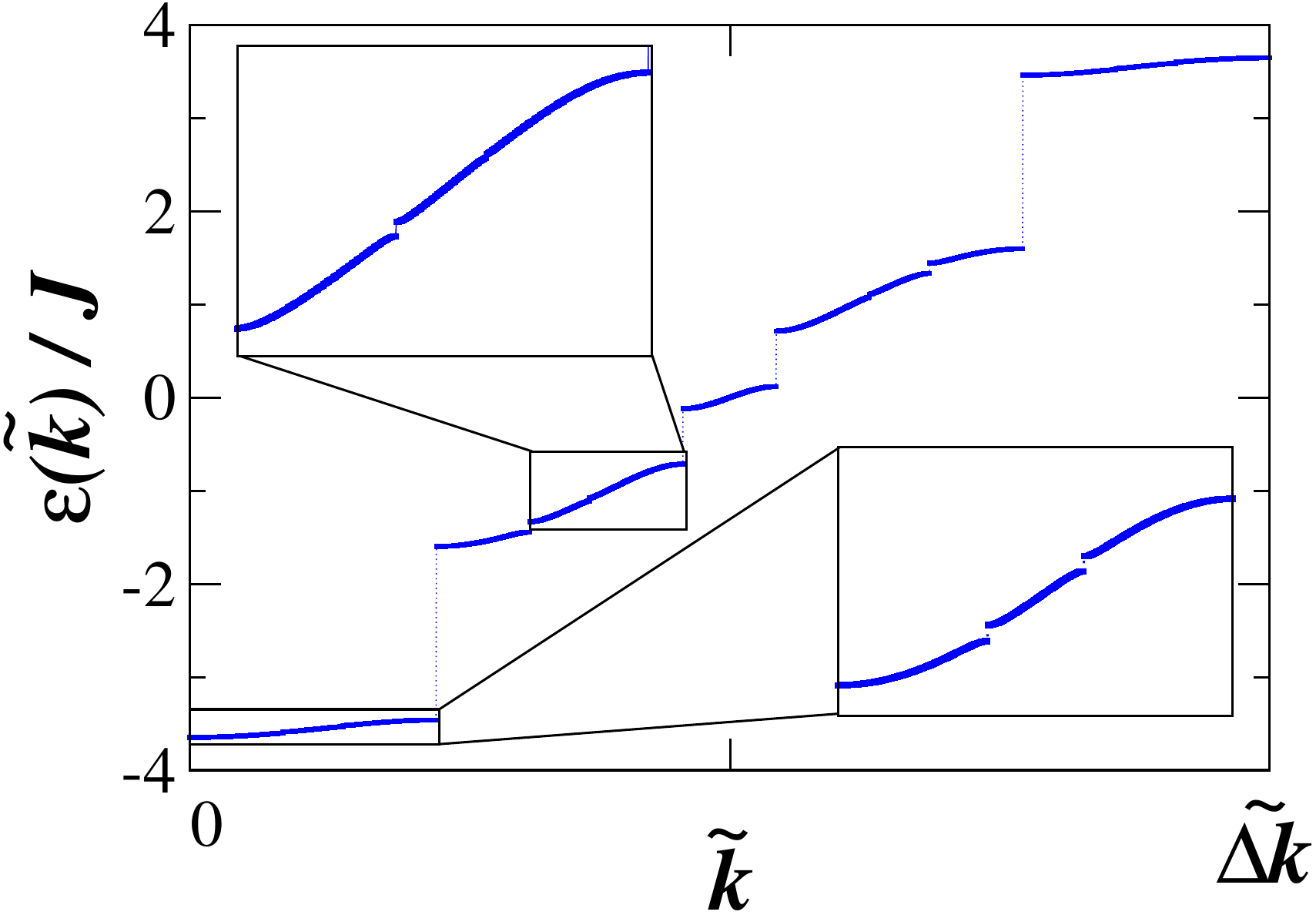}
\caption{(Color online) Dispersion relation for a single particle
in a QP potential of intensity $V_2/J=6$.}
\label{f.dispersion}
\end{center}
\end{figure} 

 According to Eq.~\eqref{e.fermiham}, the properties of the many-body 
system, and in particular its elementary excitations, are easily read 
out from the single-particle spectrum for periodic and anti-periodic
boundary conditions. Fig.~\ref{f.hardcore} shows the superfluid density
and compressibility of the hardcore-boson system as a function of 
the chemical potential and strength of the QP potential.
Both quantities exhibit a highly non-monotonic dependence on the 
chemical potential $\mu$ at all finite values of $V_2$. 
The compressibility is clearly finite for a finite single-particle 
density of states at the chemical potential,
$\rho^{(+)}(\mu)$ ($\rho^{(-)}(\mu)$) for periodic (antiperiodic)
boundary conditions depending on the number of particles as in
Eq.~\ref{e.fermiham}; in this case 
an extra particle can be added to the system by an infinitesimal
change of the chemical potential. As for the superfluid density $\rho_s$,
in a truly periodic system (namely in a \emph{commensurate} superlattice)
this quantity is directly proportional to the group velocity of the single-particle
dispersion relation at the chemical potential \cite{Rousseauetal06}, 
as it immediately emerges from the definition of $\rho_s$ as response                   
function to an infinitesimal phase twist in the operators,
$b_l \to b_l \exp{(i \delta l)}$, namely
$\rho_s = (L/2J)~\partial^2 E / \partial \delta^2 |_{\delta=0}$
($E$ is the total energy of the system).
Hence $\rho_s$ is finite when the chemical potential is inside a band 
and zero otherwise \cite{Rousseauetal06}. In the case of a 
QP potential the direct connection between superfluid density
and group velocity breaks down formally, as quasi-momentum is not 
a good quantum number any more, and the single-particle
states do not have consequently a well-defined group
velocity. Even though the new quantum number $\tilde{k}$
labeling the single-particle states is not strictly
speaking the momentum, the single-particle spectrum $\epsilon(\tilde{k})$
shows a dependence on $\tilde{k}$ which is reminiscent of that
of a system in a periodic potential (see Fig.~\ref{f.dispersion} 
for an example); in particular, close to a gap, 
the dependence of $\epsilon(\tilde{k})$ on $\tilde{k}$
is extremely weak, and it vanishes at the band edge.
If $\tilde{k}_{\rm edge}$ is the quantum number
of the edge state, and if the quantum numbers are
defined on an interval $\Delta \tilde{k}$,
the quantum number of the 
closest state at lower energy will be 
$\tilde{k}_{\rm edge}^{(-)}=
\tilde{k}_{\rm edge} - \Delta \tilde{k}/L$. 
A vanishing derivative of the energy spectrum
with respect to $\tilde{k}$ implies then
that 
\begin{equation}
\frac{\left[\epsilon(\tilde{k}_{\rm edge}) - 
\epsilon(\tilde{k}_{\rm edge}^{(-)})\right]}{\Delta \tilde{k}/L}
\to 0 
\end{equation}
for $L\to\infty$; this means in turn that 
the energy difference $\epsilon(\tilde{k}_{\rm edge}) - 
\epsilon(\tilde{k}_{\rm edge}^{(-)})$
vanishes faster than $1/L$ in the thermodynamic
limit. Consequently, when the chemical
potential sits close to the band edge, the
infinitesimal perturbation induced by a change 
in the boundary conditions from periodic to 
anti-periodic is going to mix states with an
energy difference decreasing faster than $1/L$,
and hence is it going to produce an energy change 
$E^{(-)}-E^{(+)}$
which obeys the same scaling. This then leads to 
a vanishing  $\rho_s$ as defined
in Eq.~\eqref{e.rhos}.

 On the contrary, for states corresponding
to the middle of the bands 
the perturbation induced by a change in the 
boundary conditions will cause an energy shift
scaling like $1/L$, leading to a finite $\rho_s$,
provided that such states are affected at all
by a \emph{local} perturbation \cite{explanation}, namely 
provided that they are \emph{extended} \cite{EdwardsT72}.
Hence in Fig.~\ref{f.hardcore} it is not surprising to observe 
a finite $\rho_s$ corresponding to a finite compressibility
for $V_2/J < 4$, and to observe an indentically
vanishing $\rho_s$ regardless of the chemical potential
for $V_2/J > 4$, as, according to the Aubry-Andr\'e
conjecture, the critical point marks a transition
from all extended to all localized states.
This means that the hardcore-boson system exhibits
an alternation of SF and IBI
phases for $V_2/J < 4$, while for $V_2/J > 4$
the alternation is between IBI phases
and BG phases, displaying a finite compressibility
in absence of superfluidity. 

\section{Softcore case}
\label{sec.softcore}

\begin{figure}[h]
\null\vspace{-1cm}
\begin{center}
\null\hspace*{-1.cm}
\includegraphics[
     width=70mm, angle=270]{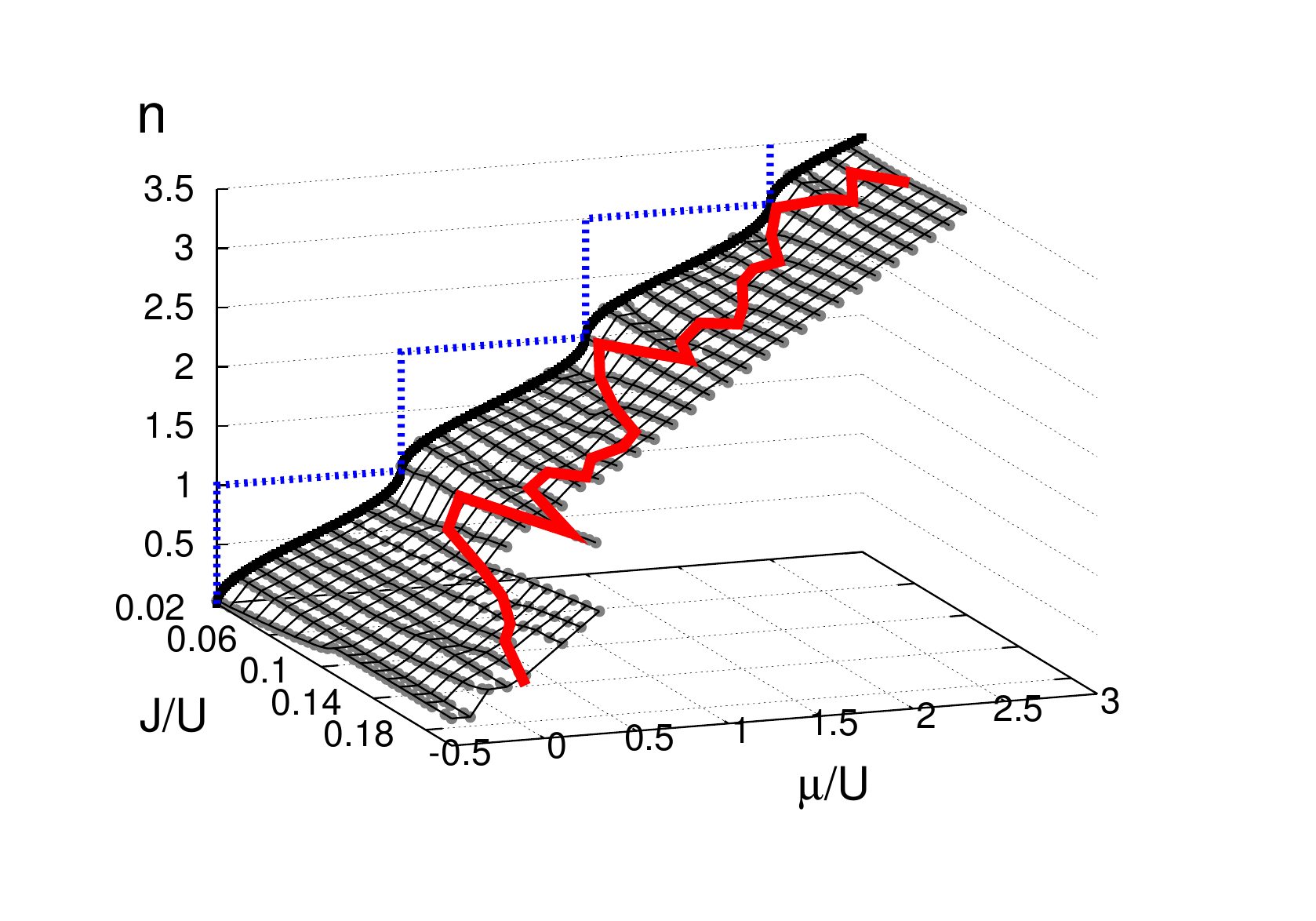} 
\caption{(Color online) Density plot for the phase
diagram of the Bose-Hubbard model in a QP potential with
strength $V_2=U$ ($L=126$). The red line marks
the boundary between the insulating (small-$J/U$) and 
the superfluid (large-$J/U$) regions. The bold black 
line is the density curve in the atomic limit 
$J\to 0$ (notice that it is reported on the 
graph minimum $J/U = 0.02$ for comparison with
the numerical data at finite $J$). The dashed blue
line shows the same curve for $V_2=0$ and
for a shifted chemical potential $\mu\to \mu-U/2$
(see text).}
\label{f.densplot}
\end{center}
\end{figure} 

 When releasing the hardcore constraint for the 
bosons, we lose the possibility of describing
the system via free fermionic quasi-particles,
and a fundamental question arises on the fate of
the phase alternation described in the previous 
section. For $V_2=0$ the phase diagram of the
one-dimensional Bose-Hubbard model in the 
grand-canonical ensemble is well
known \cite{Batrounietal90,Kuehneretal00}, 
and it features an extended SF region
at large $J/U$ ratio, and a succession of Mott lobes
with integer fillings for lower $J/U$.
Similarly to what is done in 
absence of a QP potential, we study the
phase diagram in the $(J/U,\mu/U)$ plane, 
which is also directly relevant for the 
intepretation of the behavior in a trapped
system (see Section \ref{sec.LDA}). 
 For definiteness, we choose to 
study the system with a \emph{strong} QP potential
compared to the inter-particle repulsion, 
namely we choose $V_2=U$ in Eq.~\eqref{e.hamiltonian}.
Under this condition the MI regions are 
completely destabilized, given that the Mott
gap is upper-bounded by $U$ and hence the Mott
phase is not going to survive the application
of an Hamiltonian term which is systematically
bigger than the gap. Hence the only insulating
phases surviving in the $V_2=U$ case
will generically be compressible or incompressible
\emph{incommensurate} ones,
namely insulating phases with incommensurate filling
factors, as shown in Fig.~\ref{f.densplot}. 

\begin{figure}[h]
\null\vspace*{-2cm}
\begin{center}
\null\hspace*{-3cm}
\includegraphics[
bbllx=140pt,bblly=40pt,bburx=800pt,bbury=406pt,
  width=90mm,angle=270]{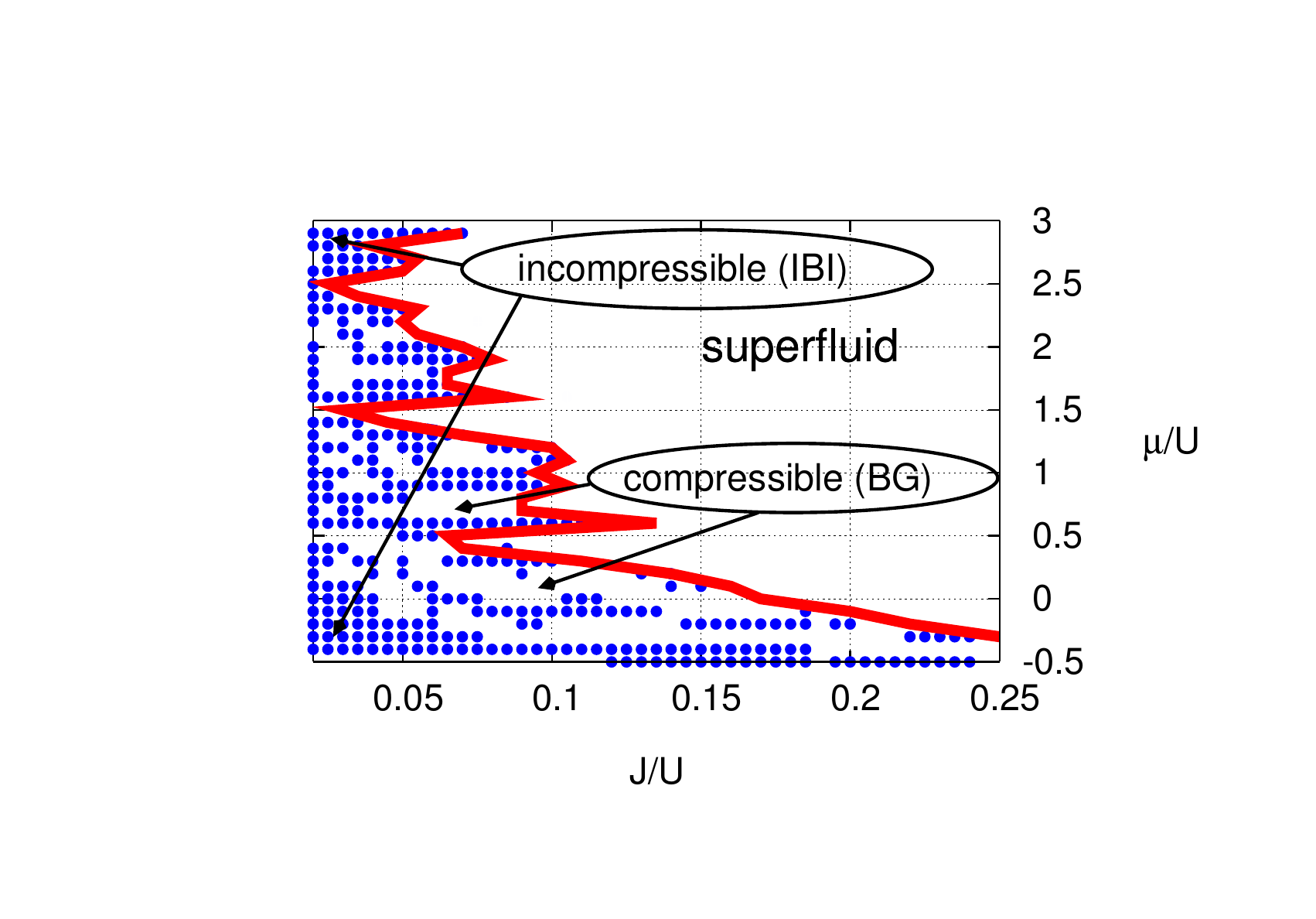} 
\vspace*{-1cm}  
\caption{(Color online) Phase diagram of the 
Bose-Hubbard model in a QP potential with
strength $V_2=U$ ($L=126$). Blue dots mark
points which are found incompressible. Other
symbols as in Fig.~\ref{f.densplot}.}
\label{f.softcore}
\end{center}
\end{figure}

 We map out the phase diagram for $V_2=U$ via 
 grand-canonical quantum Monte Carlo simulations 
 for a single lattice size, $L=126$ and for a 
 commensurate approximant $\alpha=97/126$ to the
 incommensuration parameter used in the experiment
 \cite{Fallanietal07}. The phase factor $\phi$ 
 of Eq.~\eqref{e.pot}
 is set to $0$ for simplicity. Similarly to the
 softcore case, the characterization
 of the phases is carried out by investigating the 
 compressibility $\kappa$ and the superfluid density
 $\rho_s$. The phase diagram emerging from the 
 simulations is of extreme complexity, and several
 observations are in order, as it will be discussed
 extensively in the next subsection ~\ref{ssec.finitesize}. 
 It is important to stress that what we present in Fig.~\ref{f.softcore} 
 is the phase diagram for a \emph{finite-size} system and 
 for a \emph{single realization} of the QP potential
 (namely a single value of $\phi$); we will see
 in subsection ~\ref{ssec.finitesize} how to overcome
 these limitations. 
  In the phase diagram, in absence of superfluidity,
 we observe a patchwork of compressible and 
 incompressible insulating regions.
 In Fig.~\ref{f.softcore} we indicate with a 
 blue dot a point in parameter space which is 
 found to be incompressible.
 
 \subsection{Finite-size effects}
 \label{ssec.finitesize}
  
  Before discussing the fundamental features
 of the phase diagram, it is important to stress
 the following technical point. At \emph{strictly} zero 
 temperature, the number of particles $N$ is a constant, 
 given that it is a good quantum number of the Bose-Hubbard
 Hamiltonian in an external potential; the variation
 of $N$ upon changing the chemical potential $\mu$
 proceeds in integer steps, and hence the compressibility,
 giving the slope of such curve, is simply a succession of 
 $\delta$-peaks. Given a system size $L$
 and a $\mu$ value such that the ground-state particle
 number is $N$, the energy gaps between the ground state in the 
 $N$-particle
 sector and the ground states in the sectors with $N+1$ and $N-1$ 
 particles can either scale to a constant for $L\to \infty$,
 in which case the system is \emph{incompressible} in the thermodynamic
 limit, or they can scale to 0, in which case the system
 is compressible in that limit, and the apparent incompressibility
 at finite $L$ is a finite-size artifact. 
  The Quantum Monte Carlo method
 we make use of is intrinsically a finite-$T$ approach,
 and, even for a finite-size system, it can deliver 
 a finite compressibility value 
 if $T$ is larger than the finite-size gap between
 different particle number sectors. 
 In the simulation of, \emph{e.g.}, the Bose-Hubbard model 
 without QP potential, one can rely on the (finite-temperature)
 compressibility to approximately locate the zero-temperature 
 phase boundary between the compressible SF phase and 
 the incompressible MI phase. The error that 
 one can make in that case is to ascribe to the SF phase
 that portion of the MI phase whose gap is 
 comparable to the temperature, but this is definitely a 
 tolerable error as long as the topology of the phase diagram is 
 simple (\emph{e.g.} one single boundary line dividing two 
 extended phases). 
 
  In the case under consideration, we are far from simplicity.
 In fact, Fig.~\ref{f.densplot} and Fig.~\ref{f.softcore}
 are apparently conflicting, since Fig.~\ref{f.densplot}
 seems to suggest an insulating region with a smoothly
 varying $n(\mu)$ over most of the parameter space, 
 namely an insulating region which is mostly compressible, 
 whereas Fig.~\ref{f.softcore} suggests rather the opposite,
 namely that a dominant part of the insulating phase diagram 
 is incompressible. The compressibility data
 upon which Fig.~\ref{f.softcore} is based are 
 obtained at a temperature ($T=J/96$) which is possibly
 lower than most finite-size gaps in the insulating
 region, so that the incompressible portion of the
 finite-size phase diagram is overestimated with respect 
 to that in the thermodynamic limit. Raising $T$
 is not recommendable, given that in this way 
 narrow incompressible domains in the phase diagram,
 expected in principle on the basis of the 
 hardcore limit (see Section \ref{sec.hardcore}),
 might be washed out if their gap becomes
 lower than the temperature.  Hence, given the
 tight alternation of compressible and incompressible
 regions, the finite-size or finite-temperature error 
 that one can make in overestimating
 the ones over the others is considerable.

\begin{figure}[h]
\begin{center}
\includegraphics[
     width=80mm,angle=270]{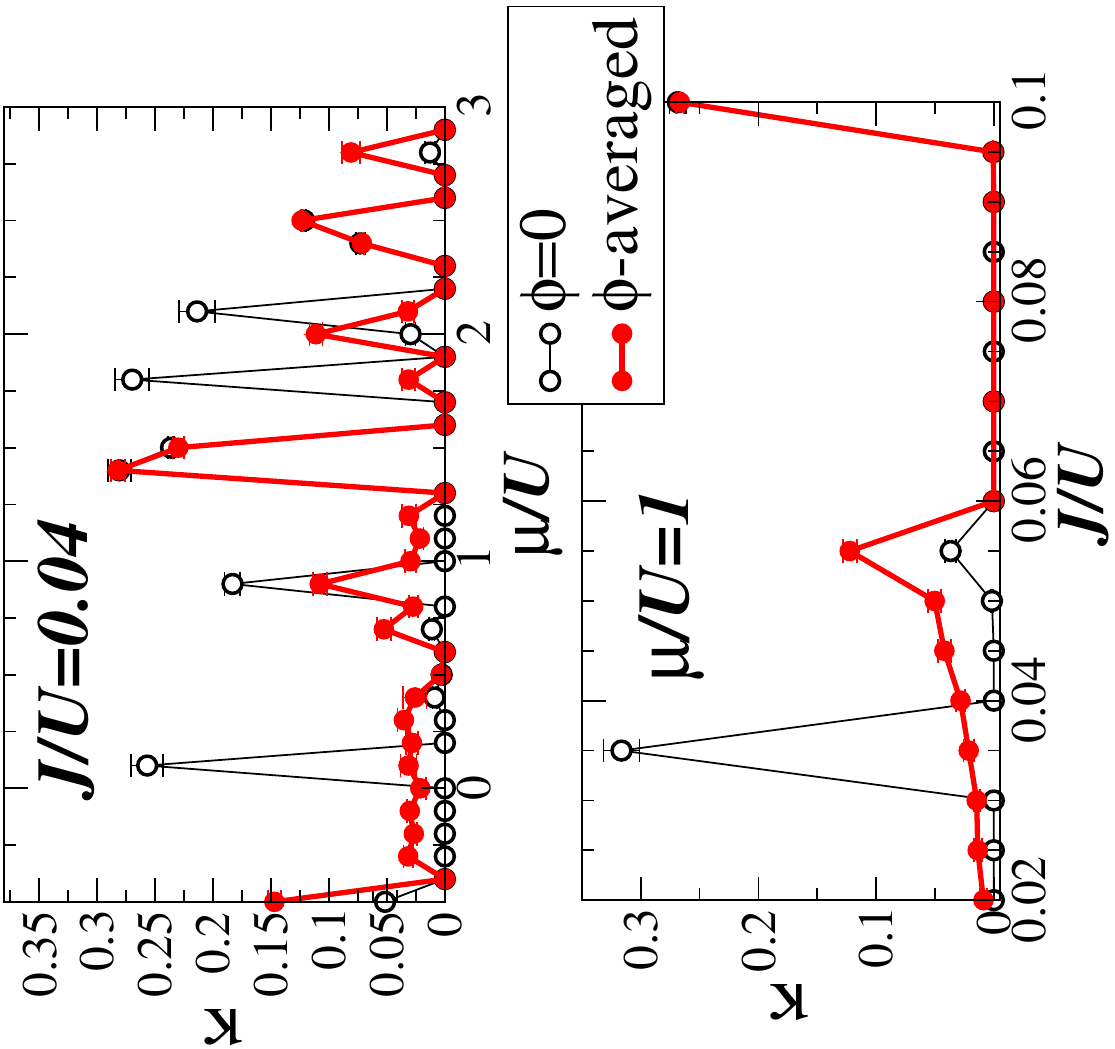} 
\caption{(Color online) Compressibility for the 
Bose-Hubbard model in a QP potential with fixed
spatial phase $\phi=0$, compared with the same
quantity averaged over $\phi$ fluctuations.}
\label{f.average}
\end{center}
\end{figure}

  One possible strategy to circumvent this 
intrinsic \emph{conundrum} of finite-size simulations
is based on the following observation. If a truly irrational
 $\alpha$ is taken in Eq.~\eqref{e.pot}, increasing the system size 
 leads to a self-averaging QP potential, namely
 the properties of the system in the thermodynamic limit 
 become independent of the phase $\phi$. This means
 that increasing the system size towards the thermodynamic
 limit at fixed $\phi$ is equivalent to averaging over the 
 phase $\phi$ on a sample of fixed size $L$, as long as $L$ 
 is chosen to be much bigger than any correlation length $\xi$ 
 in the system. Hence, away from the critical
 line between the SF phase and the insulating
 one(s) (where $\xi\to\infty$), the behavior  
 of the system in the thermodynamic limit can be obtained 
 by averaging the finite-size results over all possible values 
 of $\phi$. In practice, for a compressible region of the 
 phase diagram
 one expects that, for some values 
 of $\phi$, the finite-size gap between different $N$-sectors 
 becomes smaller than the selected temperature, so that 
 a finite compressibility is recovered on average. 
 This numerical procedure is computationally demanding, 
 and we limit it to two selected cuts
 through the phase diagram, one for fixed $J/U=0.04$
 and one for fixed $\mu/U=1$. The results are shown
 in Fig.~\ref{f.average}: they confirm the 
 expectation that the fixed-$\phi$ phase diagram 
 overestimates the incompressible regions, but at the 
 same time they confirm the presence of sizable incompressible 
 (IBI) regions, in tight alternation
 with the compressible ones (BG). Given the finiteness of 
 the grid used in the  $(J/U,\mu/U)$ plane to sample
 the phase diagram, we cannot exclude that an even 
 tighter alternation between IBI and BG regions 
 exist among the dots, possibly revealed by a finer mesh.

\subsection{Emergence of the incompressible regions}
\label{ssec.emergence}

 The tight alternation of compressible and 
 incompressible insulating regions upon varying the 
 chemical potential reveals that there are 
 several ``magic" incommensurate fillings at 
 which a \emph{quasi-particle} band is filled, similarly
 to what happens to the exact fermionic quasi-particles
 in the hardcore limit. In the compressible BG
 regions, on the other hand, varying the chemical
 potential leads to variations in the fillings,
 and all the newly added particles are effectively 
 localized by the QP potential and by the interaction 
 with the density background of the other particles. 
 Fig.~\ref{f.densplot} shows the filling of the
 lattice in the region of phase diagram we sampled
 numerically, along with the atomic limit both for the
 case $V_2=U$ and for the case $V_2=0$. It is to be
 noticed that the QP potential of Eq.~\eqref{e.pot} 
 is symmetric around zero energy, and in particular the local chemical 
 potential in its minima takes the value $\mu-U/2$; hence
 a correct comparison with the case $V_2=0$ requires
 to consider a Bose-Hubbard Hamiltonian with a shifted
 chemical potential $\mu \to \mu-U/2$. The presence of 
 the strong QP potential alters significantly the 
 density curve in the atomic limit, washing out the 
 steps associated with Mott lobes in the case $V_2=0$,
 given that the system has lost translational invariance
 and the average density can take all possible values.
 Therefore a seemingly continuous curve emerges,
 with integer densities appearing only at semi-integer
 values of $\mu/U$; this behavior reflects the broad distribution 
 of local chemical potentials in the sites of the QP potential,
 gradually occupied upon increasing $\mu$ until
 the $n$-th ``shell'' is completed when an integer
 density $n$ is reached. The density of atomic states
 in a QP potential is peaked around the energies $\pm V_2/2$
 (see Fig.~\ref{f.DOS2SL} below),
 as expected given that these are the values where the
 potential has minimum derivative, and consequently the 
 density curve has maximum slope around $\mu = (m +1/2)U$ 
 (with $m$ integer). 
  Hence the atomic limit picture suggests a \emph{finite}
 compressibility for all chemical potentials, and intuitively
 one would expect the compressibility to stay finite
 \emph{a fortiori} in presence of quantum fluctuations,
 turned on through the hopping term.
 This would be generally true in the case of a 
 \emph{truly random} potential, on average over the possible
 realizations of the potential or in the thermodynamic
 limit. Yet, in the case of a pseudo-random potential, 
 one of the main quantum effects is to break the 
 continuous classical many-body spectrum 
 into \emph{bands}. This mechanism is discussed in the
 next section.

\section{True disorder \emph{vs.} pseudo-disorder}
\label{sec.truepseudo}

 In this section we describe the formation of 
bands in the quantum \emph{many-body} spectrum
of the Bose-Hubbard model in an incommensurate
potential making use of a generalization
of the atomic limit which includes the 
first quantum correction in a \emph{non}-perturbative way.
The central idea is to subdivide the QP potential
in uncorrelated wells which are assumed to be 
independent from each other in the case $V_2/J \gg 1$,
and to diagonalize the $N$-boson problem in each
of the well. We call this approach the \emph{random atomic limit}.
This approach not only reproduces
the main features of the $n(\mu)$ curve and of 
the associated many-body density of states at
low fillings, but it also nicely elucidates the fundamental 
difference between a pseudo-random potential and
a truly random one, and the fundamental mechanism
leading to BG physics. Finally it suggests a strategy
to emulate the behavior of a truly random potential 
making use of a combination of \emph{two} superimposed
QP potentials.

\begin{figure}[h]
\begin{center}
\includegraphics[
     width=60mm,angle=270]{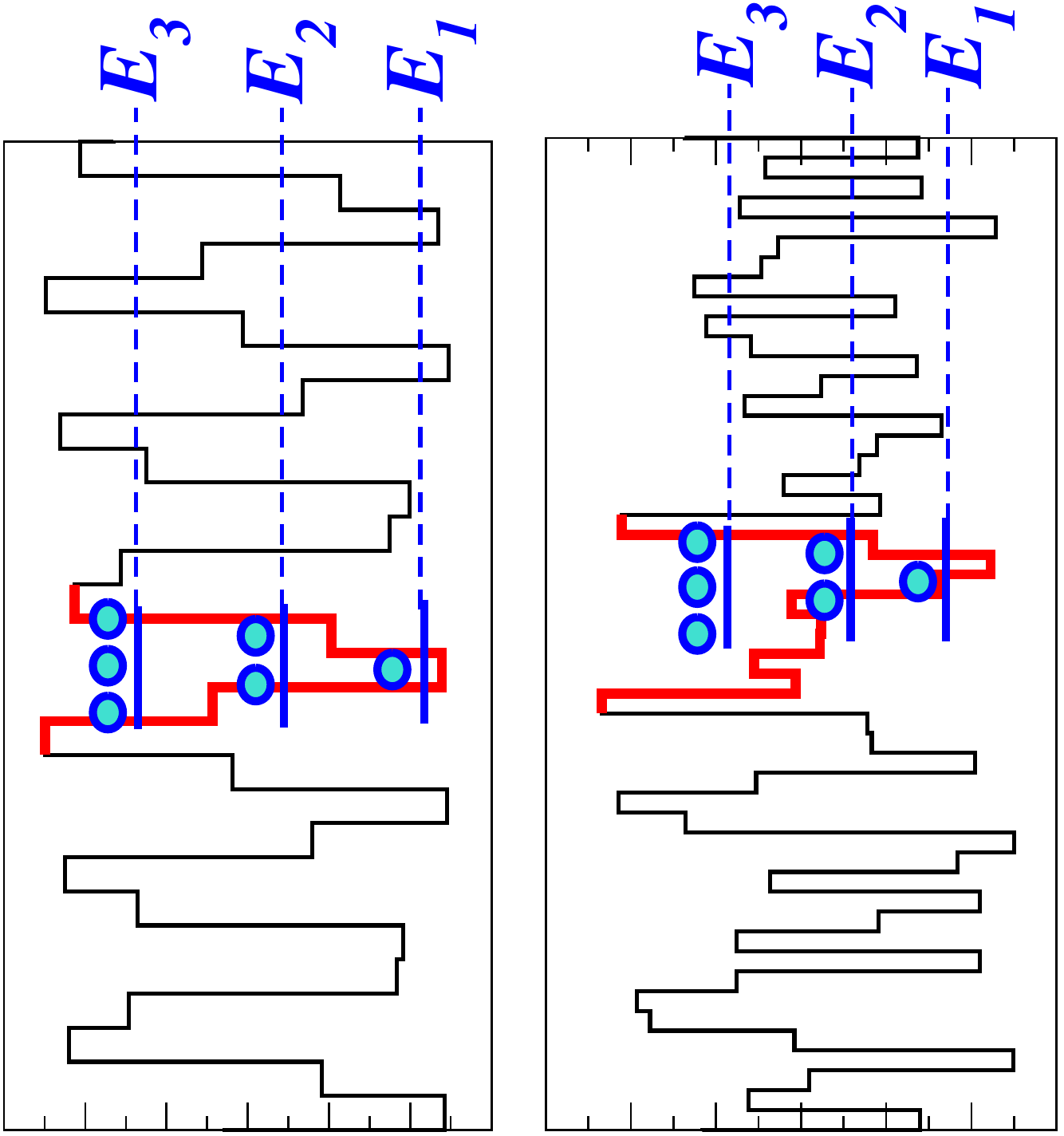} 
\caption{(Color online) Schematic depiction of the \emph{random-atomic-limit}
approach for a QP potential (upper panel) and for a truly random one (lower panel).
This approach subdivides the system into potential wells, here
marked in red, and it diagonalizes the problem of 1, 2, and 3 particles
confined in each well, extracting the corresponding ground-state
energies $E_1$, $E_2$, and $E_3$. Such energies are then used
to reconstruct the density of states of the system for low filling.}
\label{f.randomatom}
\end{center}
\end{figure} 

\subsection{Random atomic limit}
\label{ssec.RAL}

 We propose here an approximate treatment of the 
strongly interacting Bose-Hubbard model in an intense
QP potential, $U=V_2\gg J$, in the case of low fillings. 
The idea is that the QP potential subdivides the 
lattice into potential wells between which tunneling
is negligible when not assisted by the interaction
(namely for low fillings) and hence inter-well
coherence can be neglected within a good approximation. 
 Hence the Bose-Hubbard model in the grand canonical
 ensemble can be treated as follows:
1) isolate each potential well, cutting the chain
 at the location of the maxima delimiting the well
 itself, and diagonalize the Bose-Hubbard Hamiltonian 
 in this potential for $N=1,2,...$ particles,
 finding the ground state energies $E_1$, $E_2$, ...;
2) due to the QP potential and to the interaction $U$, 
 adding a second particle to the well costs more energy
 than adding the first, and so for the third, namely
 $E_1 < E_2-E_1 < E_3-E_2-E_1 <$ ...~. Hence one can build
 an effective single-particle density of states (DOS) 
 $\rho^{(1)}(E)$ 
 by accumulating the energy costs for the addition
 of a single particle to each well $\epsilon_1=E_1$, 
 $\epsilon_2=E_2-E_1$,
 $\epsilon_3=E_3-E_2-E_1$, etc. We moreover
 normalize the DOS to unity, namely:
 
 \begin{equation}
\rho^{(1)}(E) = 
\frac{1}{N_{\rm states}}
\sum_{i=1,2,3,...} \sum_{\epsilon_i}
\delta(E-\epsilon_i)
\end{equation}
 
 The filling for a given $\mu$
 is simply found by integrating $\rho^{(1)}(E)$
 up to $\mu$, similar to the case of non-interacting
 fermions. 
 
 \begin{figure}[h]
\begin{center}
\includegraphics[
     width=55mm,angle=270]{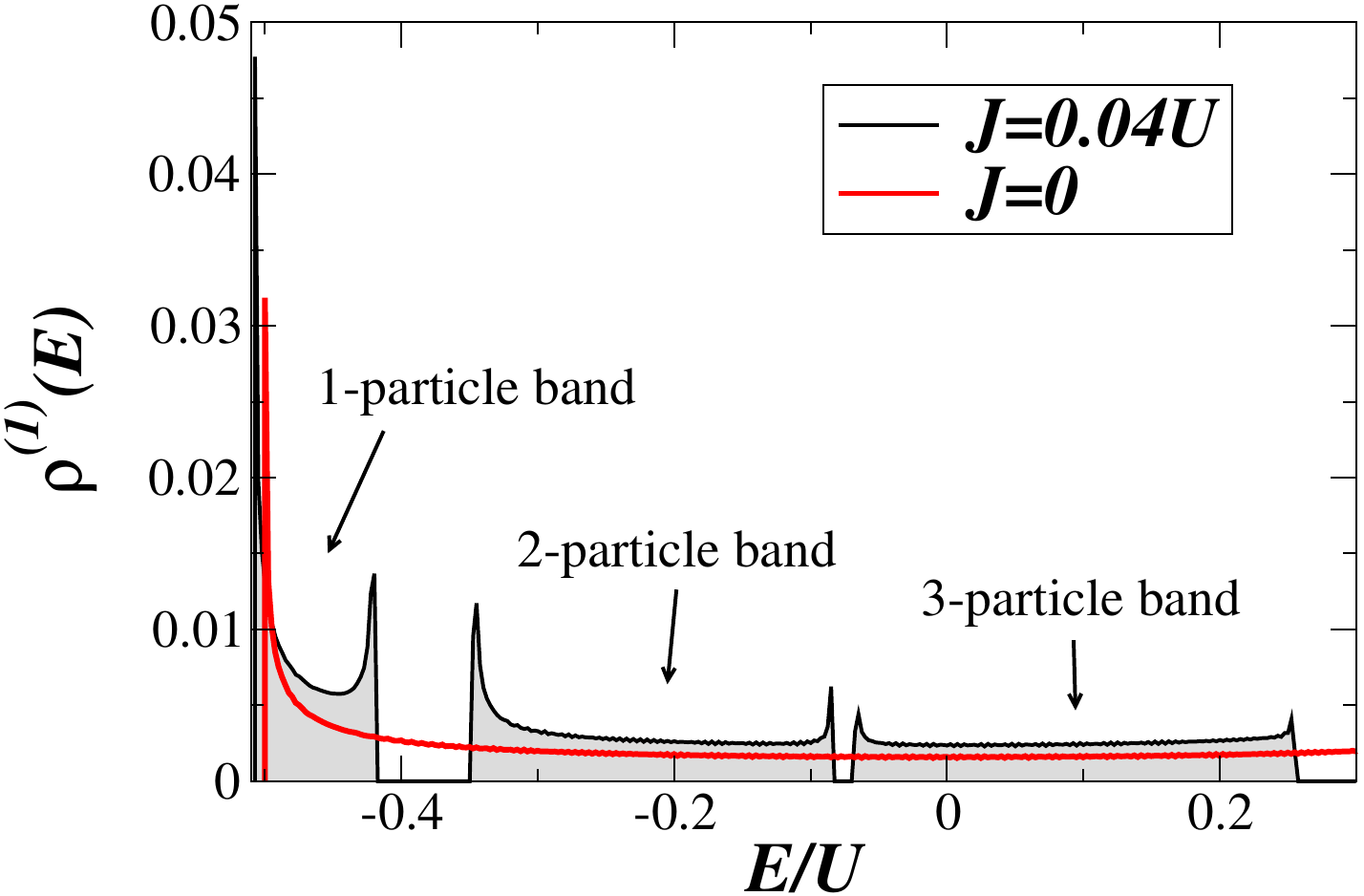} 
\caption{(Color online) Random-atomic-limit result
for the effective single-particle density of states 
of a Bose-Hubbard model in a QP potential with
 $V_2=U$, and over a large system 
size ($L=10^6$). The quantum case of $J/U=0.04$ is 
compared with the classical limit $J=0$.}
\label{f.DOS2SL}
\end{center}
\end{figure} 
 
   For $J=0$, and in the case $U=V_2$
  and $n<1$, any new particle added to the ground state
  of the system occupies the least energetic site
  that remains empty. Hence the DOS 
   $\rho^{(1)}(E)$ in that case simply
  reflects the distribution of the on-site 
  energies in the  potential, which is a continuous
  one in the interval $[-U/2,U/2]$.
  The quantum corrections
  to it, obtained for up to $N=3$ particles per well,
  are shown in Fig.~\ref{f.DOS2SL}; the calculation
  was made for a large
  sample of the QP potential considered 
  so far ($J/U=0.04$), such that self-averaging
  of the system properties is guaranteed. We observe
  the striking feature of the opening of \emph{gaps}
  in the DOS, defining effective bands 
  which are populated by the gradually added particles.
  It is important to stress that the states associated
  to these bands are localized, and that the nature of 
  these bands is determined by the many-body physics inside
  each well. In particular the first band is associated
  with states in singly occupied wells, and the finite energy 
  cost to add an extra particle to any of 
  the well in the system causes the appearence of a 
  gap to the next band. The potential-energy
  cost of adding an extra particle is completely
  compensated for by the gain in chemical potential,
  hence the only remaning energy cost which prevents
  particle addition is purely quantum in nature, and 
  it comes from the loss of kinetic energy in the first particle
  when adding a second one to the well. The same reasoning
  applies when adding a third particle, which 
  gives rise to a further band, etc. In periodic systems,
  the emergence of Bloch bands is due to the 
  strong scattering of
  traveling matter waves with wavevector at the edges 
  of the Brillouin  zone; here we have on the contrary
  the appearence of bands
  associated with tightly localized particles, and band
  gaps emerging due to strong correlation among the
  particles.  
 
\begin{figure}[h]
\begin{center}
\includegraphics[
     width=55mm,angle=270]{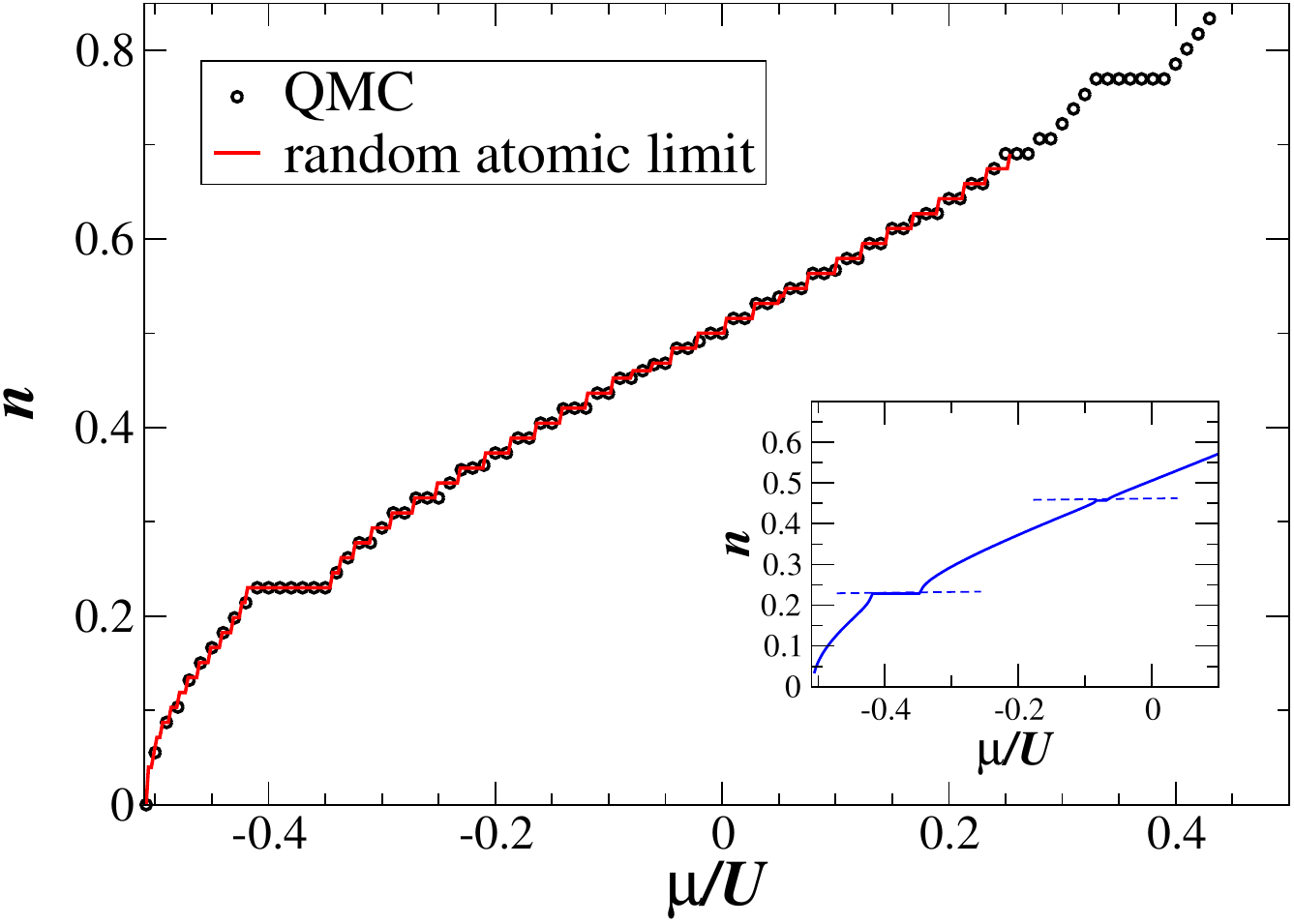} 
\caption{(Color online) Density curve for the Bose-Hubbard
model in a QP potential with $\alpha=97/126$ and $V_2=U$,
on a system size $L=126$. In the inset: random-atomic-limit
calculation on a big system size ($L=10^6$).}
\label{f.QMCvsrandomatom}
\end{center}
\end{figure}

   The opening of gaps in the effective single-particle
  DOS leads to plateaus in the $n(\mu)$ curves,
  and hence to incompressible phases at special, incommensurate
  fillings. The $n(\mu)$ curve obtained in the random atomic
  limit is compared with the QMC results at $J/U=0.04$ and 
  for low fillings in Fig.~\ref{f.QMCvsrandomatom}. We observe that 
  the random-atomic limit calculation reproduces very well
  all the fine details of the $n(\mu)$ curve from QMC: 
  1) it perfectly captures the large plateau which
  corresponds in Fig.~\ref{f.DOS2SL} to the gap
  between the 1-particle band and the 2-particle band, occurring
  at a filling $n=0.229...$;
  2) moreover it also correctly captures the finite-size
  plateaus due to the discrete jumps in the particle
  number in the $n(\mu)$ curve. Here the significance of 
  finite-size issues in the determination of 
  the phase diagram of the system is particularly
  evident: in fact for the $L=126$ chain the finite-size 
  gaps are completely masking the second plateau
  in the $n(\mu)$ curve at density $n =0.457...$, which  
  on the contrary is 
  clearly exhibited 
  in the random atomic limit calculation on a large
  system size (inset of Fig.~\ref{f.QMCvsrandomatom}).

\begin{figure}[h]
\begin{center}
\includegraphics[
     width=75mm,angle=270]{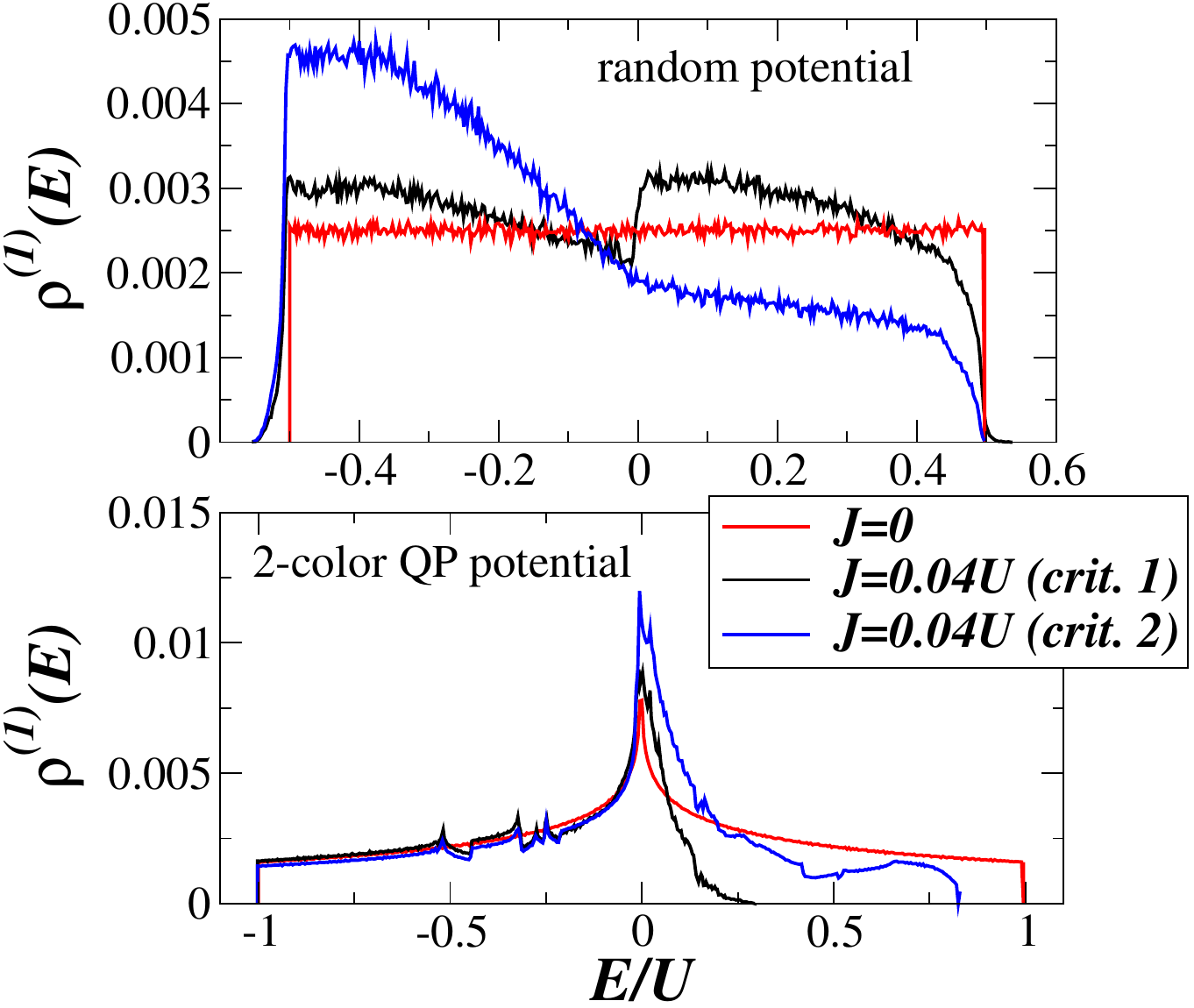} 
\caption{(Color online) Density of states from the random atomic limit
for a truly random potential (upper panel) and for a 
2-color QP potential (lower panel). The random potential 
takes values over the interval $[-U/2,U/2]$, while
the 2-color QP potential has $V_2=U$. For both
systems $L=10^6$. Results from two different criteria
for the identification of the potential wells (see text)
are compared with the classical limit ($J=0$).}
\label{f.3SLvsrand}
\end{center}
\end{figure}  

\subsection{From pseudo-disorder to true disorder: the two-color quasi-periodic
potential}
\label{sec.threecolors}  

 The discussion of the previous section allows to clearly 
 identify the origin of the incompressible behavior in the
 insulating regime of the QP potential. At some
 specific fillings the addition of an extra particle
 requires to locally squeeze the particles
 already present in the QP potential wells, and this 
 squeezing costs a finite amount of kinetic energy
 which translates into a gap over the ground state.
 Actually the energy cost for the addition of a single
 particle can be arbitrarily lowered if the potential
 wells are made arbitrarily large, given that in this
 way the kinetic energy cost for the particles
 already present can be minimal. In a well of 
 characteristic size $l$ a particle
 has a confinement energy $\sim \hbar^2/2ml^2$;
 the addition of a second particle, which is
 ideally impenetrable to the first one, 
 roughly lowers the effective space available to 
 the latter to $l/2$ and hence it increases
 its confinement energy to $\sim 4\hbar^2/2ml^2$.
 Hence the energy increase  
 scales to 0 as $l^{-2}$ when increasing
 the well size. 
 
  The QP potential has a characteristic
 length scale given by the quasi-period of the 
 QP potential, namely $l=2\pi/(k_{\rm L}-k_{\rm QP}) = 
 (1-\alpha)^{-1}$,
 coming from the beating between the 
 incommensurate cosine potential at wavevector 
 $k_{\rm QP} = 2\pi\alpha$ and the underlying lattice 
 (formally at a wavevector $k_{\rm L} = 2\pi$). 
 Potential wells
 cannot exceed this length scale, and therefore the
 confinement energy scale is bounded from below.
 This means that gaps to particle addition
 at particular incommensurate fillings are
 unavoidable in the system. Moreover, working 
 with a fixed number of particles at the same
 special filling, a gap opens to particle-hole
 excitations: intra-well excitations are
 clearly gapped due to the finite size of the 
 well, and particle-hole excitations which 
 cause the transfer of a particle from a 
 well to another are also gapped, due to 
 the chemical potential mismatch between 
 particle and hole.

  One fundamental feature of a truly random
 potential, on the other hand, is the absence of
 an upper bound for the extent of potential
 wells, and hence the possibility of always 
 adding a particle to the system at an infinitesimal
 energy cost, or, at fixed filling, to introduce
 a particle-hole excitation at arbitrarily low
 energy. This feature, based
 on the statistics of rare regions (large wells),
 is at the heart of the gaplessness of the BG 
 phase \cite{Fisheretal89,FreericksM96,
 Roscilde06}. To corroborate this statement
 quantitatively, we apply the random-atomic-limit
 approach described in the previous subsection 
 to a truly random potential with a random 
 on-site energy evenly distributed over the interval
 $[-U/2,U/2]$. We use two different criteria
 to identify the potential wells:
 \emph{(criterion 1)} a potential well is delimited by two successive
 local maxima with positive height; 
 \emph{(criterion 2)} a potential
 barrier is identified with a local maximum whose height over
 the previous local minimum is larger than $U/2$,
 and we hence identify a well with a region 
 between two such barriers (see Fig.~\ref{f.randomatom}).
 Criterion 1 is less restrictive than criterion 2,
 and it surely underestimates the size of the 
 potential wells. Yet, in the case of the
 QP potential of Eq.~\eqref{e.pot} both criteria would lead
 to the same identification of the potential wells
 as the one used in the previous subsection. 
 As shown in Fig.~\ref{f.3SLvsrand}, 
 regardless the criterion the resulting effective
 single-particle DOS $\rho^{(1)}$ associated
 with many-body intra-well states is continuous
 for a truly random potential,
 and all gaps are removed by the absence of an
 upper bound in the well length $l$. 
 
  From the point of view of the cold-atom experiments,
 realizing optically an ideal white-noise potential 
 requires to superimpose a large
 number of standing-wave components with different wavevectors 
 and with the same weight; this translates into the need 
 of a large amount
 of lasers at different frequencies and with the same 
 intensity, which is extremely demanding. 
 Hence an interesting question, both at the fundamental 
 level and at the practical one, is the following:
 can we mimic \emph{de facto} the physics of a system
 of bosons in a truly random potential by just using
 a finite number of superimposed standing waves, or,
 in more suggestive terms, \emph{how many colors do we
 need for a pseudo-random potential to call it disorder?}
 
  As discussed above, our goal is to realize a pseudo-random
 potential with the minimal amount of Fourier components giving
 rise to a continuous many-body spectrum, namely not
 exhibiting incompressible IBI phases. The guiding principle
 is that of realizing a potental whose spatial features
 are richer than those of a simple QP potential, such
 that larger wells are realized giving rise to intra-well
 excitations at lower energy, and such that states with 
 different number of particles in different wells 
 become quasi-degenerate, giving rise to low-energy inter-well
 particle-hole excitations. 
 
 \begin{figure}[h]
\begin{center}
\includegraphics[
     width=80mm,angle=270]{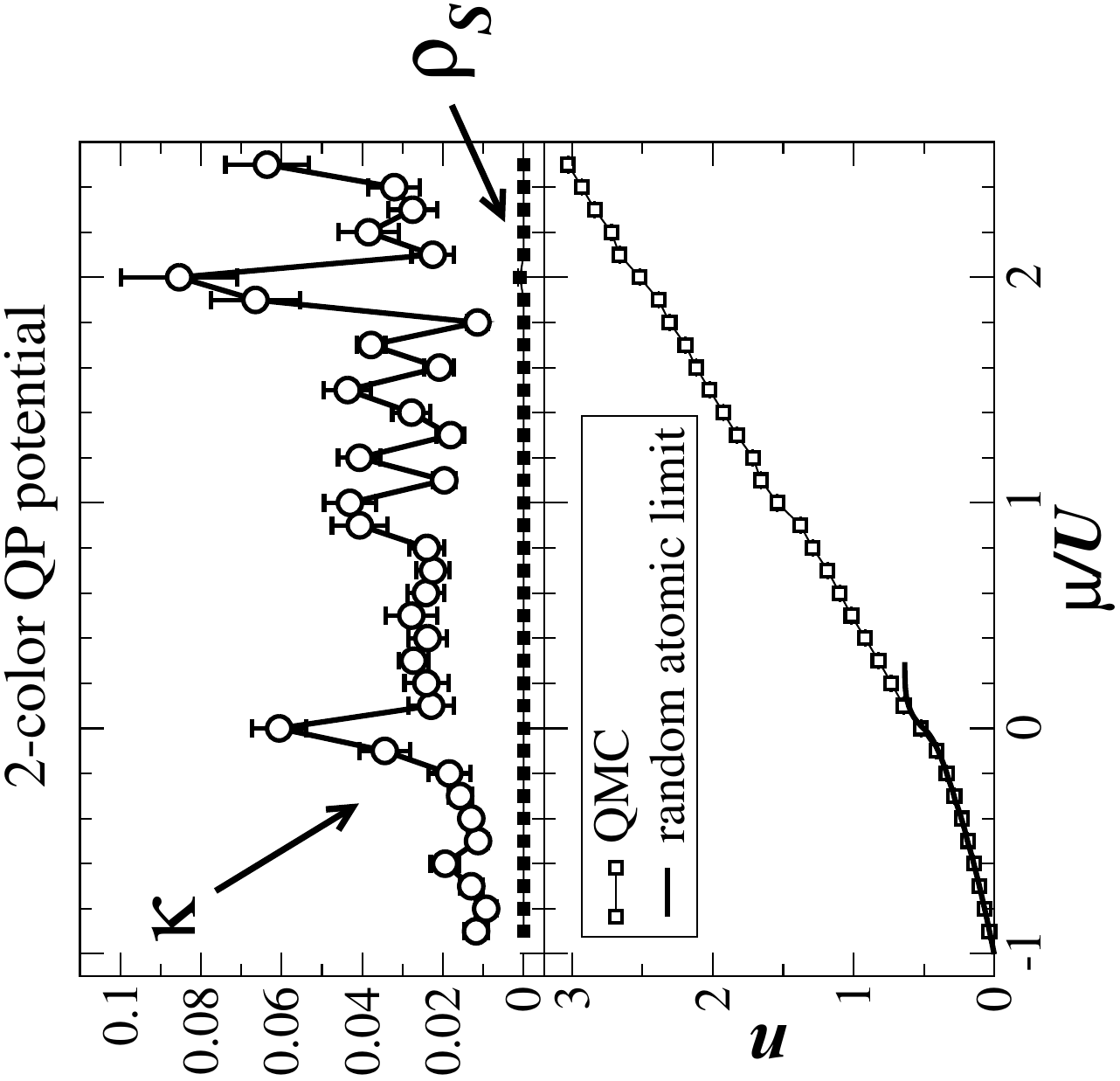} 
\caption{Compressibility and superfluid density (upper panel) 
and total density (lower panel) for the Bose-Hubbard model
in a two-color QP potential, Eq.~\eqref{e.3SL}, on a $L=126$ lattice
with $\alpha = 97/126$, $\alpha' = 71/126$, $V_2=U$, and 
$J/U=0.04$.
The QMC results are averaged over independent fluctuations of the
spatial phases $\phi$, $\phi'$.}
\label{f.3SL}
\end{center}
\end{figure} 
 
  In this perspective we explore the most straightforward
 generalization to the QP potential so far considered,
 namely we add a second incommensurate component to the
 QP potential of Eq.~\eqref{e.pot}, having the same intensity as the 
 first one, thus realizing a \emph{2-color QP potential}
 \begin{equation}
 g^{(2)}(i;\alpha,\phi,\alpha',\phi') = 
 \cos^2(\pi\alpha~i+\phi)+\cos^2(\pi\alpha'~i+\phi') - 1
 \label{e.3SL}
 \end{equation}
  
 In what follows we take $\alpha'=71/126=0.56349..$. 
 Fig.~\ref{f.3SLvsrand} shows the random-atomic-limit 
 calculation for such a potential on a large sample
 size (which makes the particular values of the 
 phases $\phi$ and $\phi'$ irrelevant). For both
 criteria of well identification we obtain a gapless
 single-particle DOS, namely no incompressible
 phase at low filling, opposite to the case of a
 simple QP potential. This encourages us to explore
 the behavior of the system over a larger filling
 interval, making use of QMC simulations. 
 Fig. \ref{f.3SL} shows the compressibility, superfluid
 density and total density for the Bose-Hubbard model in a 2-color
 QP potential for a large $\mu/U$ range; the results
 shown are averaged over independent fluctuations of the 
 $\phi,\phi'$ phase, hence exploiting the whole statistics
 associated with the potential of Eq.~\eqref{e.3SL},
 and removing finite-size effects, as discussed in 
 Sec.~\ref{sec.softcore}. Noticeably, the density 
 curve at low fillings ($n\lesssim 0.5$) agrees very 
 well with that predicted by the random-atomic-limit
 calculation. Furthermore, over the larger grid of chemical
 potentials and fillings which is accessible to 
 QMC (we went up to $n\approx3$) we find a finite
 compressibility everywhere, generally coexisting with 
 insulating behavior (only around $\mu/U\approx2$ a tiny
 superfluid density shows up). Hence this particular
 \emph{pseudo-disordered} potential features
 a continuous BG phase \emph{without} intermediate IBI 
 regions. This means that the Bose-Hubbard model in a 2-color 
 QP potential seems to capture some of the fundamental 
 physical features of the classic dirty-boson 
 problem \cite{Fisheretal89}. It remains to be seen 
 whether the transition from SF to BG in such 
 a potential belongs to the same universality
 class as that of the system in a truly random potential,
 and whether the absence of incompressible regions
 in the phase diagram applies to the whole parameter
 space: we leave these questions open for future 
 investigations.

\section{Trapped system and local-density approximation}
\label{sec.LDA}

 In the following sections we discuss the behavior of a
 system of bosons in an incommensurate optical superlattice
 in presence of a parabolic trapping potential, to 
 make contact with the experimental setup of current
 optical lattice experiments \cite{Greineretal02, Fallanietal07}.
  Hence we consider the Hamiltonian  
  \begin{equation}
  {\cal H}_{\rm trap} = {\cal H}_{\rm 0} + V_t \left(i-i_0\right)^2 
  \end{equation}  
  where $V_t$ is the strength of the trapping potential 
  and $i_0$ is the center of the trap.
  
 In absence of an incommensurate potential, the behavior
 of the Bose-Hubbard model in a trap is generally connected
 to that of the bulk system via a \emph{local-density approximation}
 (LDA) \cite{Wesseletal04,Bergkvistetal04,Kollathetal04}. 
 Here we discuss how that approximation holds only
 partially in presence of an incommensurate potential,
 and how the behavior of the trapped system has to be 
 regarded in its own respect.

\begin{figure}[h]
\begin{center}
\includegraphics[
     width=90mm,angle=270]{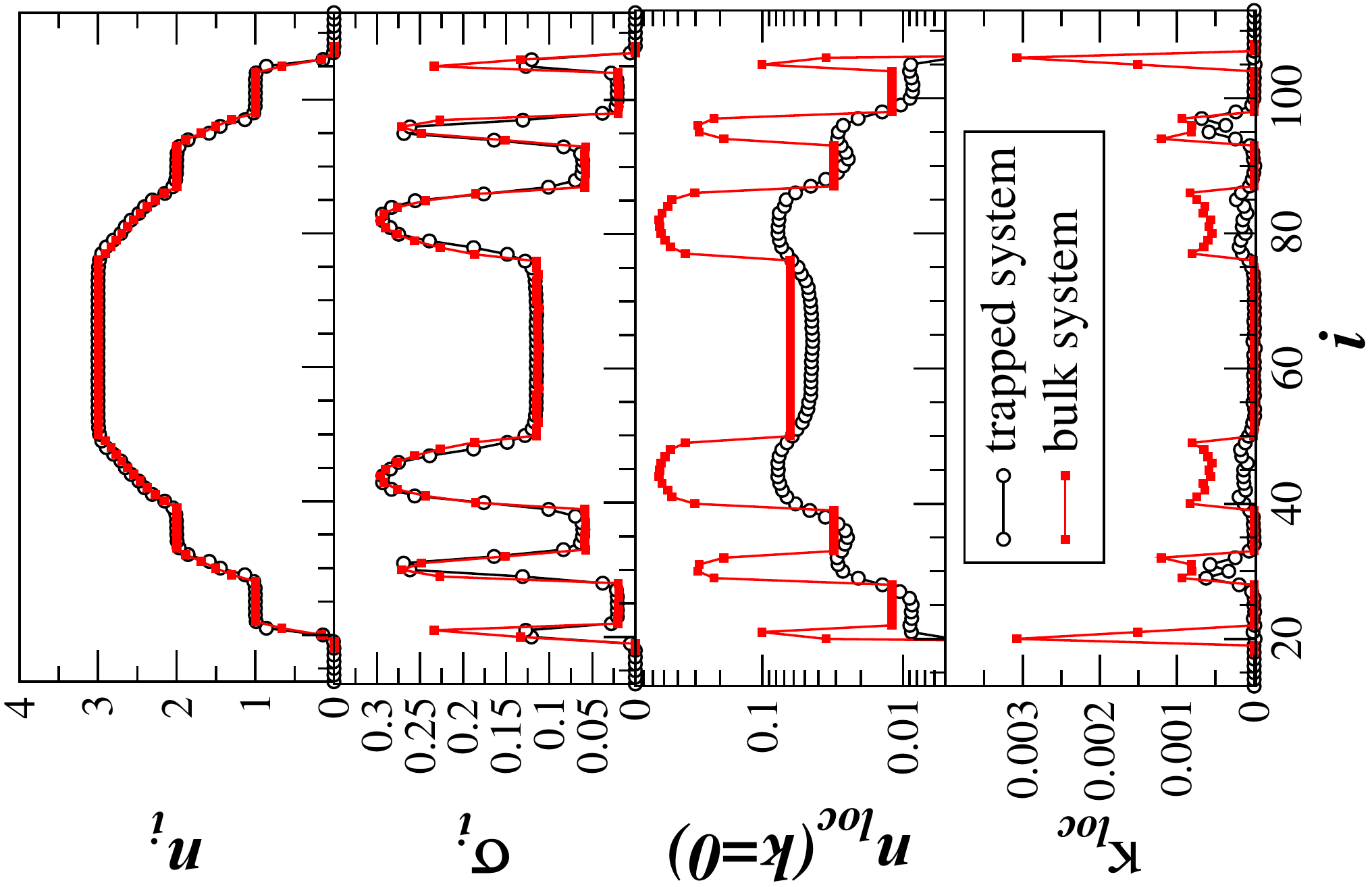} 
\caption{(Color online) Local quantities for a Bose-Hubbard
system in a trap with parameters $U=20 J$, 
$V_t=0.0014~U$,  and $\mu_{\rm trap}= 2.5~U$ 
(from top to bottom panel : density, on-site particle-number fluctuations, 
local coherent fraction, and local compressibility), compared 
with the same quantities for a bulk system with a chemical
potential matching the local one in the trap. }
\label{f.LDAnopot}
\end{center}
\end{figure} 

\subsection{Bose-Hubbard model without QP potential}

 In presence of a smoothly varying potential, namely
 under the assumption that $V_t \ll J, U$, recent works 
 \cite{Wesseletal04,Bergkvistetal04,Kollathetal04}
 have shown that a local-density approximation fully applies
 to the standard Bose-Hubbard model 
 (Eq.~\eqref{e.hamiltonian} with $V_2=0$), namely the 
 local properties in the trap, as \emph{e.g.} the density, 
 can be directly related to those of the homogeneous system
 in the grand-canonical ensemble by matching the local
 chemical potential in the trap, 
 $\mu_i = \mu_{\rm trap} - V_t \left(i-i_0\right)^2$, to that of
 the homogeneous system. A similar calculation is shown
 in Fig. \ref{f.LDAnopot}, where the local density 
 $\langle n_i \rangle$ and the on-site particle number fluctuations 
 \begin{equation}
 \sigma_i = \langle n_i^2 \rangle - \langle n_i \rangle^2
 \end{equation}
 are compared for a trapped system and a bulk system. 
 A good agreement between the data for the trapped
 system and those for the bulk system is observed 
 away from the crossover regions between a locally
 SF and a locally MI region, while in the crossover
 regions the data for the trapped system have generally
 less sharp features with respect to those for the homogeneous 
 one. This can be intuitively 
 understood by noticing that finite-size effects
 are typically most pronounced close to critical
 boundaries between different phases, where discontinuities
 and singularities are rounded off by the finiteness
 of the system. In the trapped system the finite-size
 effects are not only given by the overall trapping potential,
 but most significantly by the finiteness of those regions 
 whose local chemical potential corresponds to critical 
 boundaries between MI and SF in the homogeneous case.

  Despite the finite-size corrections,
 the above results lead typically 
 to interpret the various regions in the trapped
 system as \emph{locally superfluid}
 or \emph{locally Mott insulating} depending on 
 the corresponding phase at the same chemical
 potential in the phase diagram of the bulk system. 
 This conclusion is particularly confirmed when
 looking at the \emph{local condensate fraction}
 \begin{equation}
 n_{{\rm loc},i}(k=0) = \frac{1}{L} 
 \sum_j \langle b_i^{\dagger} b_j \rangle
 \end{equation}
 and at the \emph{local compressibility}, introduced in  
 Ref. \onlinecite{Wesseletal04}, 
 \begin{equation}
 \kappa_{\rm loc,i} = \frac{\beta J}{L}  
 \sum_j (\langle n_i n_j \rangle
 - \langle n_i \rangle \langle n_j \rangle). 
 \end{equation}
 
 Notice that in the homogeneous bulk system
 with global condensate fraction 
 $n(k=0)$ 
 and global compressibility $\kappa$, we have 
 that $n_{{\rm loc},i}(k=0)= n(k=0)/L$ and 
 $\kappa_{{\rm loc},i} =  \kappa /L$. 
 From Fig.~\ref{f.LDAnopot} we see 
 that the behavior of both $n_{{\rm loc},i}(k=0)$
 and $\kappa_{{\rm loc},i}$ in the trapped
 system follows qualitatively that of the global 
 quantities in the bulk system.
 In particular 
 it is important to notice that  both
 $n_{{\rm loc},i}(k=0)$ and $\kappa_{{\rm loc},i}$ 
 are actually non-local quantities, and 
 their final values depend not only on the 
 local chemical potential at point $i$
 but also on the properties at other points in 
 the trap, and in particular on the total number of 
 particles in the trap. Thus a close agreement
 with the bulk data is generally not expected,
 in particular whenever the correlation
 functions  $\langle b_i^{\dagger} b_j \rangle$
 and $(\langle n_i n_j \rangle
 - \langle n_i \rangle \langle n_j \rangle)$
 become long-ranged in the bulk system, namely
 in the SF phase and at the SF-to-MI boundary,
 respectively.
 Nonetheless $n_{{\rm loc},i}(k=0)$ and 
 $\kappa_{{\rm loc},i}$ capture the short-range
 behavior of those correlators in the trapped
 system around the point $i$; therefore their local enhancement 
 expresses the fact that phase coherence 
 between the points in those regions is strong,
 which is typical of a locally superfluid islands, 
 and that particle number fluctuations in those
 regions are strongly correlated, which is a signature
 of the existence of low-energy and long-wavelength 
 density modes. Conversely a vanishing local 
 compressibility manifests a local gap to particle-hole
 excitations, and it is typically accompanied by a 
 lower local phase coherence. These two combined
 observations form the basis to the LDA 
 interpretation of local MI and 
 SF phases in trap.  This interpretation
 is fundamental to conclude that the MI
 phase is realized at all in the trapped system,
 with a halo of SF
 surrounding it; indeed only an infinite repulsion $U$
 and an infinitely steep trap would allow to realize 
 a \emph{pure} MI in the trap.


\begin{figure}[h]
\begin{center}
\includegraphics[
     width=90mm,angle=270]{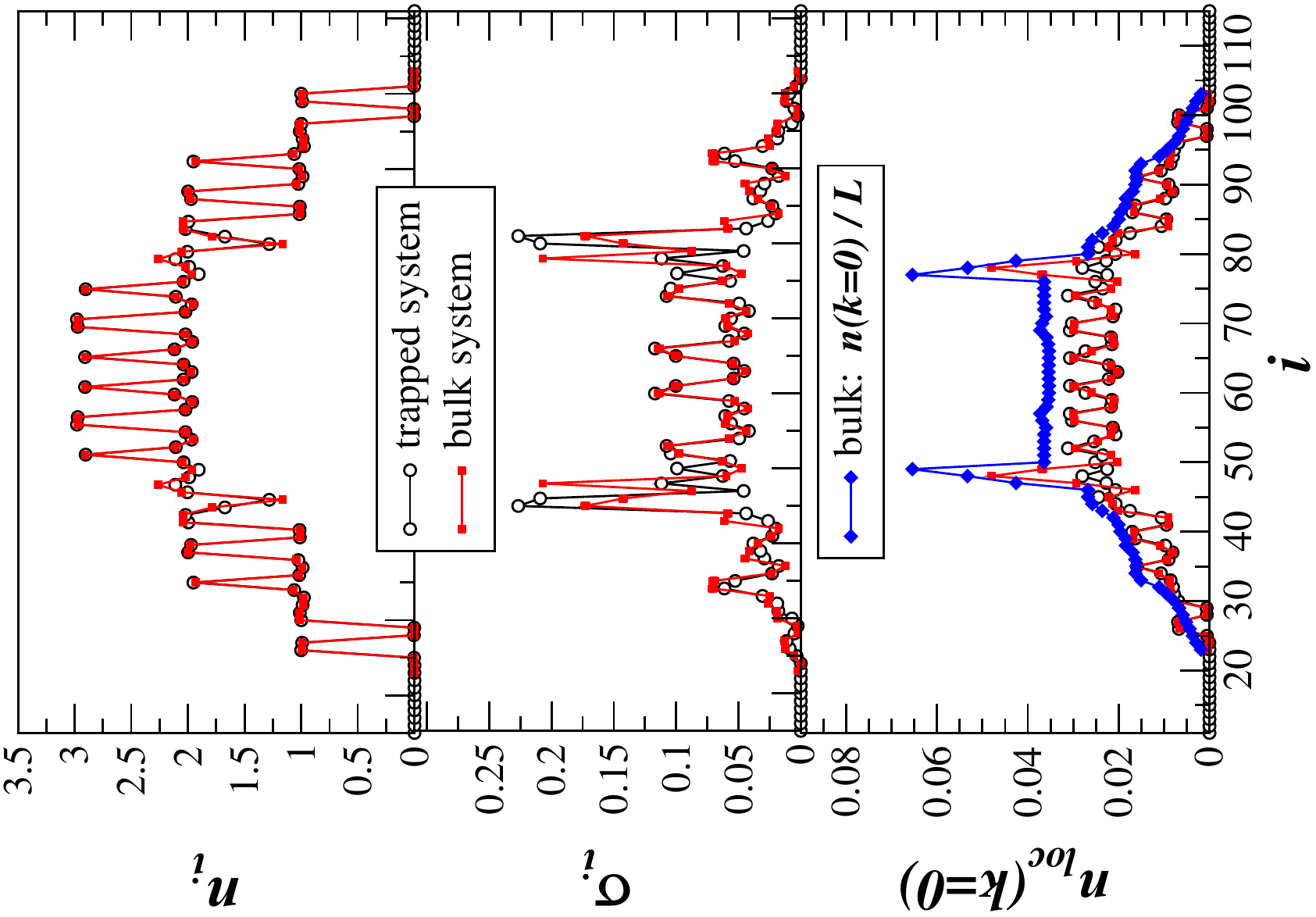} 
\caption{(Color online) Local density, particle-number fluctuations
and coherent fraction (from top to bottom) for a Bose-Hubbard
system in a trap  plus QP potential, with parameters $U=30J$, 
$V_t=0.0014~U$,  $\mu_{\rm trap}= 1.8~U$, and $V_2=U$.
The above quantities are compared to those of the 
bulk system with matching \emph{local} potential 
(Eq.~\eqref{e.bulktrap}).}
\label{f.LDA}
\end{center}
\end{figure}

 \begin{figure}[h]
\begin{center}
\includegraphics[
     width=90mm,angle=270]{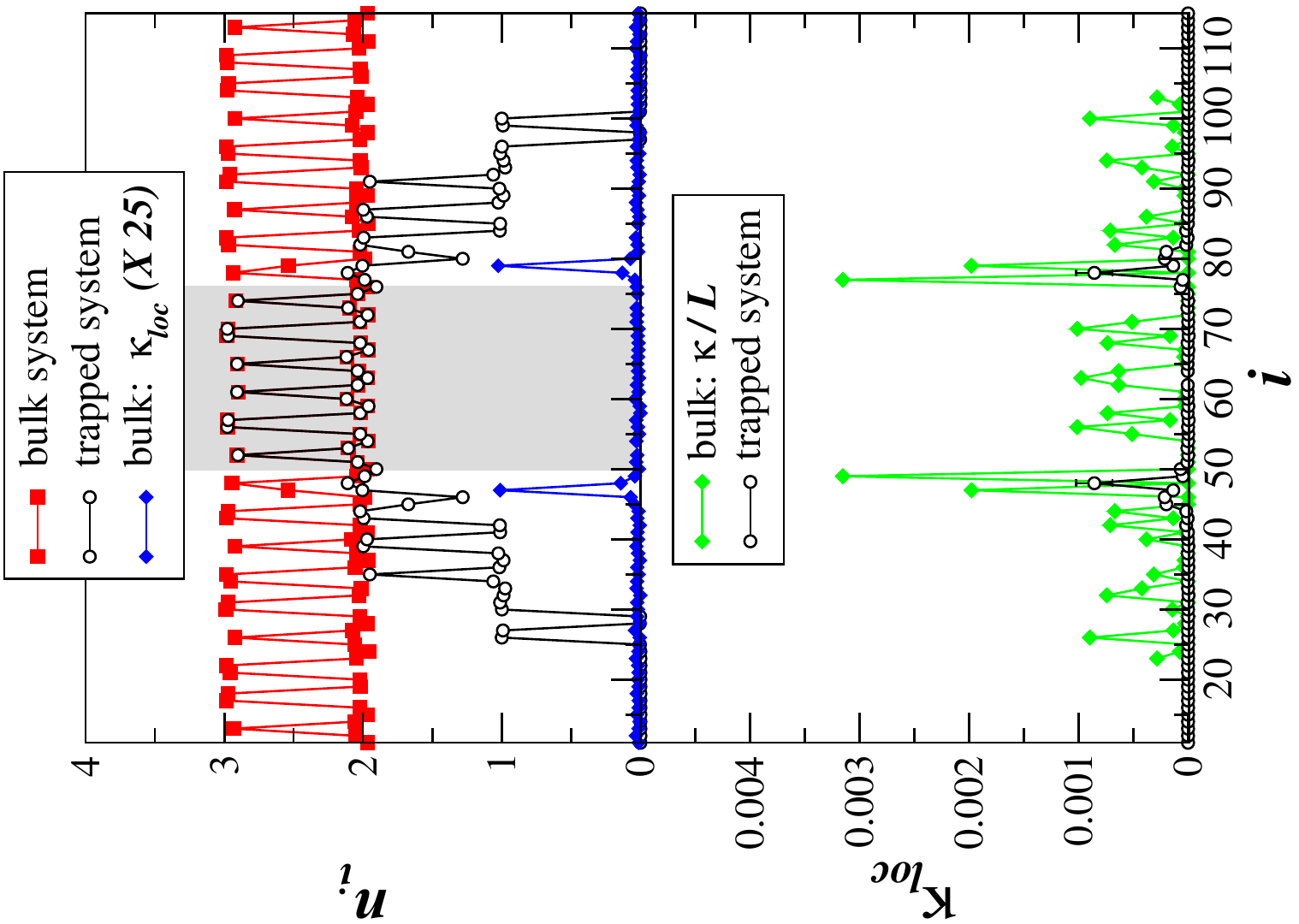} 
\caption{(Color online) Upper panel: local density of the trapped 
system compared
with local density and local compressibility of a bulk system with 
chemical potential $\mu_{\rm bulk} = 1.8~U$. The region where
trapped and bulk density profiles match is emphasized by the shaded 
area (notice that the chemical potentials $\mu_{\rm trap}$
and $\mu_{\rm bulk}$ match perfectly only at the
trap center, but the variation of $\mu_{\rm trap}$ 
due to the trapping potential is slow around the center).
The regions exhibiting a finite local compressibility
in the bulk system are seen to fall out of the region
of matching densities, and consequently the trapped
system is locally incompressible (see following panel). 
 Lower panel: local compressibility of
the trapped system compared with the global compressibility
of the bulk system. All other parameters as in Fig.~\ref{f.LDA}.}
\label{f.LDA2}
\end{center}
\end{figure}

  \subsection{Bose-Hubbard model with QP potential}
\label{ssec.failLDA}

  When introducing the incommensurate cosine potential as in 
 Eq. \eqref{e.hamiltonian}, the picture becomes 
 significantly more complex.
 We will henceforth limit our discussion to the case 
 $V_2 = U$ already discussed in Sections \ref{sec.softcore}
 and \ref{sec.truepseudo}, 
 namely the case in which the MI phase is completely
 removed from the phase diagram and substituted by a variety
 of incommensurate gapped and gapless insulating phases. 
 
 In presence of an incommensurate cosine potential, even the
 bulk system becomes inhomogeneous, and hence the LDA
 has to be rephrased as the approximation relating the 
 properties of a point in the trapped system to those
 of a point in a bulk system experiencing the same local
 chemical potential, resulting from the global chemical
 potential \emph{plus} the cosine potential; in mathematical
 terms, we have to identify a site $i_{\rm trap}$ 
 of the trapped system with a site $i_{\rm bulk}$ of
 the bulk system such that 
 \begin{eqnarray}
  \mu_{\rm loc} &=& \mu_{\rm trap}+ V_t \left(i_{\rm trap}-i_0\right)^2 + 
 V_2~g(i_{\rm trap};\alpha,\phi) \nonumber \\
  &=& \mu_{\rm bulk}+ V_2~g(i_{\rm bulk};\alpha,\phi)
 \label{e.bulktrap}
 \end{eqnarray}
 where $\mu_{\rm trap}$ and $\mu_{\rm bulk}$ are the global 
 chemical potentials for the trapped and the bulk system,
 respectively.
  Fig.~\ref{f.LDA} compares the results for the 
  bulk and the trapped system following the prescription
  of Eq.~\eqref{e.bulktrap} in the case of a strongly
  interacting system in a deep incommensurate potential,
  $U/J = V_2/J = 30$. With these parameter values the
  bulk system lies in the insulating region 
  for a broad range of chemical potentials. 
  In what follows we consider
  $\mu_{\rm trap}/U = 1.8$ so that the local chemical
  potential in the trapped system, which is upper-bounded
  by the above value, experiences a superfluid
  region of the bulk phase diagram 
  (Fig.~\ref{f.softcore}) only around the value $\mu_{\rm loc}/ U = 1.5$.
  
   Despite the high inhomogeneity
  of both the bulk and the trapped system, we generally observe 
  a good quantitative agreement between their respective data for
  the local average density, local density fluctuations
  and local coherent fraction. For the interval of chemical 
  potentials spanned by $\mu_{\rm loc}$ only two nearby
  critical points are present in the bulk phase diagram
  around $\mu_{\rm loc}/U = 1.5$, so that, away from those
  two points, the system is deep in an insulating region
  with a short correlation length, and finite-size effects
  are therefore not severe.   
  The problematic aspect of LDA in presence of the QP
potential is nonetheless the subsequent interpretation of the 
trapped system as locally exhibiting a phase of the bulk
system. In particular, what is the meaning of a \emph{local Bose
glass}? 

In this respect, the examination of the local compressibility
compared with the global one is illuminating. As shown in 
Fig.~\ref{f.LDA2}, the \emph{local} compressibility of 
the trapped system is in strong disagreement with the 
\emph{global} one,
and it is generally vanishing when the global one
is not. This is not at all surprising in a strongly inhomogeneous 
system, in which the global 
compressibility is due to the local response of disconnected 
regions to the variation of the chemical potential,
and it is related to the existence of localized low-energy 
particle-hole excitations. Although a region in the trap might
be at a chemical potential which would correspond to
a BG phase in the bulk phase diagram, it is 
very likely that such region does \emph{not} correspond
to the region of the bulk system that is hosting the 
locally quasi-gapless excitation and which is hence
exhibiting a finite local compressibility. This is 
indeed the case for what depicted in Fig.~\ref{f.LDA2},
where we show the local compressibility of the bulk
system for a chemical potential corresponding 
in the trap to a locally \emph{incompressible} region:
as observed there, the bulk system has a finite
compressibility at that chemical potential, 
coming from the local compressibility of regions 
which do \emph{not} correspond to the ones reproduced
in the trap (shaded areas). Hence it is evident
how the finite-size effects induced by the trap are 
drastically altering the local behavior of the  
system with respect to that of the bulk one.

 \subsection{Structure of the particle-hole excitations}
 \label{ssec.ph}

  The above results require to discard the LDA picture
 of the trapped system locally mimicking the behavior
 of the bulk system when it comes to the BG phase.
 At the same time, the very picture of a local 
 realisation of collective quantum phases in a trapped 
 system might be a limiting point of view.
 As pointed out in Section \ref{sec.truepseudo}, 
 the existence of gapless
 particle-hole excitations originating from 
 rare regions in the system is at the core of the BG phase: 
 \emph{e.g.} accidentally
 degenerate regions separated by a potential barrier, 
 for which tunneling creates a pair of quasi-degenerate
 symmetric/antisymmetric superposition states similarly
 to the double-well problem; or rare regions with a locally
 uniform or periodic potential, which host local low-energy
 excitations.

\begin{figure}[h]
\begin{center}
\includegraphics[
     width=90mm,angle=270]{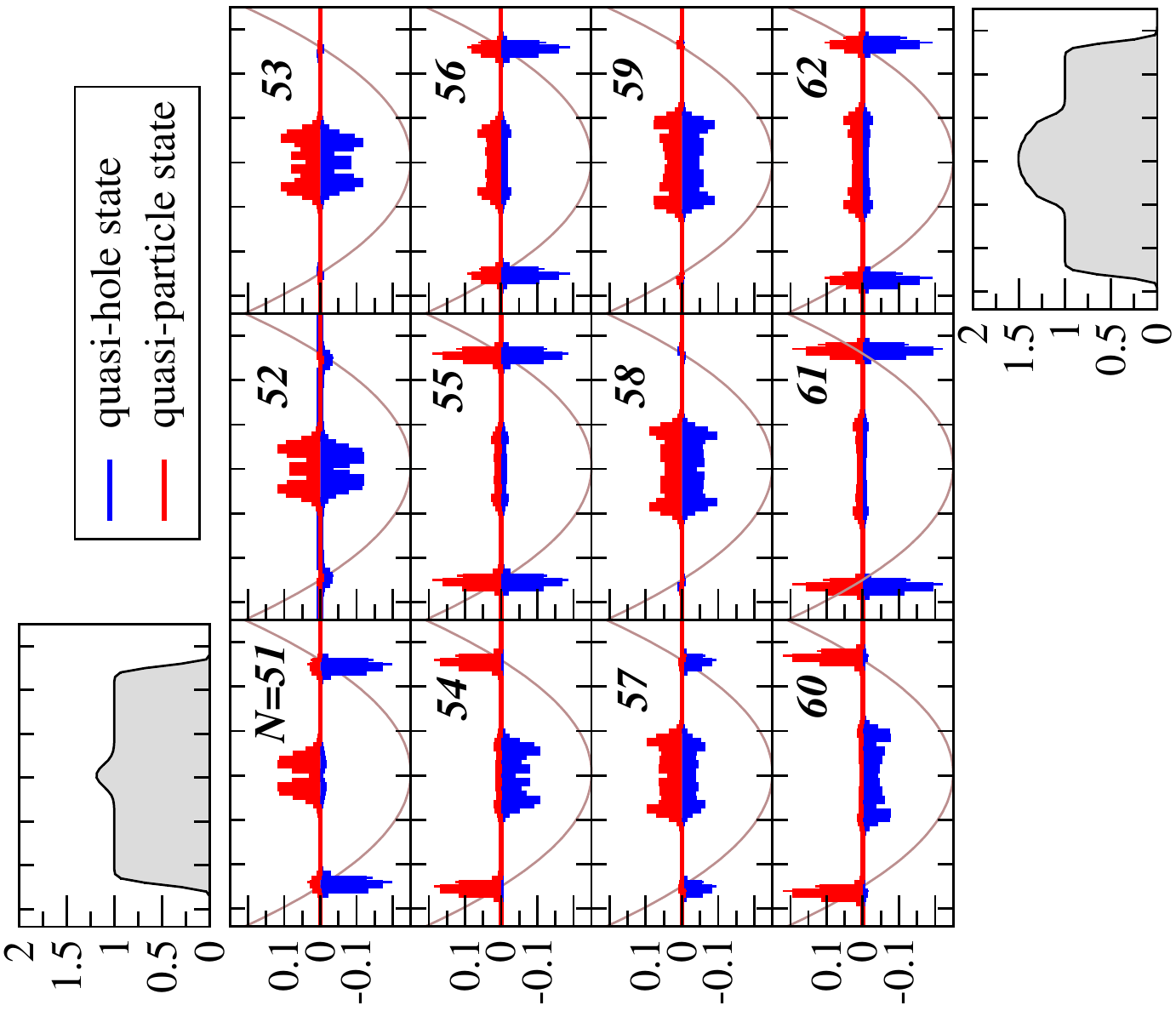} 
\caption{(Color online) Central panels: quasi-particle
($|\psi_{\rm quasi-p}^{(N)}(i)|^2$)
and quasi-hole ($-|\psi_{\rm quasi-h}^{(N)}(i)|^2$)
densities for the Bose-Hubbard model in a trap at various
particle numbers $N$, and with 
parameters $U=20 J$, $V_t=0.0014~U$. A rescaled
picture of the parabolic trap is reported in each 
panel for reference. 
Upper and lower panels:  total 
density profile associated with the minimum ($N=51$) and 
maximum ($N=62$) number of particles shown here.}
\label{f.ph-nopot}
\end{center}
\end{figure}

\begin{figure}[h]
\begin{center}
\includegraphics[
     width=45mm,angle=270]{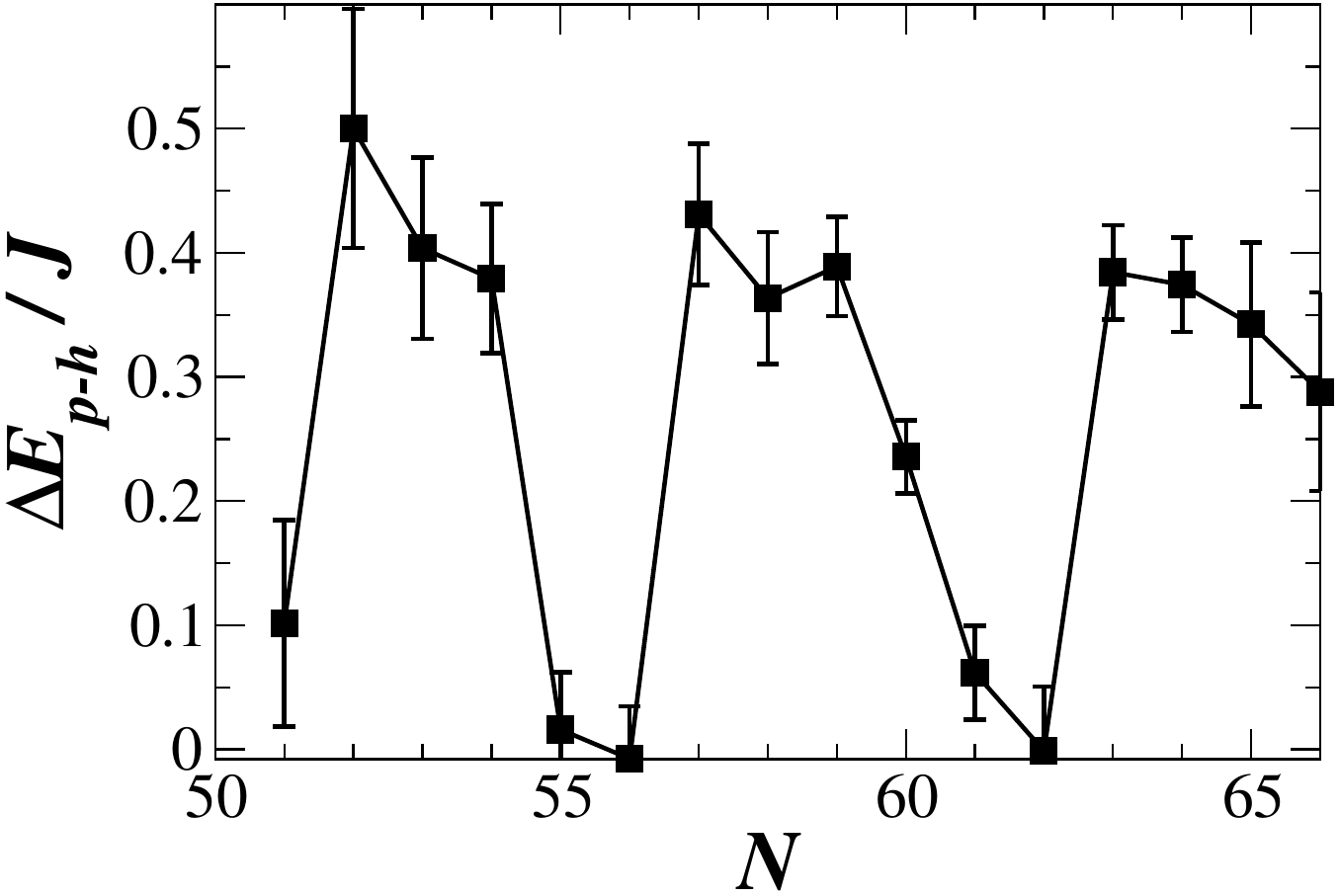} 
\caption{Particle-hole gap $\Delta E_{\rm p-h}$
for the Bose-Hubbard model in a trap. Parameters
as in Fig.~\ref{f.ph-nopot}.}
\label{f.Eph-nopot}
\end{center}
\end{figure}
 
  As we have seen, the inhomogeneity of the QP potential
  is what causes a trapped system to fail
  in reproducing locally the global properties
  of a bulk system at the same chemical potential. 
  At the same time it is conceivable that, in presence
  of the trap, accidental degeneracies might occur between local 
  quasi-particle states living at \emph{different} 
  trapping potentials,
  namely accidental degeneracies which are not   
  observed in the bulk system. To directly investigate
  this possibility, we have run a canonical Monte Carlo
  simulation in which we add particles
  progressively to the system. We then define \emph{effective} 
  quasi-particle and quasi-hole states $\psi^{(N)}_{\rm quasi-p}$
  and $\psi^{(N)}_{\rm quasi-h}$ such that their associated
  density distribution corresponds to the density variation
  of the system by adding/removing a particle, respectively:
  \begin{eqnarray}
  \left|\psi_{\rm quasi-p}^{(N)}(i)\right|^2 &=& n_i^{(N+1)} - n_i^{(N)} \\
  \left|\psi_{\rm quasi-h}^{(N)}(i)\right|^2 &=& n_i^{(N)} - n_i^{(N-1)}.
  \end{eqnarray}
  Looking at the spatial structure of the quasi-particle and
  quasi-hole state corresponding to the same particle 
  number $N$, we generally obtain information about the structure 
  of the particle-hole excitations of the system at particle
  number $N$.
    
   In fact, starting from $N-1$ particles, the energy cost of adding an 
   extra particle is given by the kinetic energy plus the interaction 
   of the extra particle with the background of the $N-1$ particles 
   already present in the system, and analogously for adding
   an extra particle to go from $N$ to $N+1$.
    If the density variation associated with growing
   the particle number from $N-1$ to $N$ happens
   in a different spatial region with respect to the variation
   caused by growing the number from $N$ to $N+1$, 
   then the quasi-hole state 
   $\psi_{\rm quasi-h}^{(N)}$
   and the quasi-particle state $\psi_{\rm quasi-h}^{(N)}$
   are non-overlapping. In presence of 
   short-range interactions as in the Bose-Hubbard model, 
   this implies that the addition of the $N$-th
   particle does not influence the background with
   which the $(N+1)$-th particle interact, which implies,
   in more coincise terms, that the quasi-particle and
   the quasi-hole are not interacting. Thus, working
   at particle number $N$, the creation of a quasi-hole
   in state $\psi_{\rm quasi-h}^{(N)}$ and of a quasi-particle
   in state $\psi_{\rm quasi-p}^{(N)}$ represents the true
   lowest-energy particle-hole excitation of the system, 
   with energy $\Delta E_{N} = \Delta E_{\rm p-h} = 
   E^{(0)}_{N+1} + E^{(0)}_{N-1} - 2E^{(0)}_N$. 
   
    If, on the contrary, the quasi-particle and quasi-hole 
   states for particle number $N$ do overlap, then in 
   principle the energy $\Delta E_{\rm p-h}$ needs 
   a correction coming form the interaction 
   in order to be identified with the energy
   of the elementary particle-hole excitation. Yet, if 
   $\Delta E_{\rm p-h}$ is found to be small ($\ll U$ for 
   definiteness), then the interaction
   between the $N$-th added particle with the background of $N-1$
   particles, entering in $E^{(0)}_N-E^{(0)}_{N-1}$ difference,
   has essentially the same energy as that of 
   the interaction between the $(N+1)$-th particle and the
   background of $N$ particles, entering in 
   $E^{(0)}_{N+1}-E^{(0)}_{N}$ difference; this in turn
   implies that the $N$-th particle minimally alters
   the interaction between the $(N+1)$-th particle and
   the remaining $N-1$, namely the quasi-particle and the 
   quasi-hole states are again effectively independent,
   and they are moreover essentially degenerate.
   If not even the above condition on $\Delta E_{\rm p-h}$
   is satisfied, 
   we can still generally assume that, in presence of repulsive
   interactions, $\Delta E_{\rm p-h}$ represents an upper bound 
   to $\Delta E_{N}$,\cite{upperbound}
   and that the spatial structure of the
   true particle-hole excitations is approximately reproduced
   by that of $\psi_{\rm quasi-h}^{(N)}$ and 
   $\psi_{\rm quasi-p}^{(N)}$.
          
   In the case of a trapped system without QP
  potential, the quasi-particle and quasi-hole states
  are shown in Fig.~\ref{f.ph-nopot} for a selected 
  range of particle numbers $N$, and the associated
  particle-hole gap $\Delta E_{\rm p-h}$ is shown
  in Fig.~\ref{f.Eph-nopot}. We generally observe
  that the quasi-particle/quasi-hole states live
  in the locally superfluid regions, as expected from
  the information on the local compressibility. For the 
  particular density we have chosen in Fig.~\ref{f.ph-nopot}
  the system exhibits two such regions, at local filling 
  $n<1$ on the wings and at filling $n>1$ in the center.
   We generally observe
  three different types of particle-hole pairs, with
  $\Delta E_{\rm p-h} \ll U$ in all cases:
  1) pairs where quasi-particle and quasi-hole 
  both live in the center of the trap --- these are the excitations
  which more closely mimic those of a finite-size SF
  system; 2) pairs where quasi-particle and quasi-hole 
  both live in the wings of the trap --- these states
  correspond to the 
  symmetric and antisymmetric superpositions of the left- and right-localized
  states in the wings, and the particle-hole gap $\Delta E_{\rm p-h}$
  is minimal in this case, because the large barrier
  provided by the intermediate MI region
  leads to quasi-degeneracy 
  (see for instance the case of $N=55,56,61,62$);
  3) particle-hole pairs with a particle
  living in the wings and the hole in the center or viceversa.
  The excitations 2) and 3) are \emph{specific} of the trapped
  system; in particular it is noteworthy that quasi-degeneracy 
  exists between states related to different trap regions,
  hence the picture of local phases described in the previous
  paragraph does not extend to excitations.

\begin{figure}[h]
\begin{center}
\includegraphics[
     width=90mm,angle=270]{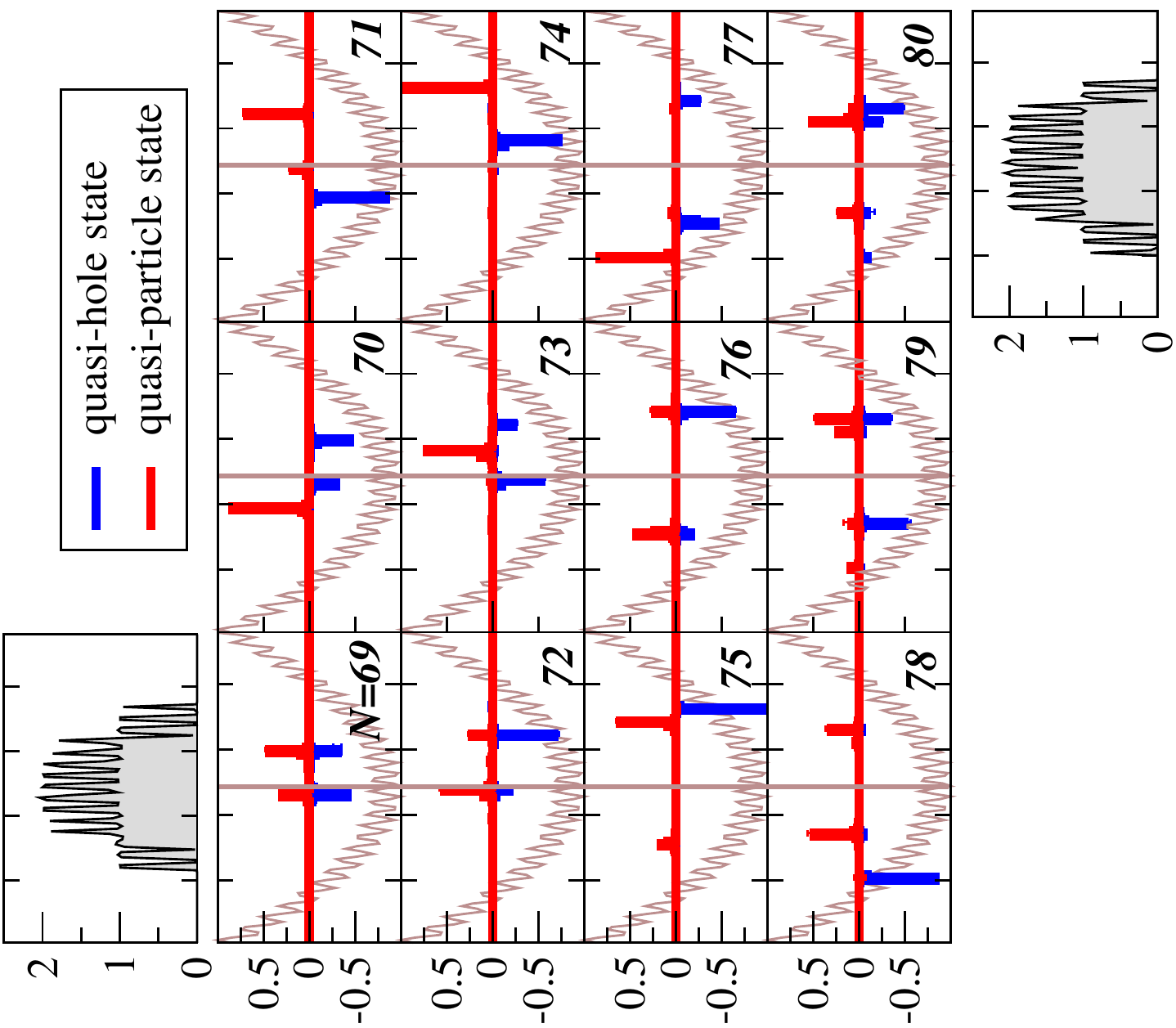} 
\caption{Central panels: quasi-particle
($|\psi_{\rm quasi-p}^{(N)}(i)|^2$)
and quasi-hole ($-|\psi_{\rm quasi-h}^{(N)}(i)|^2$)
densities for the Bose-Hubbard model in a trap plus
QP potential at various
particle numbers $N$, with $V_2=U$. Other parameters
as in Fig.~\ref{f.ph-nopot}. A rescaled
picture of the parabolic trap plus the QP potential 
is reported in each panel for reference, together with
the trap axis. 
Upper and lower panels:  total 
density profile associated with the minimum ($N=69$) and 
maximum ($N=80$) number of particles shown here.}
\label{f.ph}
\end{center}
\end{figure}

\begin{figure}[h]
\begin{center}
\includegraphics[
     width=45mm,angle=270]{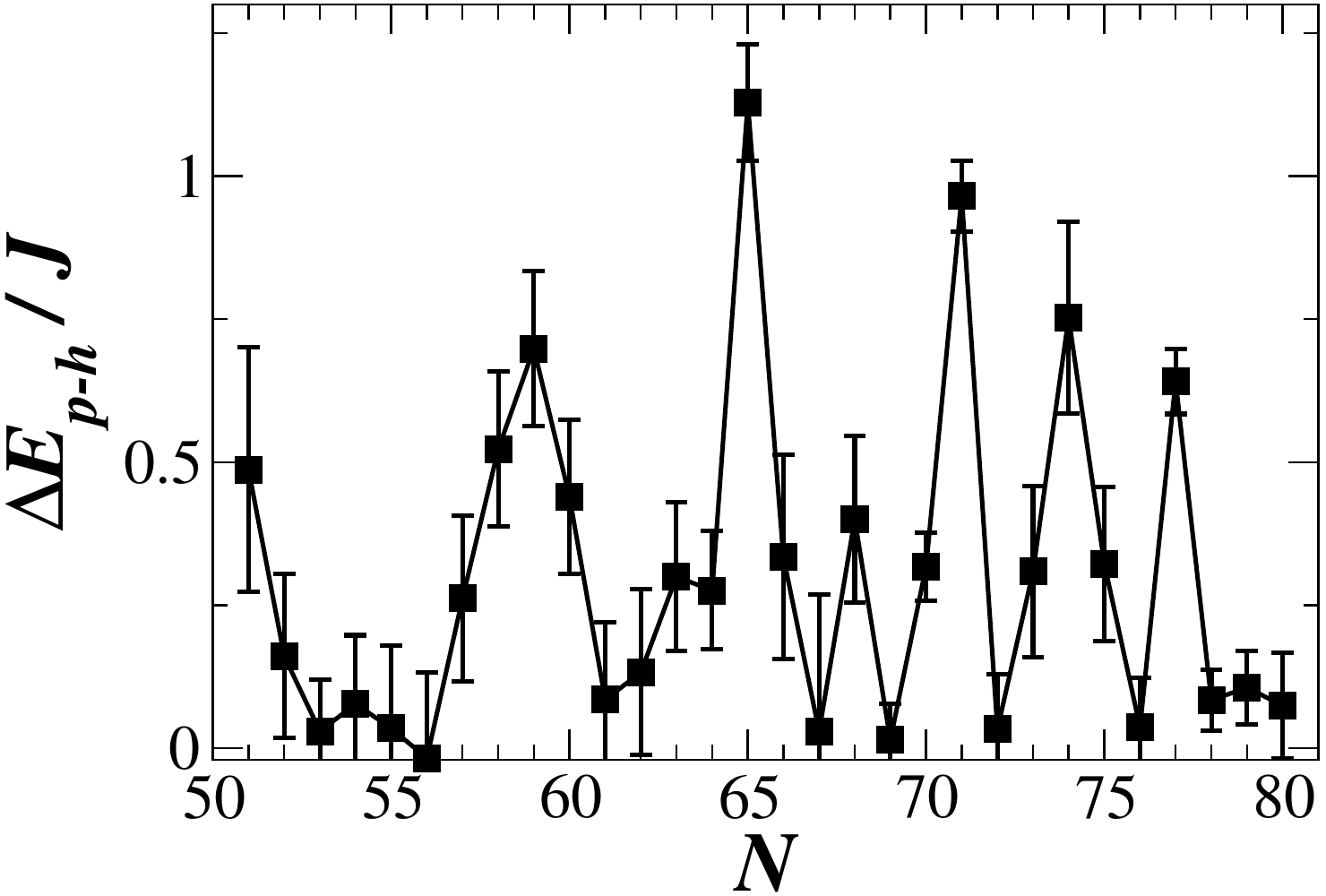} 
\caption{Particle-hole gap $\Delta E_{\rm p-h}$
for the Bose-Hubbard model in a trap plus QP potential. 
Parameters as in Fig.~\ref{f.ph}.}
\label{f.Eph}
\end{center}
\end{figure}

   In presence of a strong QP potential with $V_2=U$, 
  the structure of the excitations
  changes drastically, although some of the features 
  observed before persist, namely the cross-talk between 
  different trap regions. Fig.~\ref{f.ph} shows the spatial
  structure of quasi-particle/quasi-hole states, and 
  Fig.~\ref{f.Eph} the particle-hole gap.
  It is evident that the strong QP potential leads to tight
  localization of the added particles/holes in the system,
  typically around a few lattice sites; delocalization of 
  the quasi-particle/quasi-hole state onto two disconnected
  regions is due to the symmetric-antisymmetric combination
  of quasi-degenerate localized states (as before, these
  combinations give rise to the lowest particle-hole gaps,
  as \emph{e.g.} for $N=69,72,76$, etc.). 
   Furthermore, quasi-particle
  and quasi-hole states are very often \emph{decoupled}
  spatially (see for instance the cases $N=70,71,73,74, 75,77,78$).
  This means that the lowest-energy particle-hole excitations
  are not at all confined to equipotential regions,
  but they can be associated with accidental 
  quasi-degeneracies of quasi-particle and quasi-hole
  states between regions at quite different trapping 
  potentials. Hence the notion of ``local (quasi-)gaplessness"
  is too restrictive, and the picture of a local
  BG, which we have seen to break down in the 
  previous discussion, can be superseded by the 
  possibility of having a \emph{global} trapped BG,
  in which low-energy particle-hole excitations, albeit 
  corresponding to the transfer of a quasi-particle between two 
  localized states, can also occur between distant points 
  in the trap. 
  
\begin{figure}[h]
\begin{center}
\includegraphics[
     width=80mm,angle=270]{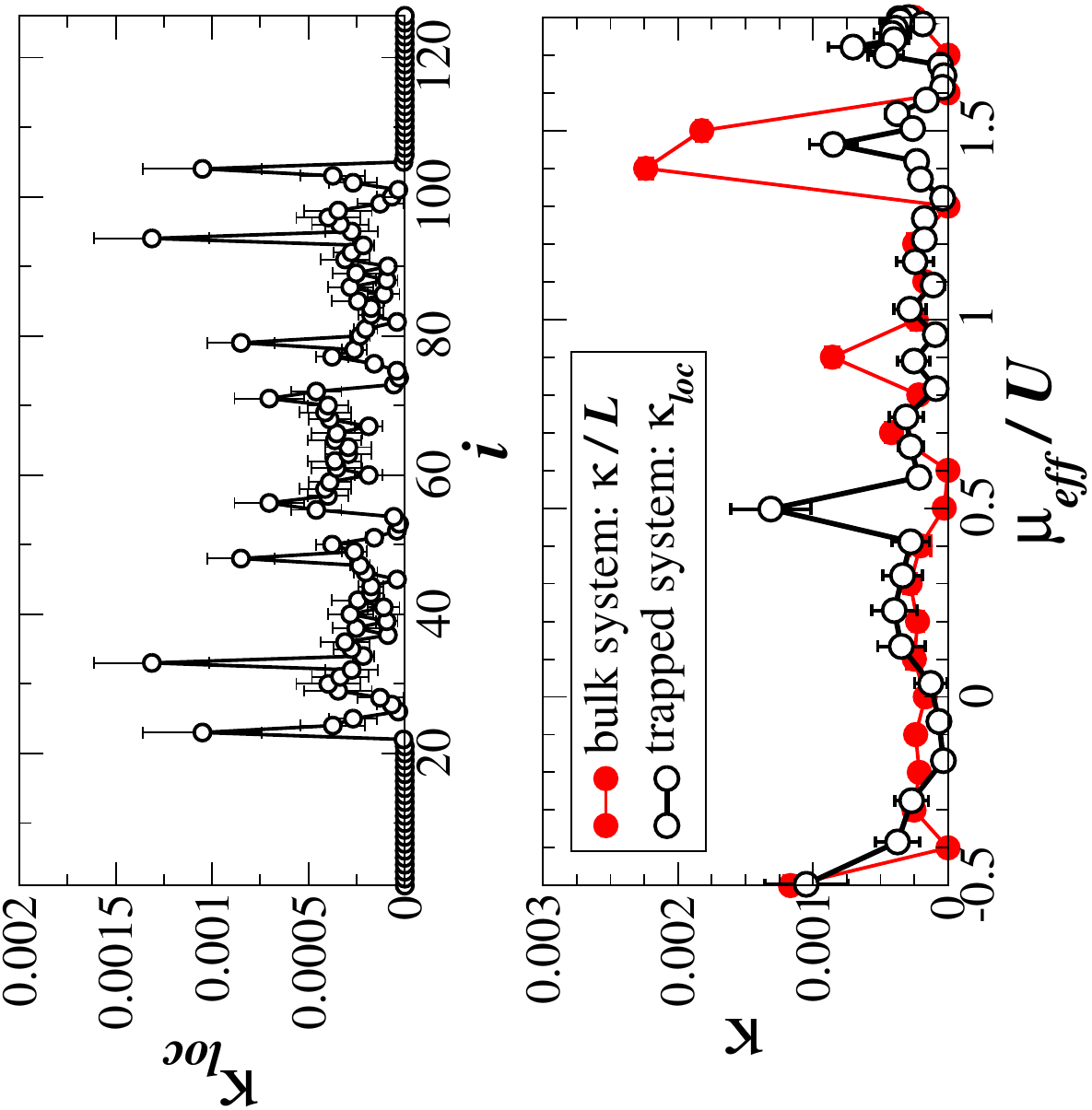} 
\caption{(Color online) Upper panel: phase-averaged local compressibility 
for the Bose-Hubbard model in a QP potential, with parameters
$U=25 J$, $\mu_{\rm trap}=1.8 ~U$, and $V_t=0.0014~U$. 
Lower panel: the phase-averaged local compressibility of the trapped
system is compared to the phase-averaged global
compressibility of the bulk system (see also Fig.~\ref{f.average}) 
by matching the average chemical potentials:
here $\mu_{\rm eff} = \mu_{\rm trap} - V_t(i-i_0)^2$
for the trapped system and $\mu_{\rm eff} = \mu_{\rm bulk}$
for the bulk system.}
\label{f.phiavtrap}
\end{center}
\end{figure}

  \subsection{Bose glass in the trap: 1-color vs. 2-color QP potential} 
  \label{ssec.2color}

   In Section~\ref{sec.LDA} we have seen that for a single
  realization of the QP potential the concept local BG 
  (in the sense of the LDA) breaks down due to
  significant finite-size effects. This feature is actually
  generic of inhomogeneous systems, and we expect it 
  to hold also for a single realization of a truly 
  random potential. The concept of local BG
  can be nonetheless recovered \emph{on average}
  over the statistics of the (pseudo)-disordered
  potential. In the specific case of a QP potential,
  one can randomize its spatial phase as in 
  Sec.~\ref{sec.truepseudo}. 
  Interestingly, from the point of view of optical-lattice
  experiments,
  averaging over fluctuations of the spatial phase 
  of the QP potential is inevitable when averaging
  the results over different experimental runs. 
  In fact the 
  spatial phase can be fixed over the duration
  of a single run, but it is typically changing 
  from run to run \cite{Fallani_private}. 
  
\begin{figure}
\begin{center}
\includegraphics[
     width=50mm,angle=270]{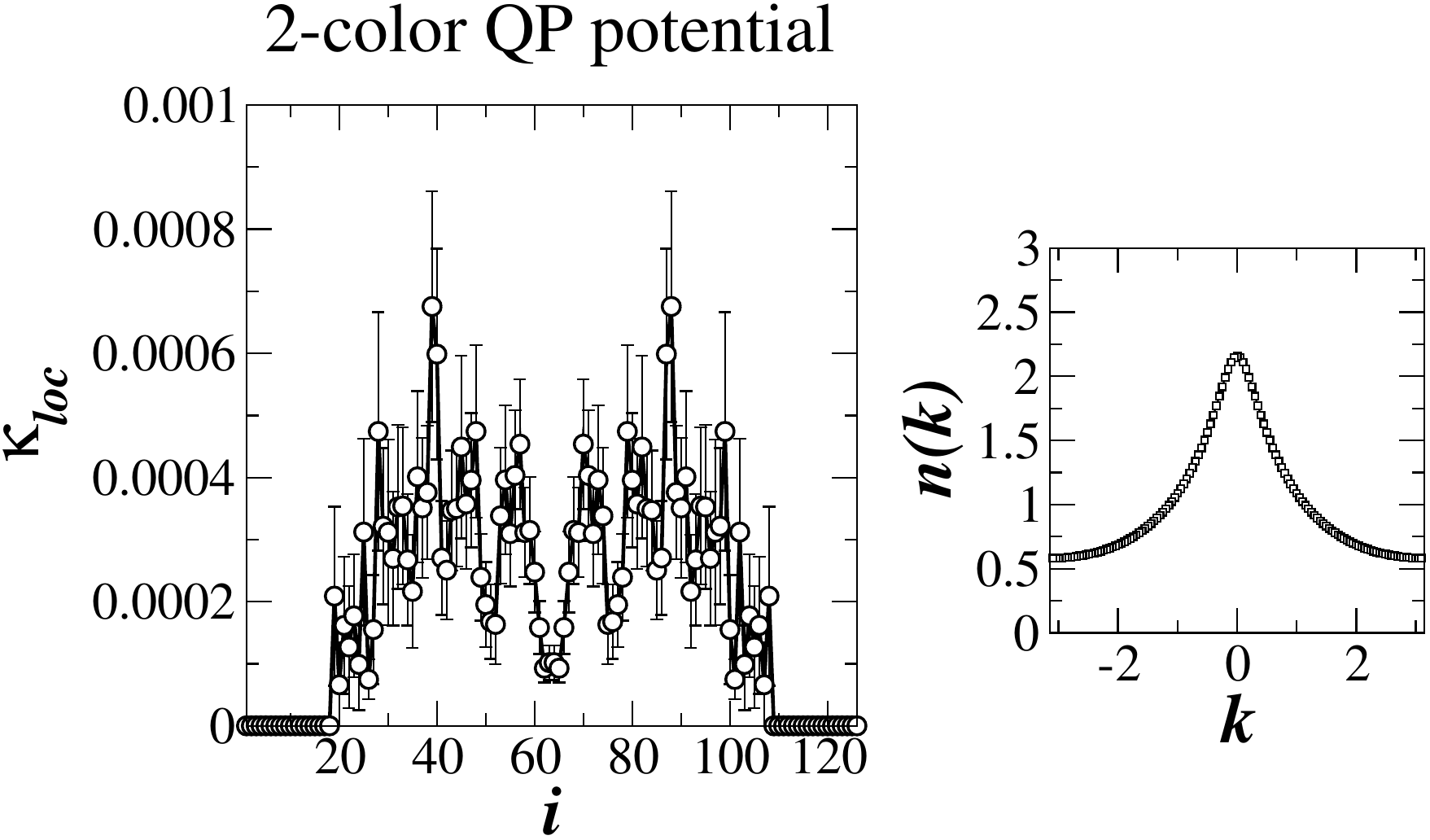} 
\caption{Left panel: local compressibility of the 
Bose-Hubbard model in a trap plus a 2-color QP potential,
averaged over fluctuations of the spatial phases
of the two colors. Here $V_t = 0.0014~U$, other parameters
as in Fig.~\ref{f.3SL}. Right panel: momentum distribution
for the same system.}
\label{f.3colortrap}
\end{center}
\end{figure}

  Randomizing the spatial phase of the QP potential
  implies that the potential
  profile can randomly ``slide'' along the trap. 
  If a region of the trap experiences a local chemical
  potential which corresponds to that of a BG
  phase in the bulk phase diagram, there is in principle
  a finite probability that, by random sliding 
  of the QP potential, a region hosting a
  gapless particle-hole excitation in the bulk system 
  will appear in the trap, so that \emph{on average} 
  that portion of the trap acquires a finite
  local compressibility. 
  Fig.~\ref{f.3colortrap} compares the data, averaged over
  $\phi$-fluctuations, of the rescaled global compressibility
  $\kappa/L$ for the bulk system (already shown in 
  Fig.~\ref{f.average}) with those of the local compressibility
  in the trap, plotted as a function of the effective
  chemical potential induced by the trap, 
  $\mu_{\rm eff} = \mu_{\rm trap}-V_t(i-i_0)^2$. This time
  we observe that, upon $\phi$-averaging,
  most compressible phases in the bulk are generally 
  mirrored by locally compressible phases in the trap,
  and even that there exist locally compressible phases 
  in the trap which correspond to \emph{incompressible} phases
  in the bulk. This suggests that the finite compressibility
  in this case is associated with low-energy excitations 
  which are specific of the trapped system, as discussed in 
  the previous section. 
     
  The above results finally suggest 
 the following  conclusion: 
 a truly random potential or a pseudo-random 
 potential, giving rise \emph{on average} to a 
 BG phase for a continuous range
 of chemical potentials in the phase diagram
 of the bulk system,
 should also be able to give rise  \emph{on average} to a locally 
 compressible phase on \emph{all} the points of 
 the trapped system. To verify this statement in 
 a particular case, we look at 
 the 2-color QP potential of Sec.~\ref{sec.truepseudo}
 in a trap. This potential is seen to satisfy 
 the above condition on the compressibility 
 over an extended range of chemical potentials 
 (see Fig.~\ref{f.3SL}). Fig.~\ref{f.3colortrap}
 shows the local compressibility of this system,
 averaged over spatial-phase fluctuations. 
 In contrast with what observed in Fig.~\ref{f.LDA}
 and Fig.~\ref{f.phiavtrap},
 we have here a finite compressibility everywhere in 
 the trap, on average over $\phi$-fluctuations. At 
 the same time the system
 exhibits a very low coherence which allows to
 rule out quasi-condensation \cite{footnote}.
 Hence we conclude that this system perfectly
 realizes a BG \emph{all over the trap}.

\section{Relevant experimental observables for the trapped system}
\label{sec.observables}

 In this section we show results for two relevant observables
that are currently accessible to time-of-flight measurements
in optical lattice experiments, 
namely the first-order \cite{Greineretal02, Fallanietal07}
and second-order \cite{Grondalskietal99, 
Altmanetal04, Foellingetal05, Spielmanetal07} 
coherence of the bosons.
 The first-order coherence contains the information
on the momentum distribution, Eq.~\eqref{e.nk}. The second
order coherence contains instead the information on the
correlations between the momentum distribution fluctuations 
(also called \emph{noise correlations}):
\begin{equation}
G(k,k') = \langle n(k) n(k') \rangle - \langle n(k) \rangle \langle n(k') \rangle.
\label{e.Gkk}
\end{equation}
What is costumarily measured in experiments \cite{Foellingetal05, Spielmanetal07} 
is actually the average
correlations between the populations at
two momenta differing by a value $q$, normalized
to the factorized correlator:
\begin{equation}
I(q) = \frac{\sum_{k\in {\rm FBZ}} \langle n(k) n(k+q) \rangle}
{\sum_{k\in {\rm FBZ}} \langle n(k) \rangle \langle n(k+q) \rangle}
\label{e.Iq}
\end{equation}
In order to evaluate $G(k,k')$ and $I(q)$ we need in general 
to be able to evaluate 4-point off-diagonal correlators
of the type $\langle b_i^{\dagger} b_j  b_l^{\dagger} b_m \rangle$
within QMC,
and to successively Fourier transform them. This is possible
within a \emph{double-directed-loop} canonical algorithm that we 
explicitly developed for this purpose, in which the evaluation of 
the above cited 4-point correlator is performed during the
update whenever the 4 ends of the two worms find themselves
at the same time slice on the sites $i$,$j$,$l$,$m$. Details
of the algorithm will be reported elsewhere.
 We investigate the evolution of the first- and second-order
coherence upon increasing the strength of the incommensurate
cosine potential. We fix the ratio $U/J$ to two different
values $U/J=5$, $10$, which give rise to two
different states at zero QP potential in a trap 
of strength $V_t/J=0.0014~U$.

\begin{figure}[h]
\begin{center}
\includegraphics[
bbllx=0pt,bblly=0pt,bburx=510pt,bbury=450pt,%
     width=65mm,angle=270]{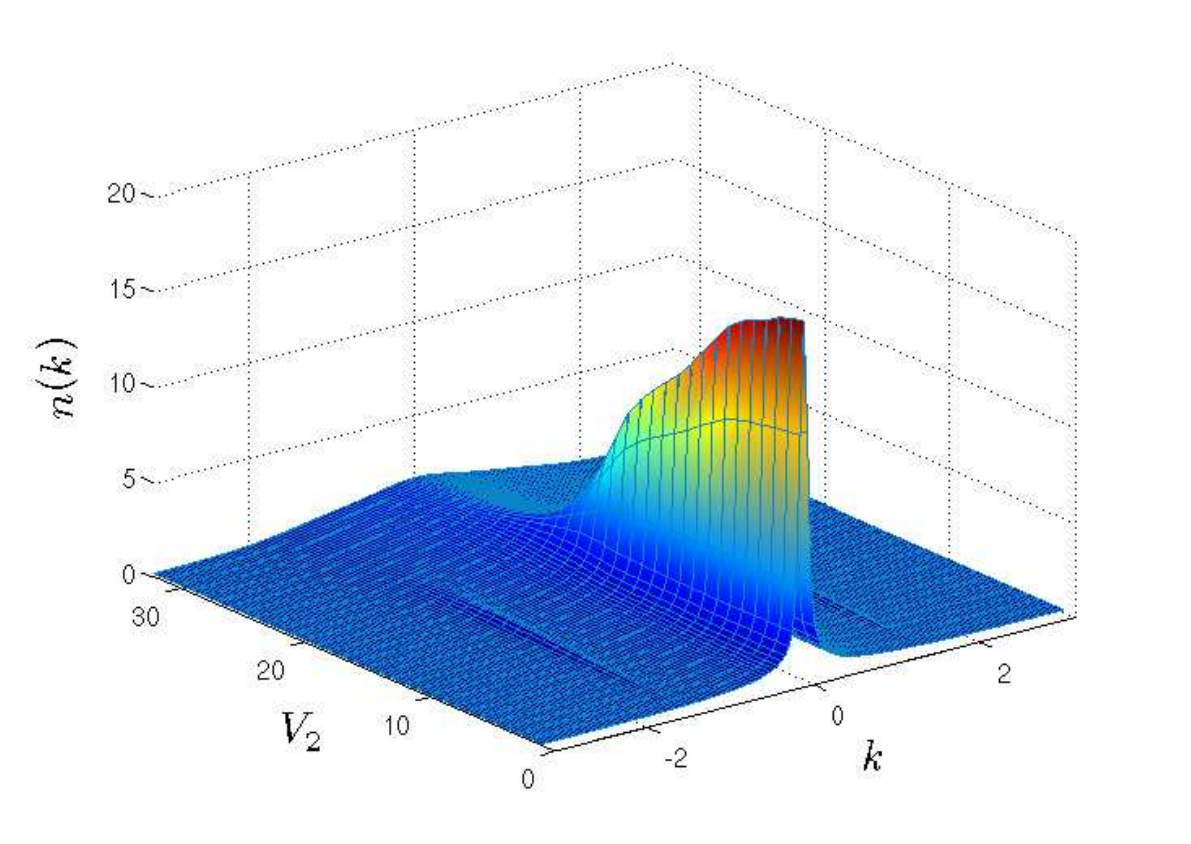}      
 \includegraphics[
     width=65mm,angle=270]{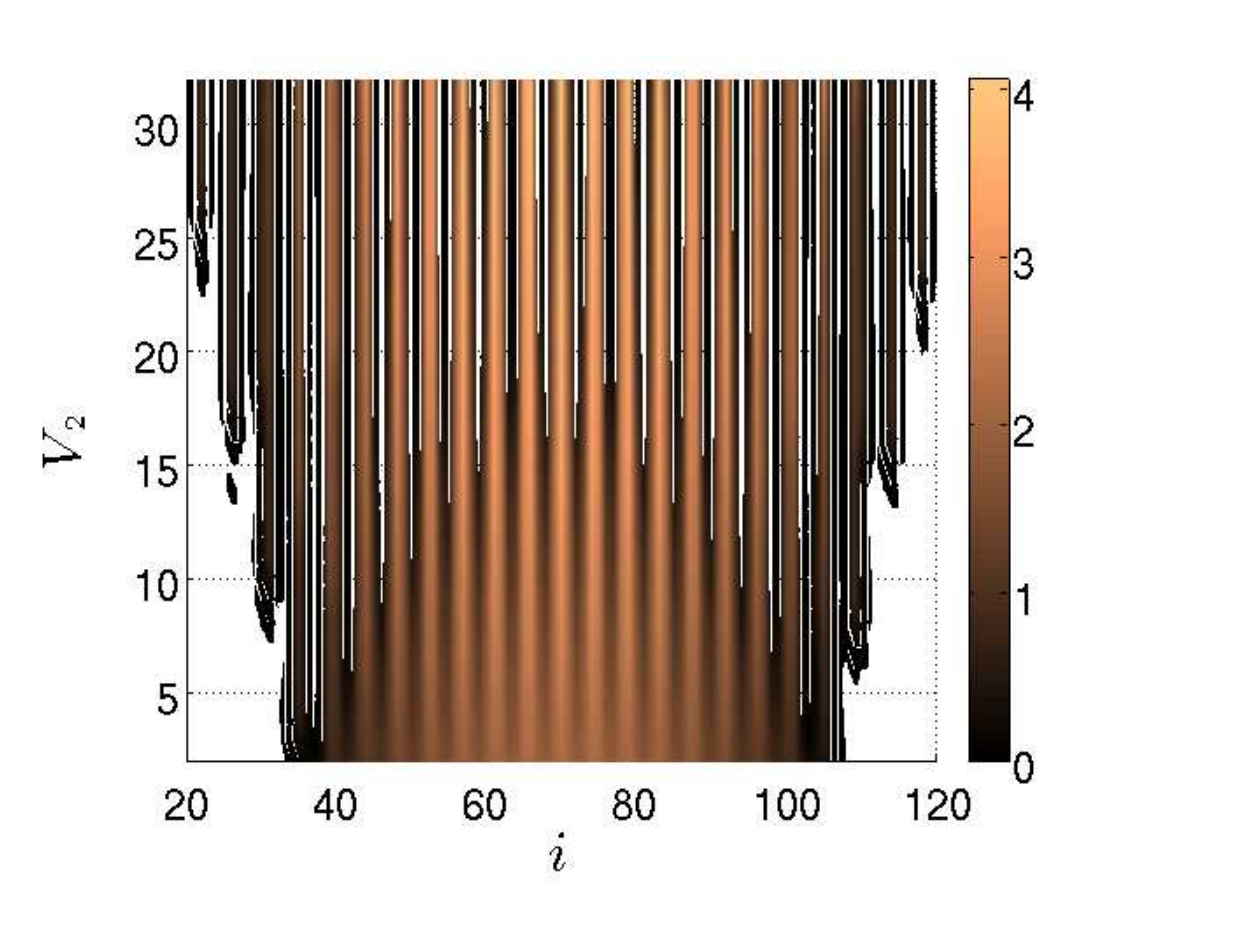}     
\caption{(Color online) Upper panel: momentum distribution for the
Bose-Hubbard model in a trap ($N=100$, $U=5J$, $V_t=0.0014~U$)
in a QP potential of increasing strength $V_2$ (here
represented in units of $J$).
Lower panel: false-color density plot for the same model;
the false colors only apply to finite densities, while
the white regions emphasize the points with zero density.}
\label{f.enkU5}
\end{center}
\end{figure}

For $U/J=5$ the trapped system with $V_2=0$ exhibits a 
high coherent fraction, and the application of the 
QP potential has the effect of
suppressing coherence, driving the system through
a superfluid-to-incommensurate-insulator crossover,
as clearly shown in Fig.~\ref{f.enkU5}. We then 
compare the evolution of the momentum distribution
with that of the density profile 
upon increasing $V_2$ (also shown in Fig.~\ref{f.enkU5}), 
which is not directly measurable in the current experiments.
We notice that the strongest suppression of
the coherent peak at $k=0$ corresponds roughly
to the value of $V_2$ at which the incommensurate
potential introduces unoccupied sites in the central
region of the trap, fragmenting the many-body 
state into droplets.

Along the superfluid-to-insulator crossover the system 
traverses a \emph{modulated superfluid}
phase, in which the coherent fraction remains significant
while the density profile acquires a strong modulation 
due to the external potential. This modulation is 
only marginally revealed in the momentum distribution
by two satellite peaks at the incommensurate beating
momenta $k_{\rm inc} = \pm 2 \pi (1-\alpha)$
(see also Ref.~\onlinecite{Rousseauetal06} for
a similar signature in commensurate superlattices).
This effect can be understood as an effective reduction 
of the first Brillouin zone, due to the longer
(quasi-)period imposed by the incommensurate 
potential; yet its signature is possibly too weak to be 
observed in current optical lattice experiments.
It is also to be observed that such signature 
becomes unobservable in the strongly insulating
regime, where the momentum distribution becomes
essentially featureless.
 
\begin{figure}[h]
\begin{center}
\includegraphics[
bbllx=0pt,bblly=0pt,bburx=510pt,bbury=450pt,%
     width=65mm,angle=270]{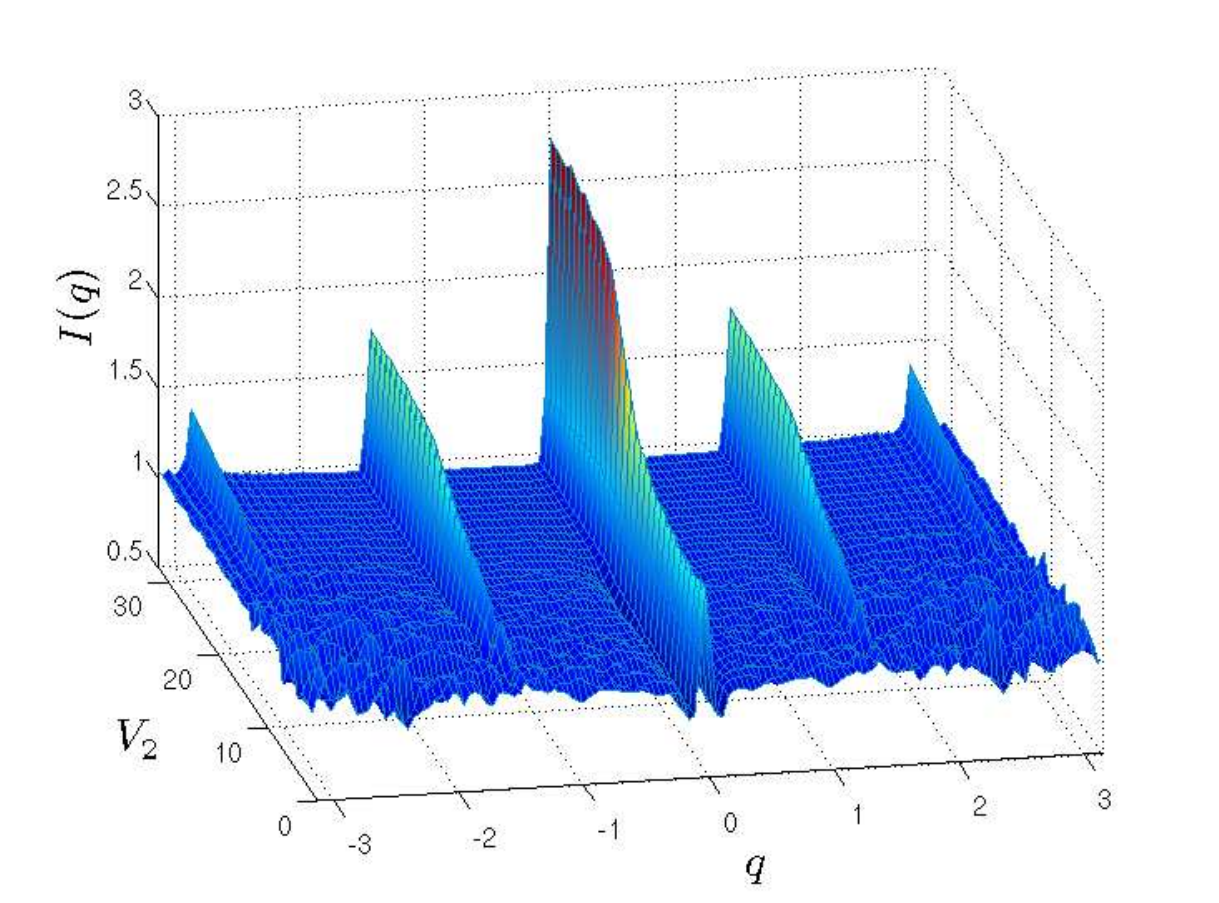} 
\caption{Noise correlations for the Bose-Hubbard
model in a QP potential; model parameters as in 
Fig.~\ref{f.enkU5}.}
\label{f.IqU5}
\end{center}
\end{figure}

 Nonetheless, as shown experimentally in Refs.~
 \onlinecite{Foellingetal05,Spielmanetal07}, 
 the fluctuations of the momentum distribution, 
 captured by Eqs.~\eqref{e.Gkk},\eqref{e.Iq}, have a momentum
 structure which reflects directly the Fourier
 transform of the density-density correlations. 
 Indeed, for a pure Fock state $|\{n_i\}\rangle$
 one can easily show that 
 \begin{equation}
 I(q) = 1 + \frac{N}{L} \delta_{q,0} + \frac{N}{L^2} S(q)
 \label{e.focklimit}
 \end{equation} 
 where $S(q) = 1/L \sum_{ij} \langle n_i n_j \rangle$
 is the static structure factor, and $N$ is the total
 number of particles. In a more general
 superposition state  
 $|\Psi\rangle = \sum_{\{n_i\}}~c(\{n_i\})~|\{n_i\}\rangle$ the 
 above relation
 does not hold; nonetheless, if the state 
 contains only one or
 a few dominant Fock components, the essential
 features of their related structure factor will be
 captured by the integrated noise correlations as in
 Eq.~\eqref{e.Iq}. Fig.~\ref{f.IqU5} shows the
 second-order coherence as a function of the intensity
 of the incommensurate potential. The central peak at
 $q=0$ shows a non-trivial evolution, with a large
 increase corresponding to the suppression of 
 coherence, and the consequent increase in the 
 fluctuations $\langle n(k)^2 \rangle$ at all momenta,
 which contribute to $I(q=0)$. But the most significant feature is the  
 appearence of satellite peaks at $q=k_{\rm inc}$
 and also at $q=2k_{\rm inc}$. The height of
 these peaks appears to saturate around the value 
 of $V_2$ ($V_2/J \sim 20$) at which the coherence
 peak in the momentum distribution gets drastically 
 suppressed, marking the fragmentation of the system into
 droplets. 

\begin{figure}[h]
\begin{center}
\includegraphics[
bbllx=0pt,bblly=0pt,bburx=510pt,bbury=450pt,%
     width=65mm,angle=270]{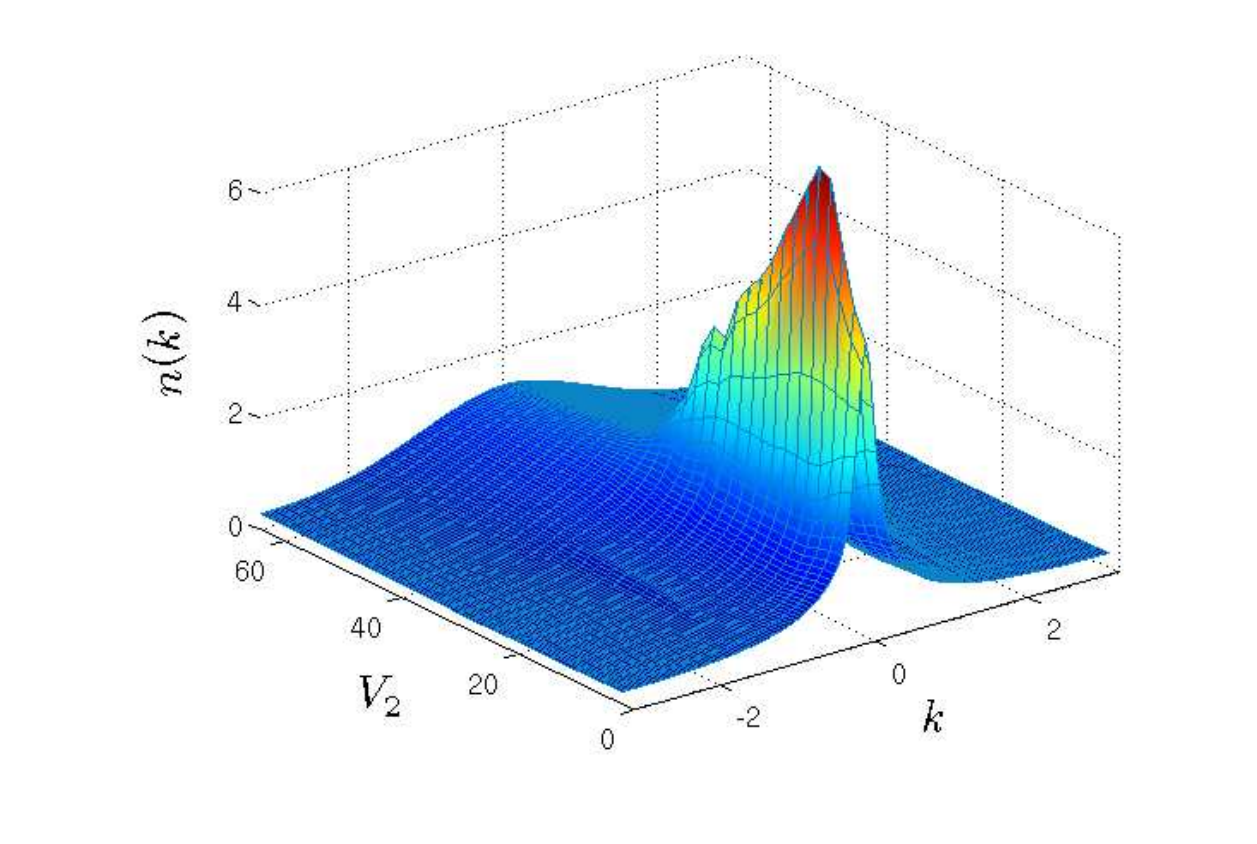} 
\caption{Momentum distribution for the Bose-Hubbard
model in a QP potential with $U=10 J$; other
parameters as in Fig.~\ref{f.enkU5}.}
\label{f.enkU10}
\end{center}
\end{figure} 

\begin{figure}[h]
\begin{center}
\includegraphics[
bbllx=0pt,bblly=0pt,bburx=510pt,bbury=450pt,%
     width=55mm,angle=270]{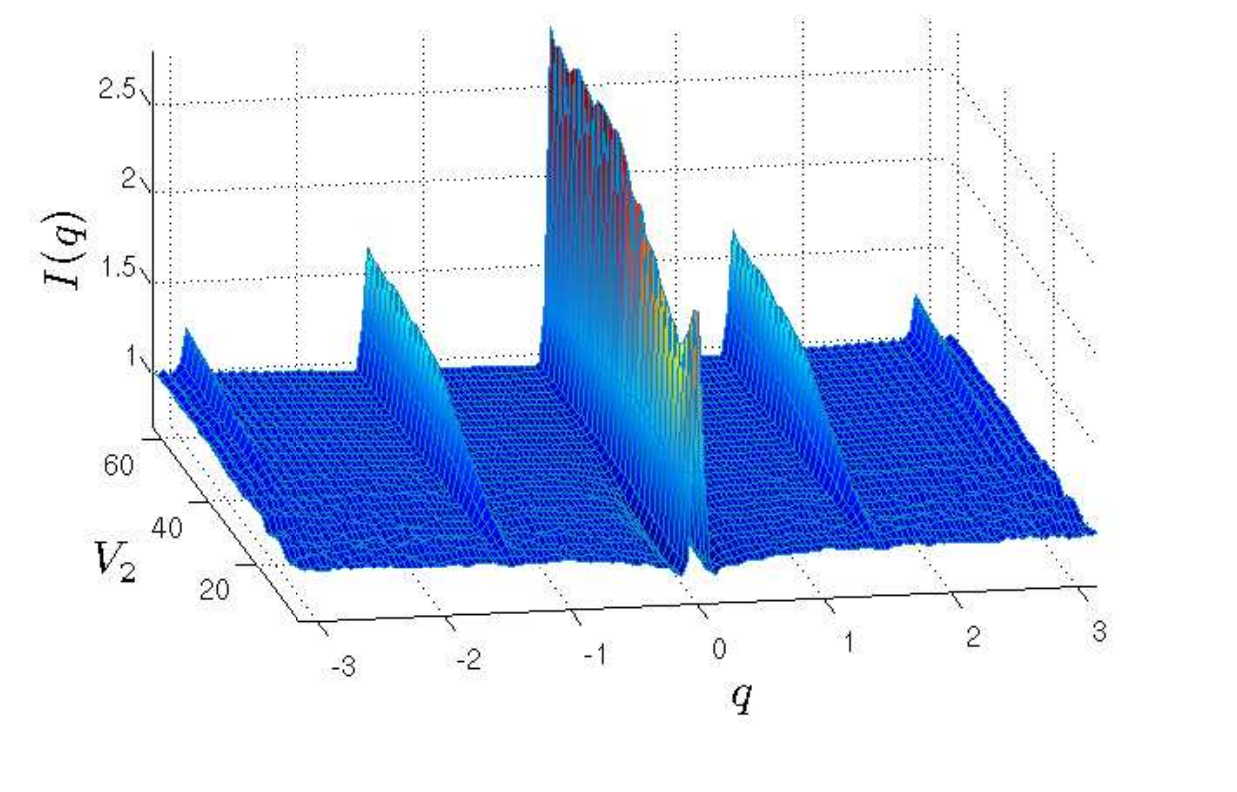} 
\includegraphics[
bbllx=0pt,bblly=0pt,bburx=510pt,bbury=450pt,%
     width=65mm,angle=270]{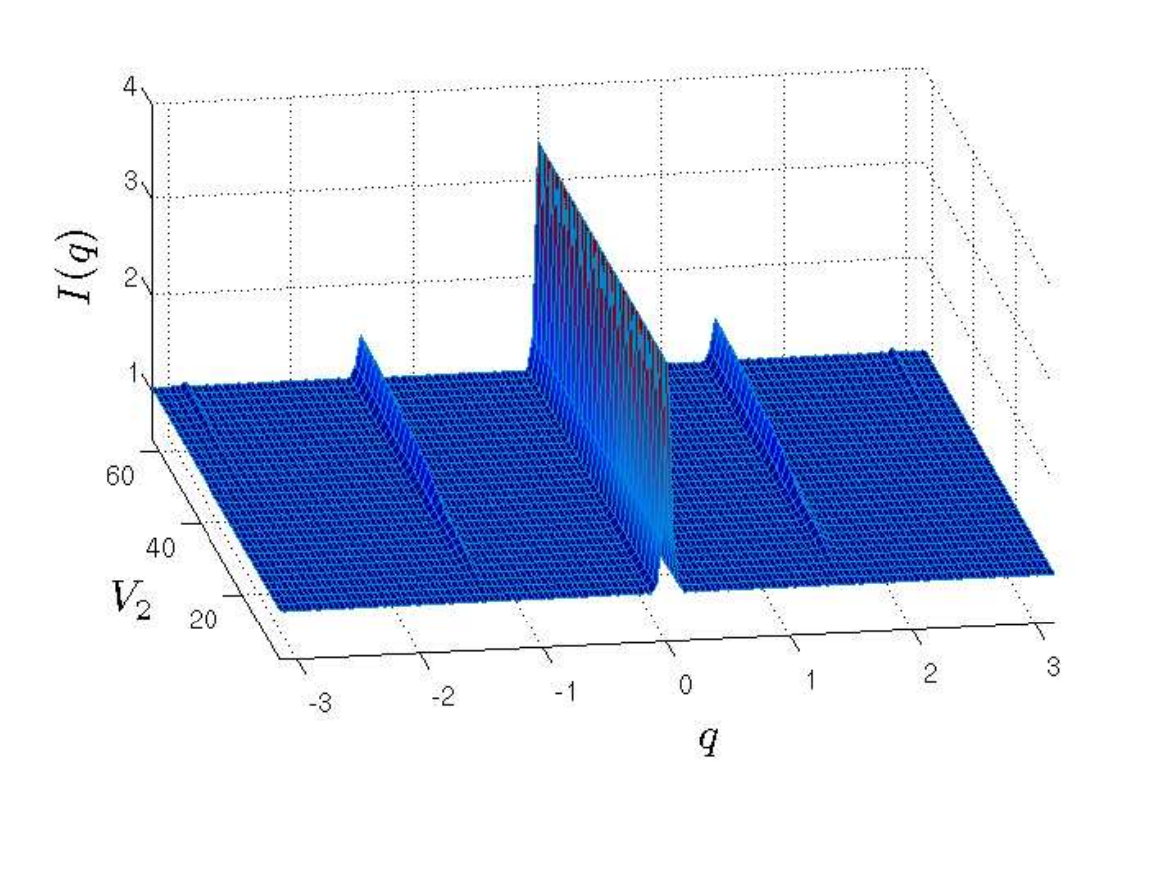} 
\caption{Upper panel: noise correlations for the
Bose-Hubbard model in a QP potential with $U=10 J$;
other parameter as in Fig.~\ref{f.enkU10}.
Lower panel: noise correlations for the classical
limit of the above model, $J=0$. Here $V_2$ is
in units of $U/10$.}
\label{f.IqU10}
\end{center}
\end{figure}

  Increasing the value of $U/J$ to 10, the system at $V_2=0$
 has a significantly lower coherence than in the case $U/J=5$. 
 Interestingly
 the application of the QP potential appears
 to \emph{increase} the coherence in the system, as shown
 in Fig.~\ref{f.enkU10}, an effect due to the simple fact 
 that MI regions are destabilized by 
 the application of the QP potential because
 they obviously do not display the right filling to minimize
 the potential energy. A similar effect of destabilization of 
 the MI by an externally applied potential has been
 previously observed both in the case of a periodic 
 potential \cite{Rousseauetal06} and in the case of 
 a random one \cite{Raghavanetal07}. A further increase in
 the QP lattice leads to a similar crossover towards
 a deep incommensurate insulating state, as seen above for
 the case $U/J=5$. The second-order coherence shown in 
 Fig.~\ref{f.IqU10} also reveals a non-monotonic
 behavior of the $q=0$ peak, where a suppression of
 momentum fluctuations are observed together
 with the enhancement of the $k=0$ peak. Satellite
 peaks at $k_{\rm inc}$ and $2k_{\rm inc}$ appear 
 for large $V_2$ as in the case of a lower $U/J$
 ratio, but it is interesting to notice that the
 momentum structure of $I(q)$ is significantly less
 pronounced than in the case of $U/J=5$, in particular
 the height of the satellite peaks has decreased with
 respect to that of the central one. To further illustrate 
 this trend we have performed an exact calculation of $I(q)$ 
 in the Fock-state limit $J\to 0$, whose results are shown in 
 Fig.~\ref{f.IqU10}, and where it appears that 
 in this limit the $q$-dependent structure of $I(q)$ is even weaker 
 than in the finite-$J$ case. Hence we can conclude that quantum
 fluctuations generally \emph{enhance} the non-trivial $q$-dependent
 part of the $I(q)$ signal with respect to the $q=0$ peak.

  In conclusion, we observe that the first- and second-order
coherence provide full information about the diagonal
and off-diagonal correlations in the system, and they
clearly reveal the crossover (or finite-size transition)
from a superfluid state to an incommensurate insulating state
upon increasing the QP potential.
The detection of the possibly unconventional nature
of the insulating state, namely the presence (albeit
local) of BG regions, is nonetheless beyond
the scope of these observables, and quantities
that directly probe the low-energy particle-hole
excitations are required, ideally
the local compressibility as discussed in Section
\ref{sec.LDA}. Future work will present an experimental
strategy to directly address this issue \cite{Roscildeprep08}.

\section{Validity of the Bose-Hubbard model in an incommensurate
potential}
\label{sec.abinitio}

 Throughout the previous sections we have worked under the assumption that
an incommensurate optical superlattice, realized in the 
experiments \cite{Fallanietal07} by superimposing 
a primary and a secondary standing wave with incommensurate
wavelength relation, 
can be fully described by a simple incommensurate cosine potential
added to the Bose-Hubbard Hamiltonian, as in  Eq.~\eqref{e.hamiltonian}. 
This means that the only effect of the secondary optical lattice
is assumed to be a shift of the local energy of the 
Wannier functions associated with the lowest Hubbard band
of the primary lattice. Yet, the frequency inside each
potential well 
of the superlattice changes in general with respect to that
of the single-color lattice, and, most importantly, 
the relative distances between consecutive wells
are also shifted \cite{Guarreraetal07}.
 These effects modify in principle
 the shape of the Wannier functions,
 leading to a modulation of the $U$ parameter, 
and the overlap between the Wannier functions associated with
two adjacent sites, leading to a modulation of the
$J$ parameter.

 Here we recall the well-known derivation \cite{Jakschetal98} of the 
 Bose-Hubbard model 
 from the most general second-quantized Hamiltonian of a set of interacting 
 bosons in an external potential, which reads
 
 \begin{eqnarray}
 {\cal H} &=& \int d^3r ~\psi^{\dagger}(\bm r) \left(-\frac{\hbar^2}{2m} \nabla^2 +
  V_{\rm opt}(\bm r) \right) \psi(\bm r) \nonumber \\
  &+& g \int d^3r ~\psi^{\dagger}(\bm r) \psi^{\dagger}(\bm r)
  \psi(\bm r) \psi(\bm r)
  \label{e.secondquant}
 \end{eqnarray}
 
  where $\psi$,$\psi^{\dagger}$ are bosonic field operators, 
  $g =  4\pi a_s \hbar^2/m$ with
  $m$ the mass of the bosons and $a_s$ the $s$-wave scattering length,
  and $V_{\rm opt}(x)$ is the optical potential applied to the atoms.
  Following the experimental setup of Fallani \emph{et al.}
  \cite{Fallanietal07}, hereafter we assume that $V_{\rm opt}$ 
  can be written as
  \begin{eqnarray}
  V_{\rm opt} (\bm r) &=& V_{||}(x) + V_{\perp}(y,z) \\
 V_{\perp}(y,z) &=& s_{\perp}
  \left[\cos^2(k_1 y) + \cos^2(k_1 z)\right]E_r  \nonumber \\
  &+& \frac{1}{2} m\omega_{\perp}^2
  \left[(y-y_0)^2+(z-z_0)^2\right] \\
  V_{||}(x) &=& s_1~\cos^2(k_1 x)~E_r 
  + s_2~\cos^2(k_2 x + \phi)~E_r \nonumber \\
  &+& \frac{1}{2} m\omega_{||}^2 (x-x_0)^2 \nonumber 
  \end{eqnarray}
  
  $V_{\perp}$ creates a strong optical lattice potential 
  with wavevector $k_1 = 2\pi/\lambda_1$
  in the $y$ and $z$ direction, defining tubes along the $x$ direction, 
  plus an overall parabolic confinement with frequency $\omega_{\perp}$. 
  Hereafter we consider a strong transverse optical lattice,
  $s_{\perp} = 40$, as obtained in recent experiments
  \cite{Fallanietal07,Stoeferleetal04}, which allows us 
  to neglect the inter-tube hopping and to consider single tubes
  independently.
  The optical potential along the $x$ direction features two optical
  lattices with incommensurate wavevectors $k_1 = 2\pi/\lambda_1$ and
  $k_2 = 2\pi/\lambda_2$, plus 
  a parabolic confinement with frequency $\omega_{||}$. 
  For definiteness we take $\lambda_1=830$ nm, 
  $\lambda_2=1076$ nm, and $m\omega_{||}^2/2 = 0.0012 E_r$
  as in Ref.~\onlinecite{Fallanietal07}.
  $s_1$, $s_2$ and $s_{\perp}$ are dimensionless amplitudes,
  and $E_r = (\hbar k_1)^2/2m$
  is the recoil energy (associated with the primary lattice).

   We then decompose the bosonic field operators onto an
  orthonormal basis of single particle wavefunctions localized around 
  the minima of the optical potential in each tube:
  \begin{equation}
  \psi(\bm r) = \sum_i w_i(x) W_i(y) W_i(z) ~b_i ~,
  \end{equation}  
  where $i$ runs over the minima. Neglecting the overlap between 
  localized wavefunctions 
  on non-adjacent minima, 
  the Hamiltonian parameters of a general one-dimensional 
  Bose-Hubbard Hamiltonian 
  \begin{equation}
  {\cal H} = \sum_i \left[\left(J_{i,i+1} b_i b^{\dagger}_{i+1} + {\rm h.c.}\right)
  +\frac{U_i}{2} ~n_i (n_i-1) + V_i~ n_i \right]
  \label{e.genBH}
  \end{equation}
  can be obtained as usual, with parameters
   \begin{eqnarray}
   J_{i,i+1} &=& \int dx~ w_{i+1}^{*}(x) 
   \left[ -\frac{\hbar^2}{2m} \frac{\partial}{\partial x} + V_{||}(x)\right]
   w_i(x) \label{e.overlap} \\
   U_i &=&  g \int dx~ |w_{i}(x)|^4 
   \int dy~ |W_{i}(y)|^4  \int dz~ |W_{i}(z)|^4 ~~~~~~\\
   V_i &=&  \int dx~ |w_{i}(x)|^2 V_{||}(x)        
   \end{eqnarray}
   
\begin{figure}[h]
\begin{center}
\includegraphics[
     width=60mm,angle=270]{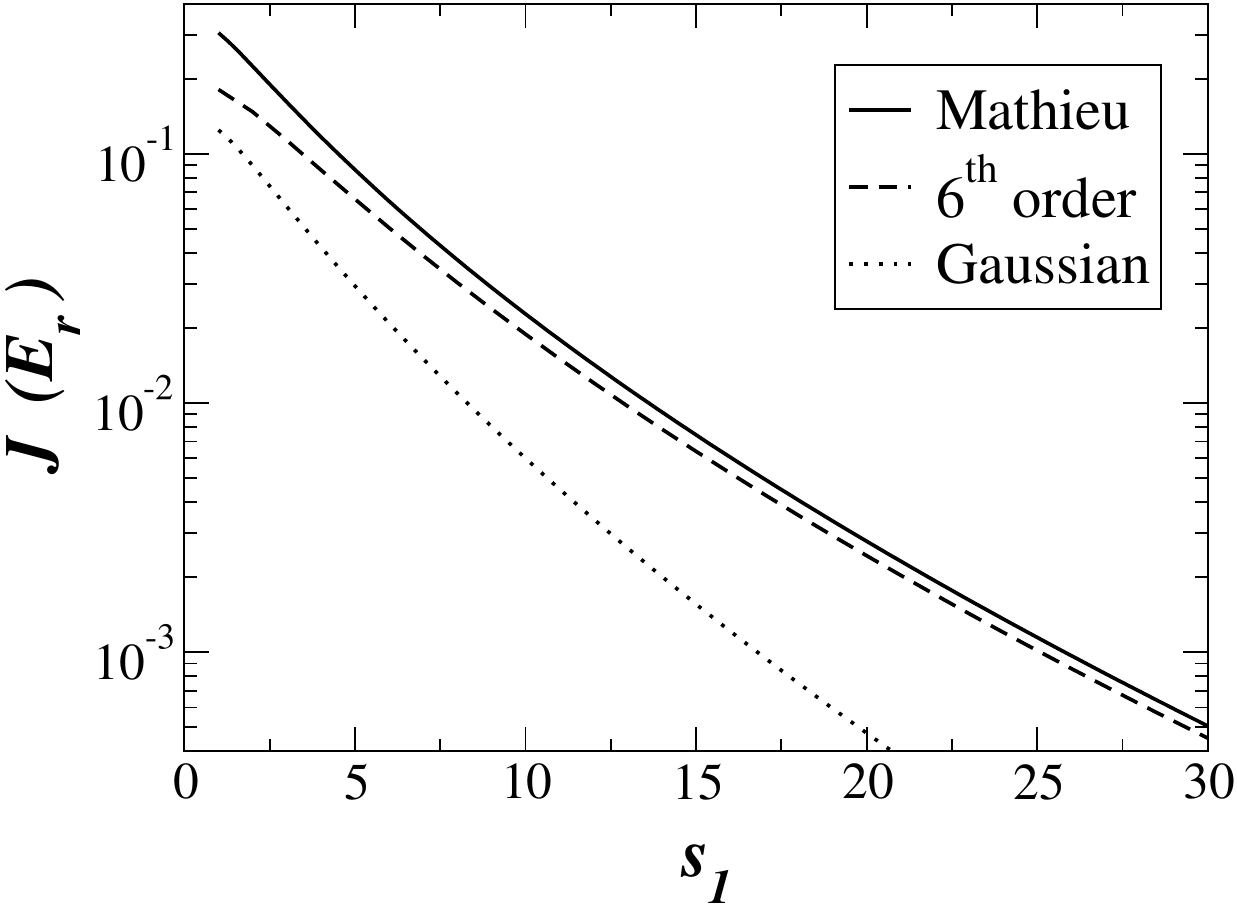} 
\caption{Hopping amplitude $J$ as a function of the 
lattice potential depth $s_1$  
for a monochromatic optical lattice (both quantities are expressed
in units of the recoil energy $E_r$). We show the estimate of $J$
obtained through the solution of the Mathieu equation\cite{Zwerger02}, 
$J/E_r = (4/\sqrt{\pi})~ s_1^{3/4} \exp(-2\sqrt{s_1})$,
through the calculation of the overlap integral Eq.~\eqref{e.overlap}
via 6$^{\rm th}$-order expansion of the optical potential
(see text), and through the simple Gaussian approximation
for the Wannier functions,
$J/E_r = (\pi^2/4-1)~s_1~\exp[-(\pi^2/4)\sqrt{s_1}]$.}
\label{f.Jestimate}
\end{center}
\end{figure}

\begin{figure}[h]
\begin{center}
\includegraphics[
     width=55mm,angle=270]{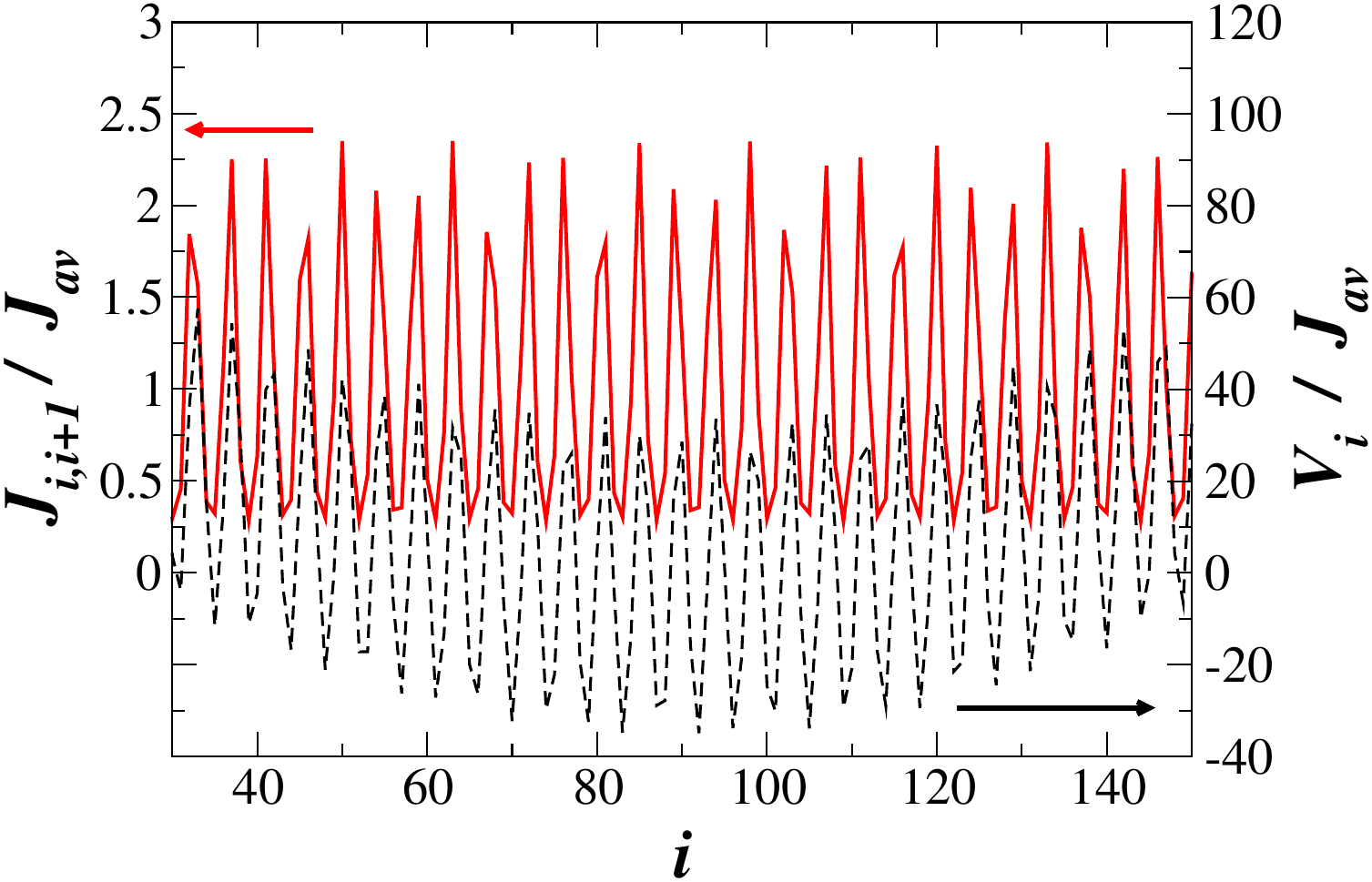} 
\caption{(Color online) Site-dependent hopping amplitude
$J_{\rm i, i+1}$ and local potential $V_i$ for
the generalized Bose-Hubbard model, Eq.~\eqref{e.genBH},
derived from the microscopic Hamiltonian for bosons in an
incommensurate optical superlattices.
Here $s_1 = 8$, $s_2=3$ and $m\omega_{\parallel}^2/2=0.0012 E_r$.}
\label{f.JV}
\end{center}
\end{figure}

\begin{figure}[h]
\begin{center}
 \includegraphics[
     width=50mm,angle=270]{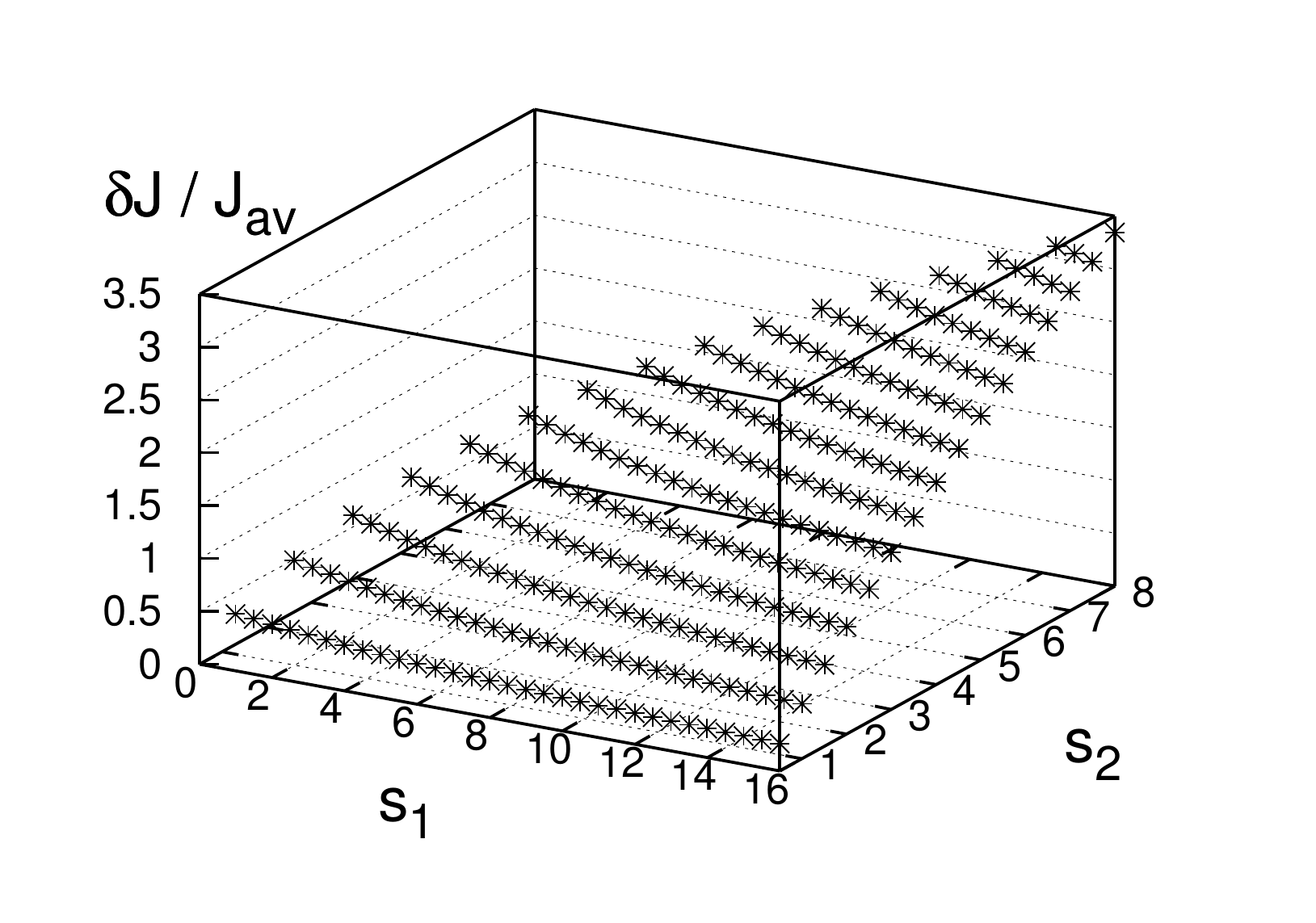} 
 \includegraphics[
     width=50mm,angle=270]{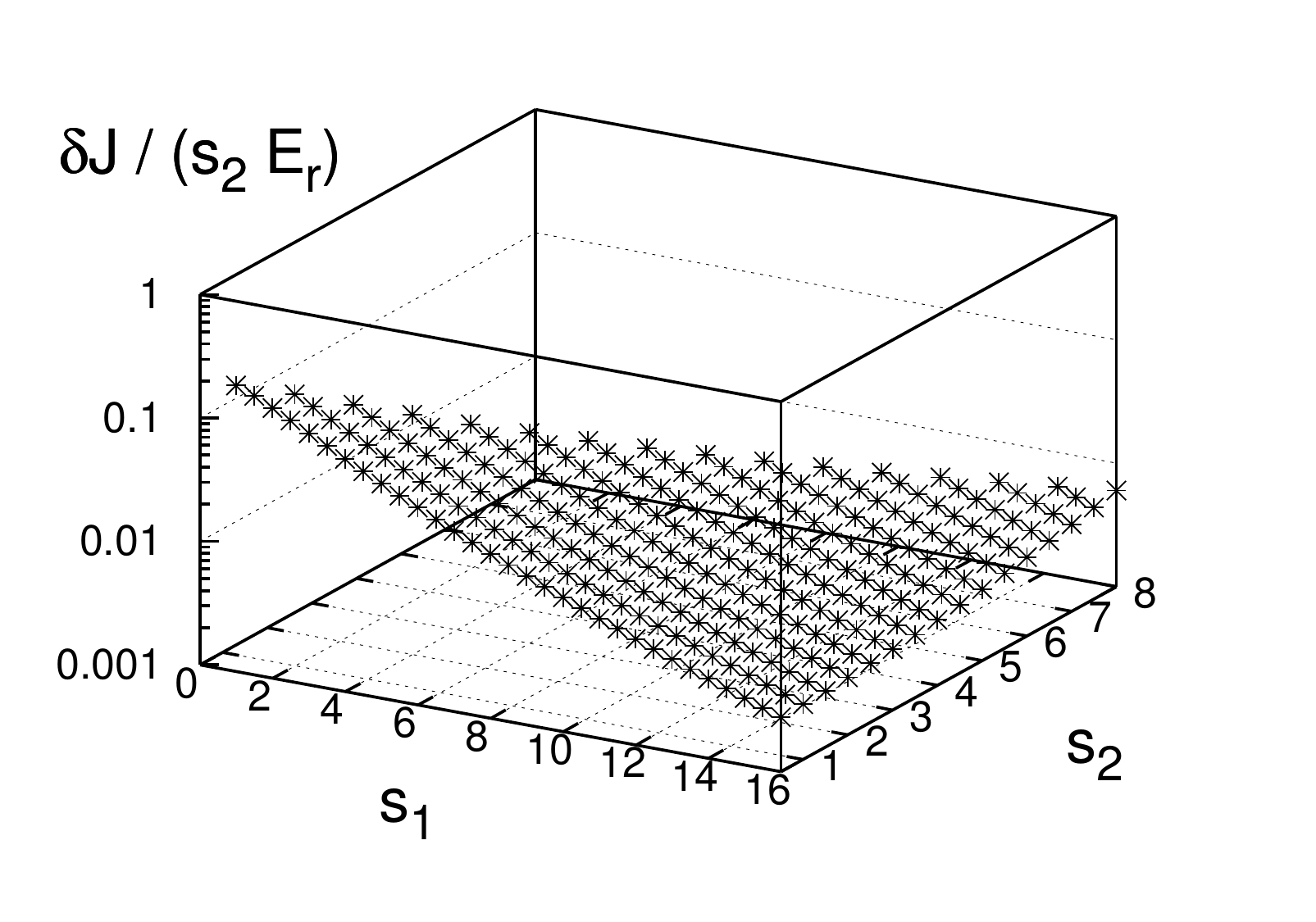}    
\caption{Upper panel: relative fluctuations of the 
effective hopping amplitude $J_{\rm i, i+1}$ for bosons
in an incommensurate superlattices, as a function 
of the intensities $s_1$ and $s_2$ of the two superimposed 
standing waves. Lower panel: ratio of the hopping
amplitude fluctuations over the potential energy 
of the secondary standing wave.}
\label{f.Jfluct}
\end{center}
\end{figure}

  The $W_i$'s are the Wannier functions associated to the lowest band of the 
  periodic potential in the $y$ and $z$ directions, which can be
  conveniently approximated by Gaussians in the limit of a deep 
  optical lattice. Along the $x$ direction Bloch's theorem 
  does not apply due to the incommensuration between the two 
  optical lattice components, so that Wannier functions cannot be 
  properly defined. Hence we take the $w_i$ to be
  the ground state of the potential obtained by a 
  polynomial expansion (up to 6$^{\rm th}$ order) of $V_{||}(x)$  
  around the $i$-th minimum, and the so-obtained set of
  localized functions is further Gram-Schmidt orthogonalized. 
  In absence of the $\lambda_2$ optical lattice, this approach
  provides hopping amplitudes $J_{i,i+1}$ in good agreement 
  with the estimate coming from the solution of the Mathieu equation
  \cite{Zwerger02}, and it improves significantly over a simple 
  Gaussian approximation for the $w_i$'s (see Fig. \ref{f.Jestimate}).
  In the case of the full incommensurate superlattice, this approach
  shows a significant site dependence of the hopping amplitudes
  $J_{i,i+1}$, with differences between different pairs
  of sites that can reach one order of magnitude for intense 
  secondary lattices. This is due to a significant
  shift in the positions of the minima of the incommensurate 
  optical superlattice with respect to the monochromatic 
  lattice, and to the exponential sensitivity of the 
  hopping amplitude to such shifts. An example of the site-dependence 
  of the hopping amplitudes for an intense secondary lattice is shown
  in Fig. \ref{f.JV}, together with the local 
  energies $V_i$. In particular, it is evident that
  the modulation of the local chemical potential and of the local hopping
  are strongly correlated, namely the amplitude of the hopping is maximum
  where the local energy $V_i$ is also maximum. Hence the modulated
  hopping competes in principle with the localization effects 
  induced by the QP $V_i$'s, given that the local kinetic 
  energy is minimized in proximity of the potential maxima. 
  Yet, looking at the magnitude of the hopping modulation with respect
  to the dominant local potential modulation, it is clear that
  one can still confidently rely on the simple Hamiltonian 
  Eq.~\eqref{e.hamiltonian} to capture the dominant low-energy features 
  of the Hamiltonian Eq.~\eqref{e.secondquant}. Fig.~\ref{f.Jfluct}
  shows the relative fluctuations $\delta J/ J_{\rm av}$
  associated with the modulation of $J_{i,i+1}$ around the average
  value $J_{\rm av}$, and their amplitude $\delta J$ compared with the 
  energy scale of the potential created by the secondary 
  lattice, $s_2 E_r$. Although the 
  relative fluctuations can be extremely big 
  and far exceed 100\%, the energy range
  spanned by those fluctuations becomes negligible 
  with respect to the energy scale of the potential,
  and hence the modulation of the hopping is not 
  expected to alter significantly the behavior of the
  system for strong secondary lattices. As for the 
  site dependence of the interaction $U_i$, we find that 
  it reaches $\sim$ 10\% for the strongest secondary
  lattice considered in Fig.~\ref{f.Jfluct}, so we discard
  it for simplicity in the following.

\begin{figure}[h]
\begin{center}
\includegraphics[
     width=55mm,angle=270]{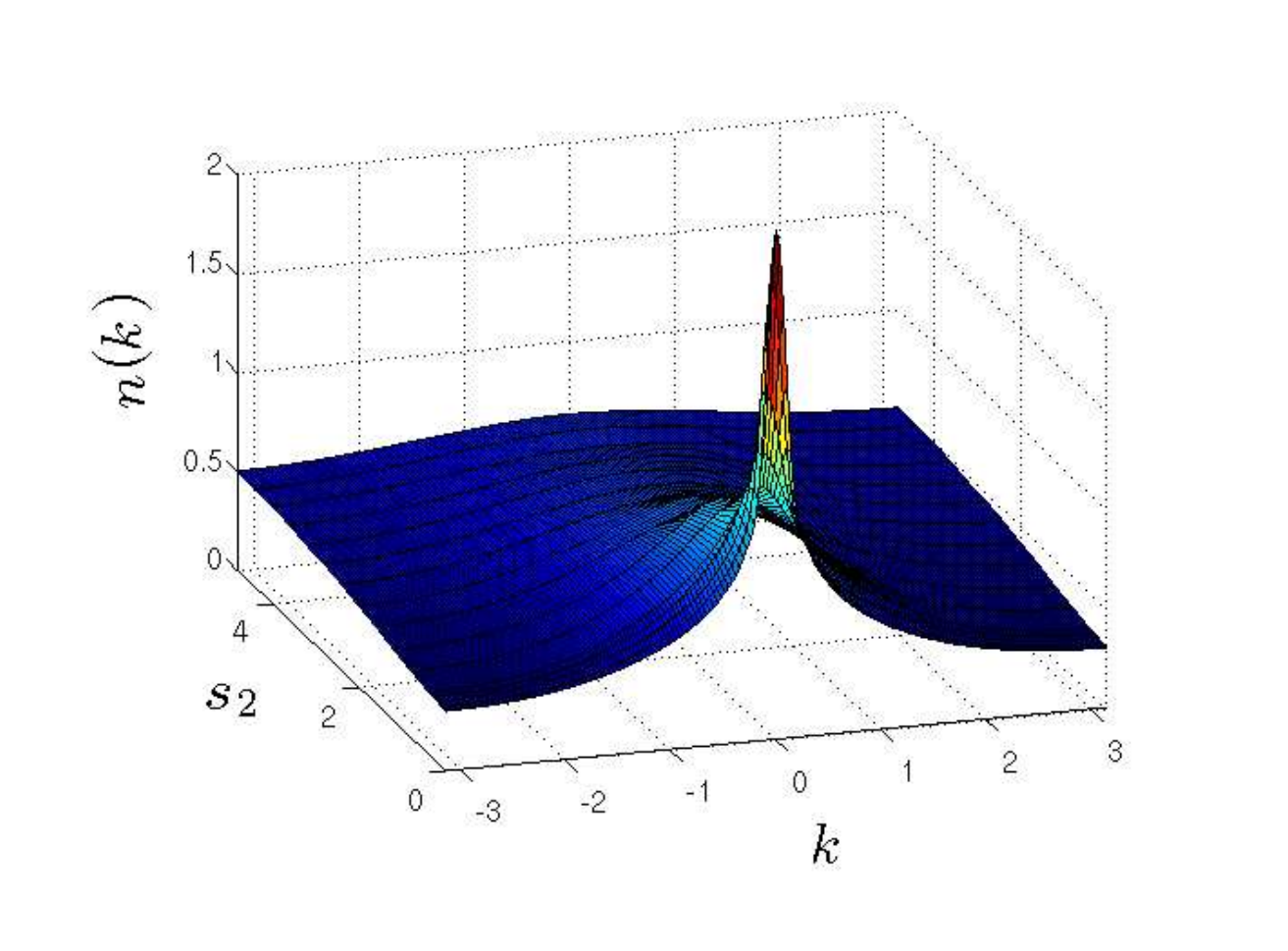} 
     \includegraphics[
     width=55mm,angle=270]{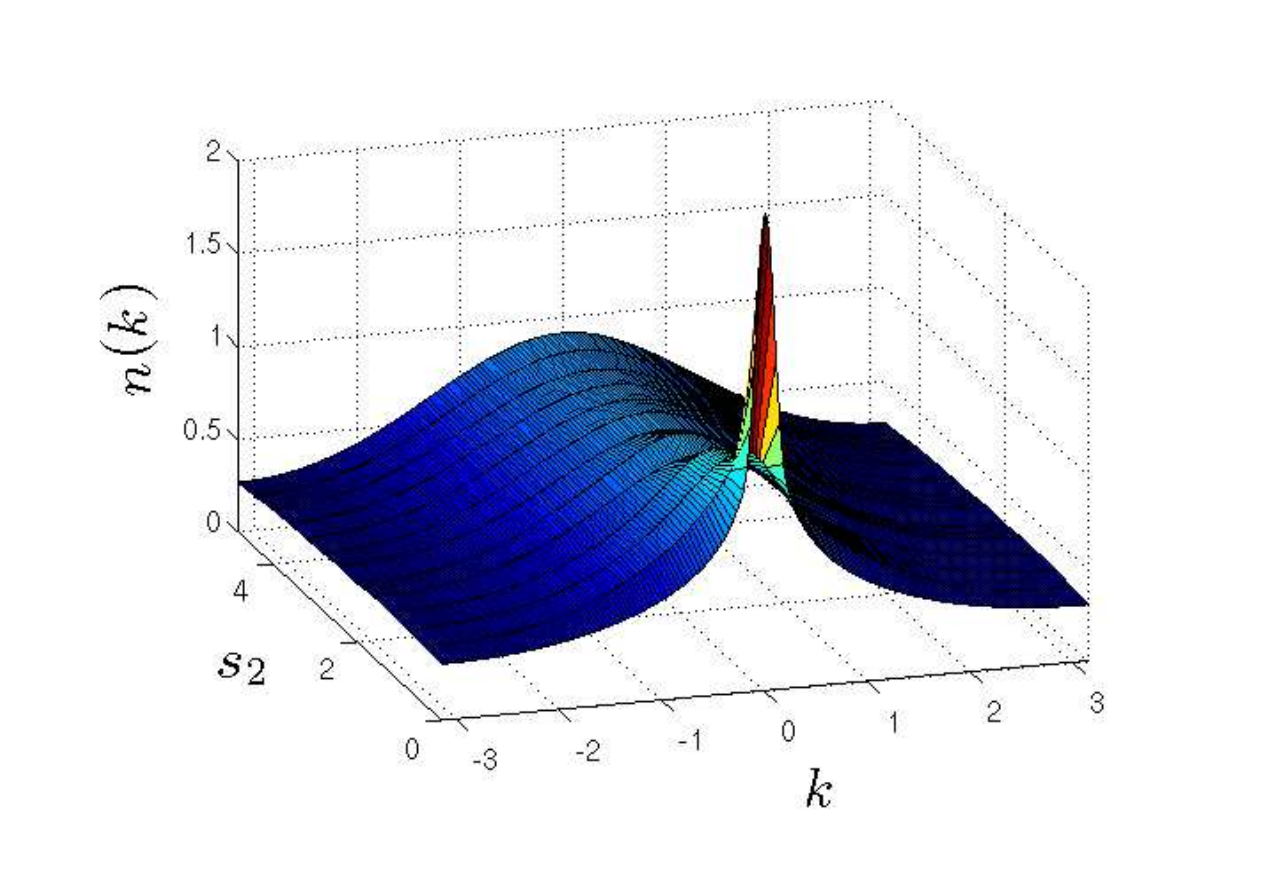} 
\caption{(Color online) Upper panel: momentum distribution of $N=100$ bosons
in an incommensurate superlattice with $s_1=8$ and variable
$s_2$ ($s_{\perp} = 40 E_r$, $a_s = 100 a_0$ as in $^{87}$Rb
with $a_0=$ Bohr radius,
and other parameters as in Fig.~\ref{f.JV}),  
calculated for the generalized Bose-Hubbard Hamiltonian
Eq.~\eqref{e.genBH}.
Lower panel: same quantity for the model Eq.~\eqref{e.genBH} \emph{without}
accounting for the hopping modulation, $J_{i,i+1}=J=$const. }
\label{f.abinitioenk}
\end{center}
\end{figure} 

\begin{figure}[h]
\begin{center}
\includegraphics[
     width=50mm,angle=270]{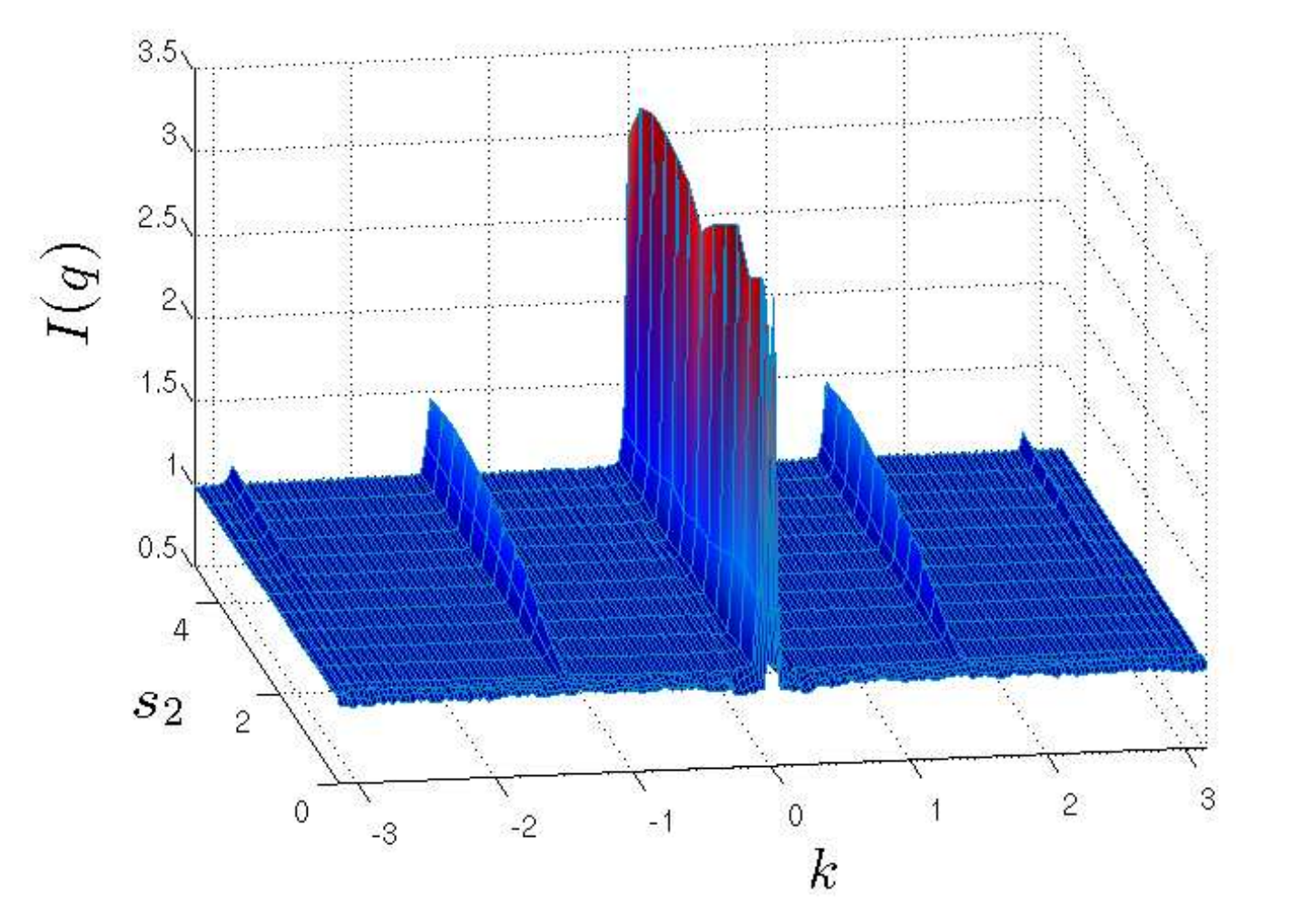} 
     \includegraphics[
     width=50mm,angle=270]{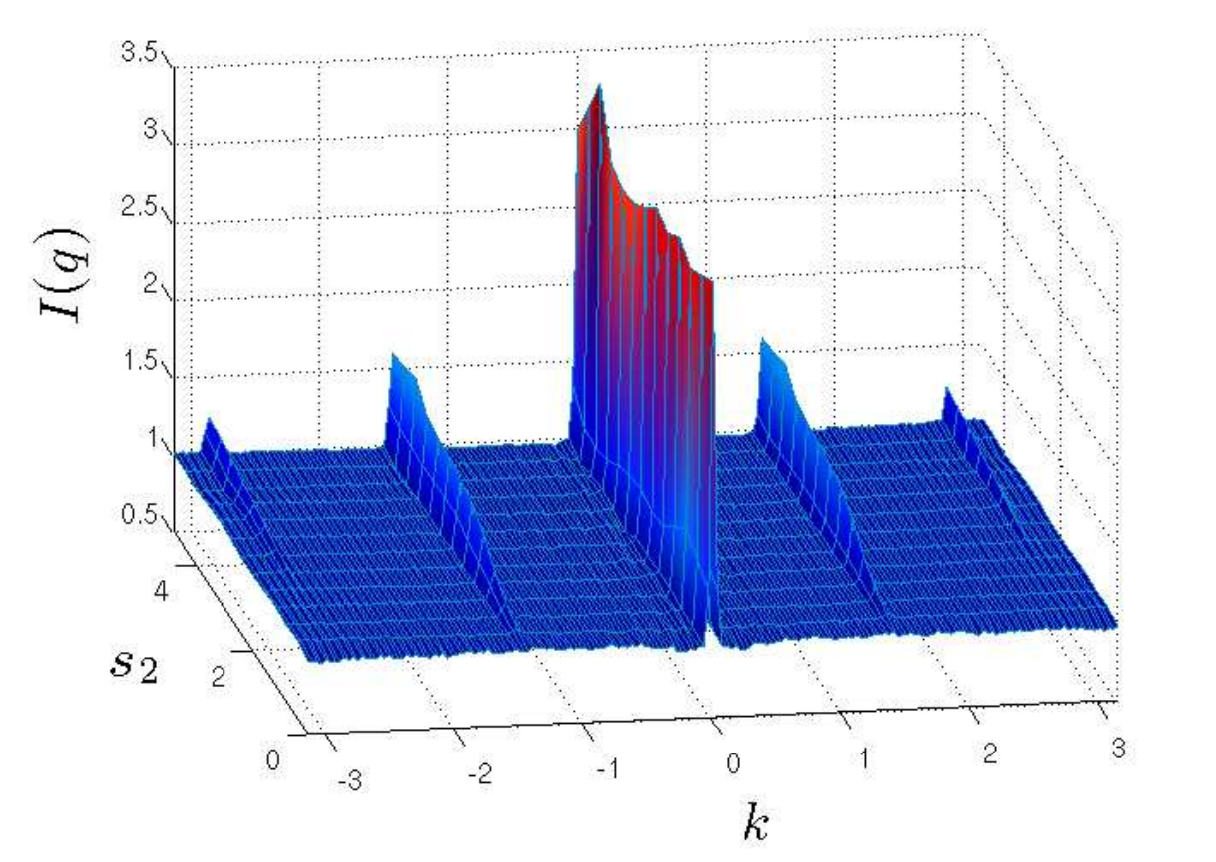} 
\caption{(Color online) Upper panel: noise correlations for  
bosons in an incommensurate superlattice, same parameters as in 
Fig.~\ref{f.abinitioenk} (upper panel).
Lower panel: same quantity for the model Eq.~\eqref{e.genBH} \emph{without}
accounting for the hopping modulation.}
\label{f.abinitioIq}
\end{center}
\end{figure}

   To quantitatively verify that the hopping modulation does 
   not significantly affect the physics of the system, we have 
  studied the first- and second-order coherence of the system with 
  the full \emph{ab-initio} Hamiltonian Eq.~\eqref{e.genBH},
  and compared the results to the case in which we neglect
  the hopping modulation, namely Eq.~\eqref{e.genBH} with
  $J_{i,i+1} = J$ = const. The results are shown in 
  Fig.~\ref{f.abinitioenk} for the momentum distribution
  and in Fig.~\ref{f.abinitioIq} for the noise correlations.
  We observe that taking into account the hopping modulation
  does not alter the most significant qualitative features
  of these observables.
  As for the momentum distribution, we observe that 
  the hopping modulation leads to a stronger suppression
  of coherence for strong QP potentials; in the case of 
  noise correlations,
  we see that the hopping amplitude leads to a reduction
  of the signal at the incommensurate momentum $q=k_{\rm inc}$,
  and even more significantly of the one at $q=2k_{\rm inc}$,
  with respect to the central peak at $q=0$. 
  For the first observation, a possible explanation is that
  the hopping modulation strongly suppresses hopping
  in the minima of the potential, and hence the residual
  coherence of the particles trapped in those minima. 
  As for the second observation, a reduction of the signal
  in the noise correlations is also compatible with a 
  reduction of quantum fluctuations inside the 
  potential wells, as observed in Section \ref{sec.observables}
  when taking the classical limit $J\to 0$ (see Fig.~\ref{f.IqU10}).

\section{Summary and Conclusions}
\label{sec.summary}
 In summary, we have presented an extensive
numerical study of the physics of interacting lattice 
bosons in a quasi-periodic (QP) potential,
in direct connection with recent experiments 
on cold bosons in incommensurate
optical superlattices
\cite{Fallanietal07}.
 
We have investigated
the exactly solvable limit of hard-core
repulsive bosons, where the system can be 
mapped onto non-interacting fermions, so that the
well-known physics \cite{AubryA79, Sokoloff85}
of one-dimensional particles
in an incommensurate superlattice directly
applies to the many-body system. There the 
emergence of gaps in the single-particle 
spectrum translates into the appearence of
a gapped incommensurate band insulator (IBI) phase 
in the many-body phase diagram, together with
the gapless Bose glass (BG) phase. This picture
survives also in the more realistic case of 
softcore interactions, for which the intricate
phase diagram is mapped out in the case of 
a strong QP potential (equalling in strength
the inter-particle repulsion). In particular,
making use of a 
generalized atomic-limit approach, 
the appearence of gaps in the many-body 
spectrum is quantitatively ascribed to
quantum finite-size effects, due to the 
finiteness of the potential wells in the 
QP potential. 

 We have furthermore discussed the behavior of
the system in a QP potential plus a parabolic
trap, as required for a meaningful comparison
to the experiments. The 
concept of a \emph{local phase}, which 
is at the heart of the local-density 
approximation (LDA),
applies successfully to a system without 
(pseudo-)disordered potentials exhibiting homogeneous
MI and SF phases, but this
concept fails to apply to the case of a
BG, whose fundamental signature, 
a finite local compressibility,
comes from localized gapless excitations 
which most likely do not appear in a single realization
of the trapped system. A finite local compressibility   
can only be attained on average over the various
local potential realizations (namely over different
spatial phase shifts). Nonetheless 
the LDA picture for a local BG phase
might be too restrictive, namely non-trivial
low-energy particle-hole excitations might appear
in the trap which are not present in the bulk 
system, due to accidental quasi-degeneracies
of particle and hole states between different
regions of the trap. 

 The above results show that the behavior
of the Bose-Hubbard model in a QP potential 
is more complex than that of more widely
investigated dirty-boson
systems \cite{GiamarchiS88,Fisheretal89}, 
and that in this respect
a QP potential applied to the Bose-Hubbard
model cannot be ascribed to the family
of short-range correlated disorder potentials,
given that correlations in such a potential are 
actually \emph{long-ranged}. 
Both at the single-particle level
and at the many-body level a QP potential leads 
to a behavior which is intermediate between that of 
a system in a random potential and that
in a commensurate periodic potential: indeed
the potential's spatial features albeit aperiodic, 
have a definite upper bound in length scale,
which introduces in turn a characteristic energy scale
in the system. Such energy scale shows up 
in the opening of gaps at incompressible 
regions of the phase diagram.
 
If a second QP potential at a different frequency is 
superimposed to the first one, preliminary results
show that the picture of a short-range correlated 
random potential is recovered,
as the resulting physics of the Bose-Hubbard model
in such a potential appears to mimic that
of the same model in a truly random potential.
In presence of parabolic
confinement, we have moreover shown that
a 2-color QP potential can realize a BG phase 
\emph{all over the trap} on
average over spatial phase fluctuations of the two
potential components.
 Hence the gap between the single-color QP potential
and an ideal (white-spectrum) random potential
can be possibly bridged by enriching the 
QP potential with uniquely one more color. 
This conclusion has immediate practical 
consequences for the ongoing effort of realizing
analog quantum simulators of correlated systems in a 
random environment within an optical-lattice setup, 
because the experimental requirement to bridge
the gap between pseudo-disorder and true
disorder could be in practice the addition 
of one or a few extra laser standing waves.     

 A final, delicate aspect is the identification of an 
 experimental signature for the occurrence
of a BG insulating phase. This issue appears
unresolved by the currently accessible observables, \emph{e.g.} through
time-of-flight measurements. This is intrinsically due to the 
fact that a BG is not characterized by correlations,
which are generally akin to those of more conventional 
insulators, but by the spatial and energetic structure of
excitations. A method to resolve the nature of 
these excitations, based on measuring the system's
response upon variations of the trapping frequency,
will be presented in a forthcoming publication
\cite{Roscildeprep08}.

   \section{Acknowledgements}
  I would like to thank G.G. Batrouni, J.I. Cirac, and S. Wessel
 for useful exchange, and particularly   
 L. Fallani, C. Fort, V. Guarrera, M. Inguscio and
 J. Lye for enlightening discussions on the experimental
 subtleties. Part of the calculations presented in this
 work have been performed on the HPC cluster at the
 University of Southern California; I thank S. Haas
 for constant support.  I would also like to thank
 the Institut ``H. Poincar\'e" in Paris for hospitality
 within the program ``Gaz Quantiques" of 2007, during
 which part of this work was completed. 
 This work is supported by the E.U. through the 
 SCALA integrated project.


\begin{thebibliography}{99}
\bibitem{explanation} The change from periodic to anti-periodic
boundary conditions can always be absorbed into the twist
of $\pi$ of a single bond in the Hamiltonian, as shown 
in Eq.~\eqref{e.fermiham}.
\bibitem{upperbound} In fact $\Delta E_{\rm p-h}$
overestimates in principle the interaction energy,
because it contains the interaction of an
extra particle with a background of $N$ particles,
instead of $N-1$ as in a system with fixed particle
number $N$.
\bibitem{footnote} More quantitatively, 
 a 1D quasi-condensate should 
 behave as a Luttinger liquid, for which the
 coherent fraction $n(k=0)$ scales with the number of 
 particles as $N^{1-K/2}$, where $K$ is the Luttinger
 exponent; the well-known 
 upper bound \cite{GiamarchiS88} for $K$ in a 
 disordered potential, $K<2/3$, imposes
 that the coherent fraction cannot grow 
 slower than $N^{2/3}$. Hence finding
 a coherent fraction $n(k=0) \ll N^{2/3}$
 is a strong indication of absence of quasi-condensation.
 In the considered example, we have $\langle N \rangle \approx 127$,
 $\langle N \rangle^{2/3} \approx 25$
 and $n(k=0) \approx 2.2$ (see Fig.~\ref{f.3colortrap}).
\end{thebibliography}
\end{document}